\documentclass{bmcart}
\RequirePackage[authoryear]{natbib}

\usepackage{amsthm,amsmath}
\allowdisplaybreaks[1]

\usepackage[utf8]{inputenc}
\usepackage{epsfig}

\begin{document}

\begin{frontmatter}

\begin{fmbox}
\dochead{Research}
\title{Technologies for supporting high-order geodesic mesh frameworks for computational astrophysics and space sciences}

\author[
   addressref={aff1},
   corref={aff1},
   email={vaf0001@uah.edu}
]{\inits{V}\fnm{Vladimir} \snm{Florinski}}
\author[
   addressref={aff2},
]{\inits{DS}\fnm{Dinshaw S} \snm{Balsara}}
\author[
   addressref={aff2, aff3}
]{\inits{S}\fnm{Sudip} \snm{Garain}}
\author[
   addressref={aff4}
]{\inits{KF}\fnm{Katharine F} \snm{Gurski}}

\address[id=aff1]{
  \orgname{Space Science Department, University of Alabama in Huntsville},
  \postcode{35899},
  \city{Huntsville, AL},
  \cny{USA}
}
\address[id=aff2]{
  \orgname{Physics and ACMS Departments, University of Notre Dame},
  \postcode{46556},
  \city{Notre Dame, IN},
  \cny{USA}
}
\address[id=aff3]{
  \orgname{Korea Astronomy and Space Science Institute},
  \postcode{34055},
  \city{Daejeon},
  \cny{Republic of Korea}
}
\address[id=aff4]{
  \orgname{Mathematics Department, Howard University},
  \postcode{20059},
  \city{Washington, DC},
  \cny{USA}
}

\end{fmbox}

\begin{abstractbox}

\begin{abstract}
Many important problems in astrophysics, space physics, and geophysics involve flows of (possibly ionized) gases in the vicinity of a spherical object, such as a star or planet. The geometry of such a system naturally favors numerical schemes based on a spherical mesh. Despite its orthogonality property, the polar (latitude-longitude) mesh is ill suited for computation because of the singularity on the polar axis, leading to a highly non-uniform distribution of zone sizes. The consequences are (a) loss of accuracy due to large variations in zone aspect ratios, and (b) poor computational efficiency from a severe limitations on the time stepping. Geodesic meshes, based on a central projection using a Platonic solid as a template, solve the anisotropy problem, but increase the complexity of the resulting computer code. We describe a new finite volume implementation of Euler and MHD systems of equations on a triangular geodesic mesh (TGM) that is accurate up to fourth order in space and time and conserves the divergence of magnetic field to machine precision. The paper discusses in detail the generation of a TGM, the domain decomposition techniques, three-dimensional conservative reconstruction, and time stepping.
\end{abstract}


\end{abstractbox}

\end{frontmatter}


\section{Introduction}
\label{sec_intro}
Objects in the universe tend to assume a spherical shape owing to the central nature of the gravitational force. Common examples include globular star clusters, stars and stellar-like objects, planets, and the larger planetary satellites. Modeling such objects' interior, surface, or atmospheric processes is most conveniently done in a spherical coordinate system because it is perfectly adapted to the shape of the object. A three-dimensional spherical coordinate system has radial distance from the center of the sphere as one of its coordinates. In a spherical \textit{polar} coordinate system the two remaining coordinates are the polar angle, or co-latitude, and the azimuthal angle. Implementing a computational mesh based on the polar spherical system incurs only a modest increase in algorithmic complexity compared with Cartesian meshes because both meshes are logically orthogonal. Unfortunately, this simplicity comes at a price: spherical polar meshes have a singularity on the polar axis where the planes of constant azimuth converge to a single line. As a result the sizes of the computational zones become progressively smaller toward the poles. A polar mesh therefore provides a very non-uniform coverage of the surface of the sphere, which is a highly undesirable property. Because the time step used in a simulation is proportional to the smallest dimension of the zone, a simulation based on a polar mesh is quite inefficient.

Polar singularities can be avoided by using a composite mesh, consisting of multiple partially overlapping patches of structured mesh, where each patch is singularity free \citep{phillips_numerical_1959, browning_comparison_1989, kageyama_yin-yang_2004, feng_three-dimensional_2010, usmanov_three-dimensional_2012}. In this approach the different meshes must be synchronized in their regions of overlap, which involves interpolation and could result in a loss of accuracy or conservation. Another approach, first introduced in the work of \citet{sadourny_integration_1968}, uses a mesh that covers the surface of the sphere without gaps or overlaps, known as a tesselation. Each ``tile'' in the tesselatation is a spherical polygon such as a triangle, a quadrilateral, a pentagon, or a hexagon. The lines connecting adjacent vertices on the sphere are usually (but not always) great circle arcs, which are geodesic lines on the sphere (hence the name, ``geodesic mesh''). A well chosen tesselation method can provide a nearly uniform coverage of the surface of the sphere which greatly improves computational efficiency.

A geodesic mesh is constructed from a regular polyhedron (Platonic solid) inscribed inside a sphere used as a template. The most common method of generating such a mesh is to project the edges of the polyhedron to the sphere and recursively subdivide each spherical polygon into smaller polygonal faces until the desired level of discretization is achieved. A cube can be used to generate a cube-sphere mesh whose faces are quadrilaterals \citep{ronchi_cubed_1996, koldoba_three-dimensional_2002, choblet_oedipus_2007, putman_finite-volume_2007, ivan_high-order_2015, ullrich_arbitrary-order_2015}. Such a mesh is topologically Cartesian within each of the six faces of the cube, requiring special treatment only in the vicinity of the eight corners. It is also possible to construct a mesh out of triangles using an octahedron \citep{feng_novel_2007}, dodecahedron \citep{nakamizo_development_2009}, or an icosahedron \citep{giraldo_lagrange-galerkin_1997, pudykiewicz_numerical_2006, bernard_high-order_2009} as the base solid. A variation of this approach uses a hexagon based dual tesselation, obtained by replacing the vertices of the triangular mesh with face circumcenters and vice versa \citep{heikes_numerical_1995a, du_voronoi-based_2003, feng_novel_2007, miura_upwind-biased_2007, florinski_magnetohydrodynamic_2013}.

Non geodesic tesselations also exist; one prominent example being the HEALPix mesh used for numerical analysis of astrophysical data on the sphere \citep{gorski_healpix_2005}. For three-dimensional problems the tesselation is extruded radially, producing a three-dimensional spherical geodesic mesh. A 3D mesh based on a geodesic tesselation has a very useful property that some of its faces (the so-called r-faces, see below) are flat, which greatly simplifies the numerical scheme. By contrast, all faces of non-geodesic meshes are curved, making such meshes less convenient for use with 3D problems.

In this paper we describe a powerful new framework for finite volume simulations on a triangular geodesic mesh (TGM) with second, third, and fourth orders of accuracy. At this time the software is developed to solve MHD problems with up to fourth order of accuracy in space and time, while conserving the divergence of the magnetic field down to machine precision. Several of the underlying numerical algorithms have been previously published and we refer the interested reader to these papers. However, implementation of these algorithms on a geodesic mesh requires a novel perspective. This is because a geodesic mesh possesses properties of both structured and unstructured meshes. A number of innovative techniques need to be brought together in order to efficiently carry out CFD type simulations on TGMs. The goal of this paper is to describe in detail the techniques that enable efficient implementation of MHD algorithms on spherical geodesic meshes.

\section{Mesh construction}
\label{sec_mesh}
The choice of spherical polygons used to tile the sphere consists of triangles, quadrilaterals, and hexagons (with a small mix of pentagons), but not all combinations result in a high quality mesh. It is desirable to have a mesh that is both highly uniform (or isotropic) and nestable. The first property demands that the faces should be approximately of the same shape and size, while the second ensures strict parent-child relationship between the recursive subdivisions, which is a critical property for domain decomposition (and hence efficient parallelization) as well as adaptive refinement. A regular polyhedron is perfectly uniform: the edges are all of equal length, the faces have the same area, and the vertex angles are the same (see the upper left panel in Figure \ref{fig_ico}). However, the very first subdivision breaks this perfect symmetry because the four daughter faces are of a slightly different shape and size. For example, in a triangular mesh shown in the lower left panel of Figure \ref{fig_ico} the daughter face in the middle of the parent face is slightly different in size from the three daughter faces at the corners. Consequently, higher division meshes are somewhat less uniform that those at lower division. This departure from uniformity is greatest near the vertices of the base polyhedron. In addition, the uniform connectivity of the mesh is violated near these singular points. As an example, consider a mesh constructed from a base hexahedron with quadrilateral faces (i.e., the cube sphere). While commonly each vertex is shared by four faces, only three meet at the eight singular points. As a result the quadrilaterals adjacent to these vertices are diamond shaped, rather than square.

The mesh described in this paper is constructed from an icosahedron and has triangular shaped faces. The upper right panel in Figure \ref{fig_ico} shows that there are twelve singular points in this mesh, where five triangles meet instead of the usual six, but the anisotropy so introduced is not as prominent because the defects are distributed over a larger number of sites. This is the reason that an icosahedron produces a superior mesh compared to a tetrahedron or an octahedron. A dodecahedron can in principle be used, but it lacks a division 0 triangular tesselation, consisting instead of pentagons, and is less convenient for practical use. A hexagonal mesh like that used by \citet{florinski_magnetohydrodynamic_2013} has good uniformity, but is not nestable.

Construction of a TGM begins with inscribing an icosahedron inside a sphere (in the rest of this paper we will always assume that the sphere has a unit radius, unless stated otherwise) and centrally projecting its edges to the surface of the sphere, see the top row of Figure \ref{fig_ico}. This projection generates a \textit{division} 0 tesselation that includes 12 vertices, 20 triangular faces, called \textit{t-faces} and 30 edges, called \textit{t-edges} (these names are chosen to distinguish them from the faces and edges oriented in the radial direction produced by the radial extrusion of the mesh that bear the prefix ``r''). For the sake of efficiency, all calculations on the sphere are performed in Cartesian coordinates using vector operations on the vertices. The input to the mesh generator consists of the coordinates of the icosahedron's vertices, vertex-vertex (VV) neighbor information, and face-vertex (FV) connectivity information.

At each division, the complete mesh connectivity information is computed and stored. For vertices, this includes the list (VV) of six neighbor vertices (five at division 0), six(five) t-edges meeting at the vertex (VE) and six(five) t-faces sharing the vertex (VF). For edges, connectivity information includes the two vertices at the ends (EV) and the two t-faces sharing this t-edge (EF). Finally, for faces we compute the list of three vertices at the corners (FV), the list of three edges (FE) and the list of three face neighbors (FF), for the total of eight connectivity tables. Table \ref{tab_connectivity} shows the order of connectivity table generation and the methods used for construction. Note that at division zero the VV and FV information is already available and steps 1 and 3 are therefore omitted. To facilitate search operations FV, FE, FF, and VF lists are ordered in the counter-clockwise direction, while the remaining tables are not ordered. None of the steps of the mesh generation process require a full search, and the algorithm is linear in the number of elements.

To produce a division 1 tesselation shown in Figure \ref{fig_ico} (bottom-left) new vertices are inserted at the midpoints of division 0 edges. These vertices are then connected with new edges (great circle arcs) that divide each spherical triangle into four smaller triangles. The process is repeated until the desired level of refinement is achieved. It can be easily verified that the number of vertices, edges, and faces in the tesselation at division $d$ are
\begin{equation}
N_v(d)=2+10\times 2^{2d},\quad N_e(d)=30\times 2^{2d},\quad N_f(d)=20\times 2^{2d}.
\end{equation}
It should be pointed out that the mesh construction algorithm described above is not restricted to icosahedral meshes, but can in principle start with any one of the five Platonic solids. Only steps 1 and 3 in Table \ref{tab_connectivity} need to be adjusted. This property permits writing highly modular geodesic mesh generation algorithms for the sphere.

The nonuniformity of the mesh can be assessed by computing the ratios between the largest and the smallest measurement of edge lengths, vertex angles, and face areas. A high quality mesh would have these ratios as close to unity as possible. Table \ref{tab_properties} documents the properties of triangular icosahedral tesselations at divisions zero through eight. Note that the ratios quickly converge to their asymptotic values. The largest face is only 30\% larger than the smallest face, so the disparity in zone sizes will not noticeably affect the time step. Figure \ref{fig_icohex} compares the geometric properties of the icosahedral TGM and the hexahedral quadrilateral geodesic mesh (QGM), also known as the gnomonic cube sphere. Shown are the edge, angle, and area largest-to-smallest ratios that should be a close to unity as possible.  One can see that the icosahedral mesh has superior uniformity of every property compared with the QGM.

The simple mesh does have a few deficiencies, mainly related to the fact that the centroids of the faces are distinct from the circumcenters, as pointed out by \citep{heikes_numerical_1995b}. Several numerical optimization algorithms have been proposed to improve the mesh, including the spring dynamics model \citep{tomita_shallow_2001} and the centroidal generation algorithm \citep{du_centroidal_1999}. Numerical optimization methods usually improve a certain mesh property at the expense of another. For example, an algorithm could trade face area uniformity for vertex angle disparity. Another problem with numerically modified meshes is that the optimization process is specific to each division and the resulting meshes lose their nestable property, i.e., become unsuitable for mesh refinement \citep{putman_finite-volume_2007}. Because we anticipate such development in the future, and because we have not observed any adverse effects from using the simple recursive mesh, it is our preferred method of construction.

The triangular tesselation is extruded radially over a number of concentric spherical layers called shells, to produce the three-dimensional TGM. The software stores the reciprocal connectivity tables for every element on the sphere (vertex, edge, or face) at all divisions, up to the maximum allowed. In addition, there are tree structures describing the parent-child relationships between the faces. For the purpose of domain decomposition, a face subdivided into higher division faces is called a \textit{sector} and a layer of consecutive shells is called a \textit{slab}. An intersection between a sector and a slab is called a \textit{block}, which is the computational unit on this mesh. Each computational zone has the shape of a truncated triangular pyramid also known as a \textit{frustum}.

Locating an arbitrary vector (i.e., finding the zone containing the vector) on the TGM follows a simple procedure valid for any nested polyhedral tesselation. Once the shell number has been determined (via a mapping function or bisection search), the vector is normalized to unity. The nearest division 0 vertex is found by computing the largest scalar product with all 12 vertices at that division. Next, the algorithm tests which of the five surrounding t-faces the vector belongs to, and then recursively tests the four daughter faces at each division. A test for the t-face interior consists of computing the triple products of the vector with two consecutive vertices (1-2, 2-3, and 3-1). If all three triple products are positive, the point belongs to the interior of the t-face with counter-clockwise vertex ordering.

Partitioning the mesh into sectors and slabs enables efficient domain decomposition and offers many opportunities for parallelization. The software framework uses MPI and MPI-derived libraries and achieves essentially linear weak scaling \citep{balsara_efficient_2019}. We will next concentrate on a single triangular block and describe its partitioning into computational zones, generating stencils, and performing reconstruction of zone based mesh variables with a desired order of accuracy.


\section{Grid blocks}
\label{sec_blocks}
The tree numbering system for the faces, edges, and vertices is too slow to be used for zone access within a sector, for which we introduce a flat, two-dimensional ``triangular addressing scheme'', or TAS. The face numbering pattern is illustrated in Figure \ref{fig_faces} which shows one block of a mesh whose sector division $d_s$ is three less than its face division ($\Delta d=d-d_s=3)$. In this example the sector has two layers of ghost zones around its interior. The numbering starts from the base vertex identified by the tesselation; the sector is always drawn in an orientation where the principal vertex is in the SW corner. The first coordinate index runs from W to E and the second index runs from SE to NW. The alternating color shading in Figure \ref{fig_faces} is used to distinguish faces with opposite orientations; many of the vector operations are performed with the opposite signs for the shaded (yellow) and unshaded (white) faces.

The number of vertices, t-edges, and t-faces in a sector with $N_g$ layers of ghost zones are
\begin{equation}
N_v=\frac{(L+1)(L+2)}{2},\quad N_e=\frac{3L(L+1)}{2},\quad N_f=L^2,
\end{equation}
where
\begin{equation}
L=2^{\Delta d}+3N_g
\end{equation}
is the length of the side of the sector. Note that the number of t-edges is three times the number of unshaded faces; it is often convenient to access the edges using a loop on unshaded faces only. The numbering scheme used for the t-edges and vertices is similar to that used for the faces. The edges are numbered in a specific order: first all NE edges, then all NW edges, and finally all S edges (relative to the respective unshaded t-face).

Figure \ref{fig_faces} draws with different colors the boundaries of the blocks of ghost zones used to exchange information with the neighboring sectors. The boundary exchange process is discussed in some detail in Section \ref{sec_impl}. Here we only mention that the grey bordered triangular regions may be absent if the block contains one or more \textit{penta-corners}, which are the vertices of the original icosahedron. These vertices have only five neighbor elements rather than six, and care must be taken to adjust stencil generation procedure and boundary exchanges between blocks near these special points. For example, if the principal vertex of the block shown in Figure \ref{fig_faces} is a penta-corner, t-faces 6, 7, 8, and 13 are absent, and the mesh must be closed along the \textit{cut line} that appears in place of the missing faces.

Grid blocks also maintain a set of local connectivity tables similar to those listed in Table 2. These tables have a very regular pattern and are much simpler to construct than the tesselation tables; all neighbors are ordered in counter-clockwise direction. The t-edge orientation is defined with respect to its unshaded neighbor face, which fixes the directions of the normal and tangent vectors on the mesh.

Each grid block needs to know the coordinates of every vertex in the local grid. Because the tesselation numbers its t-faces and vertices differently from the grid blocks, a routine is provided to assemble a list of vertices that lie in a requested sector with ghost cells in the TAS format. The convention is that the base vertex is the first vertex in the FV set of the sector. The mapping routine walks the sector, including the ghost t-faces, from W to E and from SE to NW, storing the coordinates of the vertices encountered along the path. Three step operators are defined, all relative to the base vertex of the t-face, shown in Figure \ref{fig_steps}. A type 1 step moves from the initial t-face ($t_i$) to the final face ($t_f$) in the S direction and the new base vertex ($v_f$) is to the E of the old base vertex ($v_i$) on the common edge. A type 2 step moves diagonally to the NE, and the new base vertex is opposite to the initial base vertex. Finally, a type 3 step moves to the NW, but the new base vertex belongs to the common edge. In Figure \ref{fig_steps}, the vertex moves are shown with orange arrows and the face moves with red arrows. These three operations apply to unshaded to shaded t-face movement. The shaded to unshaded step operators are algorithmically identical to those, and correspond to switching the initial and the final t-faces and vertices, and reversing the arrow directions.

The sector walk routine works as follows. From the base vertex of the sector, the code first walks to the NW until it encounters the left side of the block (t-face 25 in Figure \ref{fig_faces}). Then the code walks to the SW until it reaches the corner of the grid block (face number 1 in the grid block's numbering scheme). From there, the code makes a step to the right followed by $i$ steps diagonally (SE-NW), where $i$ is the index of the horizontal step. That way every cell in the block is visited once. Note that the alternating pattern of shaded and unshaded t-faces is broken across the cut line, and special versions of the step operators are needed to move between the faces of the same shading.


\section{Representing spherical geometry}
\label{sec_geom}
In principle, it is possible to perform all calculations on a TGM by directly using spherical geometry. We found, however, that using isoparametric mapping from a reference zone, which in this case is a right triangular (equilateral) prism, offers significant advantages. In particular, integration on spherical triangles is difficult, requiring a large number of quadrature points at higher orders \citep{beckmann_quadrature_2012}. Integration on the reference element is straightforward by comparison.

The physical zone and its reference image are shown in Figure \ref{fig_prism}. The left panel shows the physical zone that has the shape of a truncated triangular pyramid, also called a frustum. The spherical top and bottom caps are the t-faces, and the annular sides are the r-faces. The frustum therefore has three r-faces and two t-faces. The edges of the t-faces are called t-edges, and the edges connecting the bottom and top t-faces are called r-edges. There are six t-edges, three r-edges, and six vertices per zone. The vertices belonging to a t-face are numbered counter-clockwise in its connectivity tables, 1 through 3, and the t-edges of each t-face are also numbered counter-clockwise, 1 through 3. By convention, a vertex has the same index as the opposite t-edge.

A point in reference space is addressed with a coordinate triplet ($\xi$, $\eta$, $\zeta$). The bottom and the top faces of the prism lie in the planes $\zeta=0$ and $\zeta=1$, respectively. The area of the t-face in reference coordinates is $\sqrt{3}/4$, the area of the r-face is one, and the volume of the prism is $\sqrt{3}/4$. It is convenient to work with barycentric coordinates in the $\xi\eta$ plane ($\Lambda_1$, $\Lambda_2$, $\Lambda_3$), defined in Chapter 8 of \citep{zienkiewicz_the_2013} as
\begin{eqnarray}
\xi&=&\Lambda_1\xi_1+\Lambda_2\xi_2+\Lambda_3\xi_3, \nonumber \\
\eta&=&\Lambda_1\eta_1+\Lambda_2\eta_2+\Lambda_3\eta_3, \\
1&=&\Lambda_1+\Lambda_2+\Lambda_3, \nonumber
\end{eqnarray}
where $\xi_i$ and $\eta_i$, $i=1,2,3$, are the $\xi$ and $\eta$ components of the vertices of the triangle in the reference space. The barycentric coordinates are equal to the partial areas of the sub-triangles formed by the point $(\xi, \eta)$ and the three vertices of the reference triangle (please note that this is not true for the areas of the respective curved triangles). For the equilateral reference triangle the inverse of (4) is
\begin{eqnarray}
\Lambda_1&=&1-\xi-\frac{\eta}{\sqrt{3}}, \nonumber \\
\Lambda_2&=&\xi-\frac{\eta}{\sqrt{3}}, \\
\Lambda_3&=&\frac{2\eta}{\sqrt{3}}, \nonumber
\end{eqnarray}

We next introduce a set of two-dimensional linearly independent Lagrange basis functions associated with the nodal points on the curved triangular faces that fix the mapping from reference space to the physical space. It is convenient to compute the nodal point coordinates on the unit sphere; the physical coordinates are obtained simply by rescaling to the desired radial distance. We denote vectors that lie on the unit sphere with the superscript ``u''. All coordinates are factored as
\begin{equation}
\mathbf{x}(\xi,\eta,\zeta)=r(\zeta)\mathbf{x}^u(\xi,\eta),
\end{equation}
where
\begin{equation}
r(\zeta)=r_b+\zeta(r_t-r_b)=r_b[1+\zeta(\rho-1)],
\end{equation}
where $r_b$ and $r_t$ are the radial distances of the nodal points on the bottom and the top t-face, respectively, and $\rho$ is their ratio. As discussed below, this factoring enables a more efficient implementation of the reconstruction algorithm on the TGM compared with fully unstructured tetrahedral meshes. To perform integration we also require a set of curvilinear unnormalized basis vectors
\begin{equation}
\mathbf{h}_\xi=\frac{\partial\mathbf{x}}{\partial\xi}=r\frac{\partial\mathbf{x}^u}{\partial\xi},\quad\mathbf{h}_\eta=\frac{\partial\mathbf{x}}{\partial\eta}=r\frac{\partial\mathbf{x}^u}{\partial\eta},\quad\mathbf{h}_\zeta=\frac{\partial\mathbf{x}}{\partial\zeta}=(r_t-r_b)\mathbf{x}^u.
\end{equation}

Given $N$ nodal points on a triangle, there are $N$ Lagrange basis functions $\psi_i(\xi,\eta)$ that satisfy
\begin{equation}
\psi_i(\xi_j,\eta_j)=\delta_{ij},
\end{equation}
where $(\xi_j,\eta_j)$ are the coordinates of nodal point $j$ on the unit sphere. A position vector $\mathbf{x}^u$ can be represented as an expansion over the basis functions
\begin{equation}
\mathbf{x}^u(\xi,\eta)=\sum_{i=1}^N\mathbf{v}^u_i\psi_i(\xi,\eta).
\end{equation}
The coefficients in this expansion are the physical coordinates of the nodal points $\mathbf{v}^u_i$. Figure \ref{fig_elements} shows the locations of the nodes on the reference triangle. These elements use $N=3$, 6, and 10 for linear, quadratic, and cubic basis functions, respectively. The explicit formulas for the basis functions on the equilateral triangle are given in Appendix A. Note that the maps from two adjacent t-faces are continuous at the shared t-edge by virtue of the use of barycentric coordinates for their construction.

An expansion similar to (10) is used for the t-edges. Points from the surface element lying on that edge are used (see Figure \ref{fig_elements}) and the corresponding basis functions are simply restriction of the facial bases functions for one of the barycentric coordinate equal to zero. It is convenient to introduce an auxilliary variable $\delta$ that measures distance along the edge in the counter-clocksise direction. Its relation to barycentric coordinates is shown in Table \ref{tab_aux}. The basis functions $\phi_i(\delta)$ for the edges can be also found in Appendix A.

It is instructive to evaluate the disparity between the mapped surface given by Eq. (10) and the ideal surface, i.e., the unit sphere. Below we compute the error in the radial coordinate, $1-r^u$ for a mapped equilateral spherical triangle with a circumcircle radius of $5^\circ$. Figure \ref{fig_maperr} shows the error distribution for element orders one, two, and three. Obviously, the first order element with its planar faces is unable to reproduce the spherical shape resulting in a large error near the center. Switching to the second order element improves the accuracy by three orders of magnitude, while going to third order yields another factor of $\sim 20$. It is evident that both second or third order elements reproduce spherical geometry with remarkable accuracy.

It is worth mentioning that \citet{ivan_multi-dimensional_2013} have previously developed an isoparametric cube sphere model based on a cubic reference element. However, their trilinear mapping anchored at the four corners of the quadrilateral t-face is not capable of truly reproducing a spherical surface because it has only one extra degree of freedom compared with the linear map. For example, when all four vertices lie in the same plane, the trilinear map yields a surface that is flat instead of curved.


\section{Evaluation of integrals on a geodesic mesh}
\label{sec_integrals}
A finite volume scheme requires evaluating multi-dimensional integrals in the initial setup phase and during time updates of the conserved variables. This requires, at a minimum, volume and surface integrals. The use of constrained transport scheme to advance the magnetic field requires, in addition, evaluation of integrals along the edges (see Section 7 below). We will therefore define the following integral operations: volume integration over a zone, surface integration on t-faces and r-faces, and line integration on t-edges and r-edges. For a three-dimensional vector variable $\mathbf{V}$ these are defined as
\begin{equation}
\iiint\limits_\text{zone}\mathbf{V}(\mathbf{x})dV=\iiint\mathbf{V}(\xi,\eta,\zeta)(\mathbf{h}_\xi\times\mathbf{h}_\eta)\cdot\mathbf{h}_\zeta d\xi d\eta d\zeta,
\end{equation}
\begin{equation}
\iint\limits_\text{t-face}\mathbf{V}(\mathbf{x})\cdot\mathbf{dS}=\iint\limits_\Delta\mathbf{V}(\xi,\eta)\cdot(\mathbf{h}_\xi\times\mathbf{h}_\eta)d\xi d\eta,
\end{equation}
\begin{equation}
\iint\limits_\text{r-face}\mathbf{V}(\mathbf{x})\cdot\mathbf{dS}=\int\limits_0^1\int\limits_0^1\mathbf{V}(\delta,\zeta)\cdot(\mathbf{h}_\delta\times\mathbf{h}_\zeta)d\delta d\zeta,
\end{equation}
\begin{equation}
\int\limits_\text{t-edge}\mathbf{V}(\mathbf{x})\cdot\mathbf{dl}=\int\limits_0^1\mathbf{V}(\delta)\cdot\mathbf{h}_\delta d\delta,
\end{equation}
\begin{equation}
\int\limits_\text{r-edge}\mathbf{V}(\mathbf{x})\cdot\mathbf{dl}=\int\limits_0^1\mathbf{V}(\zeta)\cdot\mathbf{h}_\zeta d\zeta,
\end{equation}
where the symbol `$\Delta$' designates integration over a triangle. We will now describe our strategy for evaluating the integrals using quadrature rules. Consider a single zone in the mesh addressed with a t-face index $f$ whose top and bottom vertices lie at $r$ and $\rho r$, respectively. Further, suppose $e$ is the index of one of the t-edges of the zone, and $v$ is one of the vertices.

\subsection{Integration on r-edges}
R-edges are addressed by the vertex index with specified $r$ and $\rho$. Because r-faces are always straight, the integrals can be evaluated directly using Gauss-Legendre quadrature points. Define such set of points on the reference interval $\zeta=[0,1]$ as $Q_{1r}$. Each quadrature point $q$ has position $\zeta_q$ and weight $w_q$. Then it is evident that
\begin{equation}
\int\limits_\text{r-edge}\mathbf{V}(\mathbf{x})\cdot\mathbf{dl}\approx r(\rho-1)\sum_{q\in Q_{1r}}w_q\mathbf{V}(r\zeta'_q\mathbf{x}^u_q)\cdot\mathbf{x}^u_v,
\end{equation}
where $\mathbf{x}^u_v$ is the position of the vertex $v$ on the unit sphere and
\begin{equation}
\zeta'_q=1+(\rho-1)\zeta_q
\end{equation}
is the elevated radial position. The code uses 1 quadrature point for integrating polynomials of degrees zero and one, 2 points for degrees two and three, 3 points for fourth and fifth degree polynomials, etc.

\subsection{Integration on t-edges}
T-edges are addressed by the edge index with fixed $r$. These edges are curved (except when using linear basis functions) and the quadrature weights are therefore multiplied by the Jacobian equal to the length of the tangent vector $\mathbf{h}_\delta$. Again, designate the set of Gauss-Legendre points on the reference interval $\delta=[0,1]$ as $Q_{1t}$ (which may or may not be the same as $Q_{1r}$). Using the definition (8) we can write
\begin{equation}
\int\limits_\text{t-edge}\mathbf{V}(\mathbf{x})\cdot\mathbf{dl}\approx r\sum_{q\in Q_{1r}}w_q\mathbf{V}(r\mathbf{x}^u_q)\cdot\frac{\partial\mathbf{x}^u_e(\delta_q)}{\partial\delta},
\end{equation}
where $\delta_q$ are the locations of the quadrature points on the reference interval and
\begin{equation}
\mathbf{x}^u_q=\mathbf{x}^u_e(\delta_q).
\end{equation}
Here the subscript 'e' refers to the fact that the map specific for edge $e$ is used to evaluate the coordinate and its derivative. We use the same number of points for t-edge integration as for r-edge integration. In practice, the values of the point coordinates and tangent vectors on the unit sphere are precomputed for each t-edge at the start of a simulation for fast retrieval.

\subsection{Integration on r-faces}
R-faces are addressed by the edge index with specified $r$ and $\rho$ and approximate annular regions (trapezoids for elements of order 1). The position is specified via the $(\delta,\zeta)$ pair of coordinates. We now introduce quadrature points on the reference square $(\delta,\zeta)=[0,1]\times[0,1]$ as $Q_{2r}$. These points are conveniently computed as tensor products of the Gauss-Legendre quadrature points. The quadrature rule for r-faces can be written as
\begin{equation}
\int\limits_\text{r-face}\mathbf{V}(\mathbf{x})\cdot\mathbf{dS}\approx r^2(\rho-1)\sum_{q\in Q_{2r}}w_q\zeta'_q\mathbf{V}(r\zeta'_q\mathbf{x}^u_q)\cdot\left(\frac{\partial\mathbf{x}^u_e(\delta_q)}{\partial\delta}\times\mathbf{x}^u_q\right)
\end{equation}
with $\mathbf{x}^u_q$ given by Eq. (19).

The right panel of Figure \ref{fig_quadpts} shows the locations of the points on the reference square. On a rectangle, three points are sufficient for exactly integrating a quadratic polynomial, four for cubic, and six for quartic. However, it is our intention to maintain exact polynomial integration rules for first order elements, where the r-face is a trapezoid. Its Jacobian is linear in the $\zeta$ coordinate, and the order of accuracy is reduced by one. For this reason we use four, six, and nine point rules to integrate polynomials of degrees two, three, and four, respectively.

\subsection{Integration on t-faces}
T-faces are addressed by the face index with fixed $r$. They approximate spherical triangles (flat triangles for linear coordinate transformation). The position is specified via the $(\xi,\eta)$ pair of coordinates, and the set of quadrature points defined on a unit equilateral triangle is designated as $Q_{2t}$. Here we use the symmetric quadrature rules given in \citet{dunavant_high_1985} with quadrature point locations shown in the left panel of Figure \ref{fig_quadpts}. The integration algorithm for t-faces is
\begin{equation}
\int\limits_\text{t-face}\mathbf{V}(\mathbf{x})\cdot\mathbf{dS}\approx r^2\sum_{q\in Q_{2t}}w_q\mathbf{V}(r\mathbf{x}^u_q)\cdot\left(\frac{\partial\mathbf{x}^u_f(\xi_q,\eta_q)}{\partial\xi}\times\frac{\partial\mathbf{x}^u_f(\xi_q,\eta_q)}{\partial\eta}\right),
\end{equation}
where
\begin{equation}
\mathbf{x}^u_q=\mathbf{x}^u_f(\xi_q,\eta_q).
\end{equation}
Three, four, and six points are sufficient to integrate a quadratic, cubic, and quartic polynomial exactly on a flat triangle. The four-point rule should be avoided because it has a negative weight, and we use the six point rule at third order. These points and the normal vectors are also precomputed for each t-face.

\subsection{Integration on frustums}
A frustum can be addressed by the face index with specified $r$ and $\rho$. Defining a position requires all three reference coordinates $(\xi,\eta,\zeta)$. We arrange the quadrature points in $p$ ``planes'', where each plane corresponds to a triangular quadrature rule with a set of points $Q_{2t}$ described in the previous subsection. The planes themselves are located at $\zeta_p$ corresponding to the Gauss-Legendre points on $[0,1]$ that we designate as $P_1$ with the plane weights given by $w_p$. Then a volume integral can be evaluated as
\begin{equation}
\int\limits_\text{zone}\mathbf{V}(\mathbf{x})dV\approx r^3(\rho-1)\sum_{p\in P_1}w_p{\zeta'_p}^2\sum_{q\in Q_{2t}}w_q\mathbf{V}(r\zeta'_p\mathbf{x}^u_q)\left(\frac{\partial\mathbf{x}^u_f(\xi_q,\eta_q)}{\partial\xi}\times\frac{\partial\mathbf{x}^u_f(\xi_q,\eta_q)}{\partial\eta}\right)\cdot\mathbf{x}^u_q
\end{equation}
where $\mathbf{x}^u_q$ is given by Eq. (22). We use two quadrature planes for polynomials of degrees 0 and 1, three for polynomials of degrees 2 and 3 and four for degrees 4 and 5.

In curved spaces the total degree of the reconstruction polynomial increases significantly upon transformation to the reference coordinates. For example, a third degree polynomial in $\mathbf{x}$ on a quadratic surface element gives an integrand or degree $3^2+2=8$ in $\alpha$ and $\beta$, where the Jacobian adds two extra powers. The same polynomial on a cubic element gives an integrand of degree $3^3+4=13$. However, it is quite unnecessary to match the order of the quadrature algorithm to the resulting total degree of the polynomial in the reference space because the truncation error decreases at the rate imposed by the quadrature scheme alone. The magnitude of error depends on the details of the coordinate mapping, but the order of convergence does not.


\section{Conservative reconstruction on a geodesic mesh}
\label{sec_recon}
The TGM framework presented here is intended to be used primarily with finite volume schemes for systems of PDEs. These methods usually operate on conserved (extrinsic) physical variables associated with each zone in the mesh. Conserved variables are advanced in time using the fluxes evaluated at the zone boundaries. The fluxes may be generated by means of a Riemann solver that computes, often approximately, the self-similar wave pattern developed from an interaction of two or more constant states. The Riemann solver may be invoked for a set of points in each face, and the total flux is evaluated as the average over these points. The invocation of multiple Riemann solvers at suitably placed quadrature points within each face of the mesh contributes to the high order accuracy of the scheme.

The constant states fed to the Riemann solver are obtained via high-order spatial reconstruction of the conserved variables, which amounts to finding a functional form of the variable within each zone consistent with a given piecewise distribution at the beginning of the time step. Reconstruction is performed on a set of stencils associated with each zone (the principal zone of that stencil) that include zones in a certain proximity to the principal. We use conservative polynomial reconstruction (known in one-dimensional or directionally split applications as reconstruction via primitive functions) from multiple stencils for each computational zone.

\subsection{Stencil construction}
We now discuss the reconstruction strategy focusing on the TGM specific issues. At the start of a simulation a set of stencils is built for each computational zone. The number of zones in a stencil cannot be smaller than the number of degrees of freedom in the polynomial reconstruction, given by
\begin{equation}
D(M)=\frac{(M+1)(M+2)(M+3)}{6},
\end{equation}
where $M$ is the degree of the reconstruction polynomial. It has been argued that using the number of zones in the stencil equal to $D(M)$ does not always produce satisfactory results \citep{ollivier-gooch_high-order-accurate_2002}. For this reason we use over-determined stencils that are larger than the minimal size. With such stencils the conservative property of the reconstruction is enforced in the least squares sense.

The zones in the TGM are arranged in a regular pattern (see Figure \ref{fig_faces}) allowing us to design universal stencils valid on any mesh. The zones in a stencil are arranged in ``planes'' that correspond to different radial shells. Each plane consists of a two-dimensional t-face stencil. The principal plane contains the largest 2D stencil and the other planes contain progressively smaller stencils. Figure \ref{fig_stshapes} shows the choice of 2D stencils available in the code. In addition to the symmetric central stencil that is used in regions where the solution is smooth (top row of Figure \ref{fig_stshapes}), twelve directional stencils are defined to be used in situations where the central stencil produces a large variation due to a sharp gradient or discontinuity in the solution \citep{kaser_ader_2005}. Directional stencils can point in the in- or outward radial direction and along six directions in a plane (three forward-biased and three backward-biased, see \citet{kaser_ader_2005} for the explanation of these terms). We use three, five, and seven planes per central stencil and two, three, and four planes per directional stencil for $M=1$, 2, and 3, respectively. The code can use all thirteen stencils, but can also be run without backward stencils which nearly halves the execution time of the reconstruction step.

Contrary to the fully unstructured meshes, stencils on the TGM can be generated using pre-defined patterns and in principle need not rely on mesh connectivity information. The exception to this rule are the penta-corners, where some of the neighbors may be missing. Figure \ref{fig_7stencils} shows some examples of stencils in the principal plane that could be used for third order polynomial reconstruction. The central stencil, shown in the top panel, clearly contains a penta-corner. The middle row shows the forward and the bottom row the backward stencils. Notice how the first of the backward stencils has a different shape than the other two. If a stencil is found to be defective (i.e., contains fewer zones than required), the software will repeatedly upgrade to the next largest stencil until the order condition is fulfilled.

Consider a conservative mesh variable $\mathbf{U}$ defined via its averages over each zone $i$, $\bar{\mathbf{U}}_i$. A reconstruction of this variable in zone $i$ using $M$th degree polynomials can be written as
\begin{equation}
\mathbf{U}_i(\mathbf{x})=\sum\limits_{|\alpha|=0}^{D(M)-1}\mathbf{U}_i^\alpha\left(x^\alpha-\langle x^\alpha\rangle_i\right),
\end{equation}
where multi-index notation is used with $\alpha=(\alpha_1,\alpha_2,\alpha_3)$, $|\alpha|=\alpha_1+\alpha_2+\alpha_3$, and $x^\alpha=x_1^{\alpha_1}x_2^{\alpha_2}x_3^{\alpha_3}$. The term $\langle x^\alpha\rangle_i$ denotes the moment of zone $i$, divided by the volume of the zone, and $\mathbf{U}_i^\alpha$ is the coefficient (or mode) in the reconstruction. To enforce the conservation property $\langle\mathbf{U}(\mathbf{x})\rangle_i=\mathbf{U}^{(0,0,0)}_i=\bar{\mathbf{U}}_i$ one must formally set $\langle x^{(0,0,0)}\rangle_i=0$ in (25). The remaining moments are computed using high order quadratures given by Eq. (23). The moments are computed in Cartesian coordinates. These moments are transformed into the center of mass frame of the zone using the parallel axis theorem \citep[for details, see][]{balsara_efficient_2019} and scaled by the characteristic length determined by the dimensions of the zone.

An optimal choice of stencils should achieve a balance between accuracy and performance. To find this balance we have performed a statistical study of error in the reconstruction with cubic polynomials (i.e., at fourth order of accuracy) using only the central stencil, as a function of the number of zones in the stencil. The results, presented in Appendix B, demonstrated that where as the $L_\infty$ error can become unacceptably large for the minimal stencil, the deficiency is cured by increasing the stencil size by as little as 15\%. Past this point, both the $L_1$ and the maximum errors have a weak increasing trend previously noted in \citet{ivan_high-order_2015}. Based on these results, we introduced an adjustable parameter in the code to set the minimum number of zones in the stencil to be slightly larger than $D(M)$.

\subsection{Utilizing radial similarity}
A spherical mesh commonly has shell thickness varying with radial distance to satisfy the needs of the particular computational problem. Let us introduce a dimensionless variable $\chi\in[0,1]$ and a mapping $r(\chi)$ that satisfies $r(0)=r_\mathrm{min}$, $r(1)=r_\mathrm{max}$, where $r_\mathrm{min}$ and $r_\mathrm{max}$ are the inner and the outer boundaries of the entire simulation domain, not including the ghost shells. One example of such a mapping is a power law
\begin{equation}
r(\chi)=r_\mathrm{min}\left\lbrace 1+\left[\left(\frac{r_\mathrm{max}}{r_\mathrm{min}}\right)^{1/b}-1\right]\chi\right\rbrace^b,
\end{equation}
where $b$ is some positive real number. The interior of the simulation domain is partitioned into $L$ shells of equal width $\Delta\chi=L^{-1}$ that map physical shells of variable widths $\Delta r(r)$. Suppose the zone $i$ is indexed by shell $s$ and face $f$. In physical coordinates the zones corresponding to the same $f$ but different $s$ have different aspect ratios. For example, for the mapping (26) the zones closer to the origin will be more radially elongated than those at larger distances (for $b>1$).

One particular function of $\chi$ preserves the zone aspect ratio, such that $\Delta r/r=\mathrm{const}$. This is the exponential mapping,
\begin{equation}
r'(\chi)=r_\mathrm{min}\left(\frac{r_\mathrm{max}}{r_\mathrm{min}}\right)^\chi,
\end{equation}
\citep[e.g.,][]{koldoba_three-dimensional_2002}, that also satisfies $r'(0)=r_\mathrm{min}$, $r'(1)=r_\mathrm{max}$. One can then introduce exponential coordinates given by
\begin{equation}
x_1'=r' x^u_1,\quad x_2'=r' x^u_2,\quad x_3'=r' x^u_3,
\end{equation}
where, as before, the coordinates with the superscript `u' are measured on the unit sphere.

A conserved mesh variable $\mathbf{U}(\mathbf{x})$ is defined via
\begin{equation}
\int_{(i)}\mathbf{U}(\mathbf{x})r^2 dr d\Omega=\bar{\mathbf{U}}_iV_i.
\end{equation}
Integration over the solid angle $\Omega$ corresponds to integration on the unit sphere. Equation (29) can be rewritten in exponential coordinates as
\begin{equation}
\int_{(i)}\mathbf{U}(\mathbf{x})\frac{r^2}{r'^3}\frac{dr}{d\chi}r'^2dr'd\Omega=\frac{\Omega_f(r_{s+1}^3-r_s^3)}{3}\ln\left(\frac{r_\mathrm{max}}{r_\mathrm{min}}\right)\bar{\mathbf{U}}_i,
\end{equation}
where $\Omega_f$ is the area of face $f$ on the unit sphere. We next introduce a three-dimensional polynomial reconstruction of the quantity $\mathbf{W}=\mathbf{U}r^2/r'^3(dr/d\chi)$ in the zone $i$
\begin{equation}
\mathbf{W}_i(\tilde{\mathbf{x}}_s)=\sum\limits_{|\alpha|=0}^{D(M)-1}\mathbf{W}_i^\alpha\left(\tilde{x}_s^\alpha-\langle\tilde{x}^\alpha\rangle_f\right),
\end{equation}
where $\tilde{\mathbf{x}}_s=\mathbf{x}'/r'_s$ is the position vector expressed in the exponential normalized coordinates (ENC). Here they are normalized to the exponential distance to the bottom of the zone. The moments of any zone with a face index $f$ are
\begin{equation}
\langle...\rangle_f=\frac{3}{\Omega_f[(1+\Delta\tilde r)^3-1)]}\int_1^{1+\Delta\tilde{r}}\tilde{r}^2d\tilde{r}\int_{(f)}(...)d\Omega,
\end{equation}
where 
\begin{equation}
\Delta\tilde{r}=\frac{r'_{s+1}-r'_s}{r'_s},
\end{equation}
which is the same for all shells $s$. In the ENC the moments are independent of the shell, so the index $s$ is dropped for them. It is evident that
\begin{equation}
\mathbf{W}^{(0,0,0)}_i=\frac{r_{s+1}^3-r_s^3}{{r'}_s^3[(1+\Delta\tilde r)^3-1)]}\ln\left(\frac{r_\mathrm{max}}{r_\mathrm{min}}\right)\bar{\mathbf{U}}_i.
\end{equation}
To obtain the remaining modes a geometry matrix is computed for each three-dimensional stencil. Suppose $S_i$ denotes the set of zones comprising the stencil, and that the zone $j$ that belongs to this stencil, $j\in S_i$, $j\neq i$ is indexed by shell $\sigma$ and face $\phi$. Using the fact that
\begin{equation}
\tilde{x}_s^\alpha=\left(\frac{r'_\sigma}{r'_s}\right)^{|\alpha|}\tilde{x}_\sigma^\alpha,
\end{equation}
averaging (31) over a given zone in the stencil, which uses the ENC specific for its own shell $\sigma$, rather than the principal shell $s$, yields
\begin{equation}
\sum\limits_{|\alpha|=0}^{D(M)-1}\mathbf{W}_i^\alpha\left[(1+\Delta\tilde{r})^{|\alpha|(\sigma-s)}\langle\tilde{x}^\alpha\rangle_\phi-\langle\tilde{x}^\alpha\rangle_f\right]=\mathbf{W}_j^{(0,0,0)}-\mathbf{W}_i^{(0,0,0)}.
\end{equation}
This is a linear system for $\mathbf{W}_i^{\alpha}$. The geometry matrix on the LHS has the number of rows equal to the number of zones in the stencil, without counting the principal zone, and its column count is $D(M)-1$. The geometry matrix's coefficients only depend on the relative shell displacement in the stencil, $\sigma-s$, and are identical for any zone with the same face index because the corresponding stencils all have the same structure.

The advantage of the described scheme is that the amount of storage is significantly reduced (by the factor equal to the number of shells in the block) compared with the method that treats each zone as unique. Only a single copy of each moment and the geometry matrix are needed per t-face. This also permits us to precompute the LU decomposition or inverse of each geometry matrix and store it to perform reconstruction with a different RHS in (36) at each time step. The physical variable is then recovered via
\begin{equation}
\mathbf{U}_i(\mathbf{x})=\mathbf{W}_i(\tilde{\mathbf{x}}_s)\frac{{r'}^3}{r^2}\left(\frac{dr}{d\chi}\right)^{-1}.
\end{equation}

\subsection{Limiting the reconstruction}
The code performs reconstruction on all thirteen (or seven) stencils and stores the resulting modes for each stencil. The solutions from multiple stencils are combined in a nonlinear fashion into a single reconstruction polynomial using the weighted essentially non-oscillatory (WENO) method \citep{harten_uniformly_1987, shu_efficient_1988, liu_weighted_1994, jiang_efficient_1995, friedrich_weighted_1998, balsara_monotonicity_2000, dumbser_arbitrary_2007}. The nonlinear hybridization helps to stabilize the WENO scheme when local discontinuities develop in the flow.

Suppose there are $S$ stencils associated with face $i$, with the central stencil bearing the index 1, and the directional stencils numbered 2 through $S=7,13$. The central stencil is the most accurate and therefore carries the largest linear weight, $\gamma_1=\in[0.85,0.95]$, where as the remaining stencils have $\gamma_s=(1-\gamma_1)/(S-1)$. Suppose we need to perform a reconstruction of a scalar variable $U(\mathbf{x})$. The WENO procedure computes a weighted average of the reconstruction polynomials derived on each of the stencils with preference given to stencils achieving a smoother reconstruction (roughly speaking, having smaller absolute values of the modes $U^\alpha_{is}$ where $|\alpha|>0$ and $s=1,...,S$). The scheme is biased by the smoothness indicators that can be estimated simply as
\begin{equation}
\beta_{is}=\sum\limits_{\alpha}(U_{is}^{\alpha})^2.
\end{equation}

We have implemented plain second and third order WENO schemes and an adaptive order WENO-AO(4,3) scheme within the TGM framework. The plain WENO procedure computes the nonlinear weights as
\begin{equation}
w_{is}=\frac{\gamma_s}{(\beta_{is}+\epsilon)^2},
\end{equation}
where $\epsilon\sim 10^{-12}$ is used to avoid possible division by zero. The weights are then normalized so that they add up to unity. The normalized weights are obtained as
\begin{equation}
\bar{w}_{is}=\frac{w_{is}}{\sum\limits_{s=1}^S w_{is}}.
\end{equation}
The coefficients of the hybrid reconstruction polynomial are computed as
\begin{equation}
U^\alpha_{i,\,\text{WENO}}=\sum\limits_{s=1}^S\bar{w}_{is}U^\alpha_{is}.
\end{equation}

At fourth order of accuracy we have used an adaptive order method to avoid the excessive computational cost of performing high order reconstruction on all thirteen stencils. The WENO-AO method has been described in great detail in \citet{balsara_efficient_2016}, while its implementation on unstructured meshes was presented in \citet{balsara_efficient_2019, balsara_efficient2_2019}. Here we only discuss some specifics of its implementation on the geodesic mesh. The WENO-AO(4,3) method uses, in addition to the set of stencils used to perform third-order reconstruction, a large central stencil that we assign the index of 0 to avoid relabeling of the third-order stencils. This large stencil is used to perform reconstruction of polynomial degree 3 and carries the linear weight $\gamma_0=\in[0.85,0.95]$. For example, the third order central stencil may be the stencil shown in the fourth or fifth column of Figure \ref{fig_stshapes}, while the fourth-order stencil will be from column seven or eight of that figure. The linear weights $\gamma'_s$ of the adaptive order scheme are given by
\begin{equation}
\gamma'_0=\gamma_0,\quad\gamma'_1=(1-\gamma_0)\gamma_1,\quad\gamma'_s=\frac{(1-\gamma_0)(1-\gamma_1)}{S-1},\;s=2,...,S
\end{equation}
(note that the number of stencils used in this case is $S+1$). The smoothness indicators and nonlinear weights are obtained according to (38) and (39), respectively using $\gamma'_s$ in place of $\gamma_s$, where $s=0,...,S$. The normalized nonlinear weights are given by
\begin{equation}
\bar{w}_{is}=\frac{w_{is}}{\sum\limits_{s=0}^S w_{is}}.
\end{equation}
The coefficients of the hybrid reconstruction polynomial in the adaptive case are computed as
\begin{equation}
U^\alpha_{i,\,\text{WENO-AO}}=\frac{\bar{w}_{i0}}{\gamma'_0}\left(U^\alpha_{i0}-\sum\limits_{s=1}^S\gamma'_s U^\alpha_{is}\right)+\sum\limits_{s=1}^S\bar{w}_{is}U^\alpha_{is}.
\end{equation}
Expression (44) reduces to $U^\alpha_{i0}$ in the limit that the solution is smooth on all stencils and therefore $\bar{w}_{is}\to\gamma'_s$. This choice yields the most accurate reconstruction because it is based entirely on the large central stencil.

The reconstruction procedure is carried out in each zone lying in the interior of the block and in two more layers of ghost zones. The latter is needed by the slope flattening procedure that scales down the reconstruction coefficients within the zones lying near strong density enhancements. The stencils shown in Figure \ref{fig_stshapes} extend a distance equal to the degree of the reconstruction polynomial beyond the principal zone. As a result we use three layers of ghost zones at second order of accuracy, four at third order and five at fourth order.


\section{Constrained reconstruction of the magnetic field}
\label{sec_constr}
For MHD problems, it is essential to keep the magnetic field divergence free. The most successful technique to maintain $\nabla\cdot\mathbf{B}=0$ is the constrained transport method \citep{evans_simulation_1988, devore_flux-corrected_1991, ryu_divergence-free_1998, balsara_staggered_1999} that is based on the Yee type staggered mesh. In this approach the magnetic field is a face based variable, unlike the zone averaged mass, momentum, and total energy conserved variables. More specifically, the variable is a normally projected, face averaged value of the magnetic field that will be called $\bar{B}$, possibly with a subscript of the face where it is defined. This magnetic field is initialized using the vector potential
\begin{equation}
\mathbf{B}=\nabla\times\mathbf{A},
\end{equation}
and is updated in time via Faraday's law,
\begin{equation}
\frac{\partial\mathbf{B}}{\partial t}=-\nabla\times\mathbf{E},
\end{equation}
where $\mathbf{E}$ is the electric field and SI units are used. Integrating equations (45) and (46) requires edge based vector potential and electric field, respectively, in applying the Stokes theorem.

Let us focus on a single zone with index $i$ in the mesh. Denote by $F_i$ the set of faces that belong to this zone. The set can be further partitioned into three r-faces (set $R_i$) and two t-faces (set $T_i$). By convention, the normals $\hat{\mathbf{n}}_j$ for $j\in R_i$ are directed outward as viewed from a zone corresponding to an unshaded t-face (and hence inward as viewed from a shaded face, see Figure \ref{fig_faces}), where as the normals for $j\in T_i$ always point in the outward direction (direction of increasing $r$). Further, suppose $E_j$ is the set of edges that comprise the boundary of face $j$. For $j\in R_i$, the boundary consists of two t-edges and two r-edges; while faces $j\in T_i$ have three t-edges. The tangent vectors to the t-edges are assumed to be directed counter-clockwise relative to the unshaded face while the r-edge tangents point outward.

Using the above conventions, the face-based magnetic field initialization procedure is written as
\begin{equation}
\bar{B}_jS_j=\iint\limits_{\text{face}\;j}\mathbf{B}\cdot\mathbf{dS}=\sum_{k\in E_j}\int\limits_{\text{edge}\;k}\mathbf{A}\cdot\mathbf{dl}=\sum_{k\in E_j}\bar{A}_k l_k,
\end{equation}
where $S_j$ is the area of face $j$, $l_k$ is the length of the edge $k$, and $\bar{A}_k$ is the average over the edge $k$ of the vector potential dotted with the tangent vector to that edge. The line integral in (47) is evaluated using formulae (16) and (18). In addition, the integral divergence free condition for $\mathcal{D}=\nabla\cdot\mathbf{B}$ may be written as
\begin{equation}
\bar{\mathcal{D}}_iV_i=\sum_{j\in F_i}\iint\limits_{\text{face}\;j}\mathbf{B}\cdot\mathbf{dS}=\sum_{j\in F_i}\bar{B}_j S_j=0,
\end{equation}
where $V_i$ is the volume of zone $i$. In practice, the numerical code defines variables of zone, face, and edge types and the curl and divergence integral operations to perform ``conversions'' between the types.

Following \citet{balsara_divergence-free_2015} the model presented here uses a supplementary zone based vector variable $\mathbf{B}'$. At the start of the simulation, this variable must be initialized in each zone $i$ in some way consistent with the primary field $\bar{B}$ defined on $F_i$. One possibility is to use the least squares fit
\begin{equation}
\iint\limits_{\text{face}\;j}\mathbf{B}'_i\cdot\mathbf{dS}=\bar{B}_j S_j, \quad j\in F_i.
\end{equation}
The integral in the above equation is evaluated using (20) and (21), giving five equations (one per face) for the three unknown field components. The alternative is to initialize $\mathbf{B}'$ directly as a zone variable using the expression for the field rather than the potential. The resulting $\mathbf{B}'$ is subsequently treated like any other zone variable. In particular, it is subjected to the same volume reconstruction procedure described in the previous section. This reconstruction is not functionally divergence free, and an additional procedure, described below, is applied to obtain a constrained reconstruction. This approach represents a low computational cost alternative to a face based reconstruction.

Suppose the preliminary, non-divergence-free reconstruction, computed as discussed in the previous section, is given by
\begin{equation}
\mathbf{B}'_i(\mathbf{x})=\sum\limits_{|\alpha|=0}^{D(M)-1}\mathbf{B}_i'^{\alpha}\left(x^\alpha-\langle x^\alpha\rangle_i\right),
\end{equation}
where $\mathbf{B}_i'^{\alpha}$ are the modes. We seek a constrained polynomial reconstruction for the magnetic field $\tilde{\mathbf{B}}(\mathbf{r})$ as
\begin{eqnarray}
\tilde{B}_{1i}(\mathbf{x})=\sum_{\substack{|\alpha|=0 \\ \alpha_2,\alpha_3\leq M}}^{D(M+1)-1}\tilde{B}_{1i}^{\alpha}\left(x^\alpha-\langle x^\alpha\rangle_i\right), \nonumber \\
\tilde{B}_{2i}(\mathbf{x})=\sum_{\substack{|\alpha|=0 \\ \alpha_1,\alpha_3\leq M}}^{D(M+1)-1}\tilde{B}_{2i}^{\alpha}\left(x^\alpha-\langle x^\alpha\rangle_i\right), \\
\tilde{B}_{3i}(\mathbf{x})=\sum_{\substack{|\alpha|=0 \\ \alpha_1,\alpha_2\leq M}}^{D(M+1)-1}\tilde{B}_{3i}^{\alpha}\left(x^\alpha-\langle x^\alpha\rangle_i\right). \nonumber
\end{eqnarray}
These reconstructions have $\tilde{D}(M)=2D(M)-D(M-1)$ degree of freedoms, which is larger than $D(M)$. While the degree of the reconstruction polynomials (51) is one higher than of (50), not every additional high order mode is present. The need for the extra modes will be demonstrated shortly. We now describe the five separate constraints imposed on the magnetic field modes that ensure that the magnetic field remains divergence-free not only in the integral sense (zero total flux through all faces of a zone), but also functionally at any location within the zone.

\subsection{Constraint 1}
This step ensures that the polynomial reconstruction of the magnetic field has zero divergence everywhere in the zone. Taking the divergence of Eq. (51) and making the resulting polynomial expression equal to zero yields $D(M)$ equations of the form
\begin{equation}
\alpha_1\tilde{B}_{1i}^{\alpha_1}+\alpha_2\tilde{B}_{2i}^{\alpha_2}+\alpha_3\tilde{B}_{3i}^{\alpha_3}=0.
\end{equation}
Clearly, $\tilde{B}_1$, $\tilde{B}_2$, and $\tilde{B}_3$ modes with $\alpha_1=0$, $\alpha_2=0$, and $\alpha_3=0$, respectively, do not contribute to (52). Only the extra modes that contain powers of $x_1$ for $\tilde{B}_1$, $x_2$ for $\tilde{B}_2$, and $x_3$ for $\tilde{B}_3$ are included. For instance, at third order of accuracy ($M=2$) the extra modes present in the first equation of (51) are those containing $x_1^3$, $x_1^2 x_2$, $x_1^2 x_3$, $x_1 x_2^2$, $x_1 x_2 x_3$, and $x_1 x_3^2$, where as the second equation includes $x_1^2 x_2$, $x_1 x_2^2$, $x_1 x_2 x_3$, $x_2^3$, $x_2^2 x_3$, and $x_2 x_3^2$ terms. The remaining high order modes do not contribute to the local divergence-free conditions.

\subsection{Constraint 2}
The second constraint imposed on the reconstruction (51) is the requirement that its normal component, evaluated from any two adjacent zones sharing the face $j$ and averaged over that face must be equal to $\bar{B}_j$, namely
\begin{equation}
\iint\limits_{\text{face}\;j}(\tilde{B}_{1i}dS_1+\tilde{B}_{2i}dS_2+\tilde{B}_{3i}dS_3)=\bar{B}_j,
\end{equation}
where the integral is evaluated according to the rules (20)--(21). This is the requirement of zero divergence in the integral sense. The order of the quadrature rule need not be very high, but only sufficient to match the order of the overall scheme. For example, for a third order scheme that uses polynomials of up to third degree, we use six point quadratures on all faces.

It should be pointed out that because of (53) one constraint in (52) is redundant. This is readily demonstrated by computing the divergence of $\tilde{\mathbf{B}}$ (equation 51) analytically, integrating over the volume of the zone, and setting the integral to zero. For the sake of symmetry, we chose to discard the first equation in (52), so that system's equation count is reduced to $D(M)-1$.

\subsection{Constraint 3}
\citet{balsara_divergence-free_2015} proposed a method seeking to match, at each face, complete polynomial reconstructions of the normal component of the magnetic field. Here we use a weaker requirement that the reconstructions of the normal component should approximately match at the facial quadrature points used to perform integration on that face. This procedure nonetheless ensures a very close matching of the modes of the magnetic field within each face.

The matching procedure starts by evaluating $\mathbf{B}'_i(\mathbf{x})$ from (51) at each quadrature point of face $j\in F_i$ and projecting it onto the unit normal to the face at that point. Initially this normal component is not continuous at the zone boundaries, so there are two values of the normal component, $B_{iq}$ and $B_{kq}$, at each facial quadrature point $q$ contributed by two adjacent zones $i$ and $k$, where $j\in F_i,F_k$. The common normal component at each quadrature point $q$, $B_q$, is evaluated in two steps as
\begin{equation}
B^*_q=\bar{B}_q+\mathrm{minmod}(B_{iq}-\bar{B}_j,B_{kq}-\bar{B}_j),
\end{equation}
\begin{equation}
B_q=B^*_q-\langle B^*\rangle_j+\bar{B}_j,
\end{equation}
where $B^*$ is the intermediate value of the common normal component of the field at the interface and the angular brackets denote its average over the face $j$. The normal component of the magnetic field given by (55) is continuous and its average over the face matches the respective value of the primary variable $\bar{B}_j$. Therefore, the third set of constraints can be written as
\begin{equation}
\tilde{B}_{1i}(\mathbf{x}_q)\hat{n}_1+\tilde{B}_{2i}(\mathbf{x}_q)\hat{n}_2+\tilde{B}_{3i}(\mathbf{x}_q)\hat{n}_3\approx B_q,
\end{equation}
for each quadrature point $q$, with $j\in F_i$. The number of conditions in (56) is equal to the total number of quadrature points on all five faces of the frustum.

\subsection{Constraint 4}
Next, we demand that the divergence-free reconstruction (51) should be as close to the volume reconstruction of $\mathbf{B}'$ as possible, i.e.,
\begin{equation}
\tilde{B}_{1i}^{\alpha}\approx B_{1i}'^{\alpha},\quad\tilde{B}_{2i}^{\alpha}\approx B_{2i}'^{\alpha},\quad\tilde{B}_{3i}^{\alpha}\approx B_{3i}'^{\alpha},\quad|\alpha|<D(M).
\end{equation}
Eq. (52) is based on the observation that the initial (unconstrained) volume reconstruction is the best possible starting point for determining the constrained modes. With this condition the convergence order of the constrained reconstruction stays close to the order of convergence of the unconstrained volume reconstruction.

\subsection{Constraint 5}
In the same spirit it is desirable that the ``extra'' high order modes should be small, i.e.,
\begin{equation}
\tilde{B}_{1i}^{\alpha}\approx 0,\quad\tilde{B}_{2i}^{\alpha}\approx 0,\quad\tilde{B}_{3i}^{\alpha}\approx 0,\quad |\alpha|\geq D(M).
\end{equation}

Table \ref{tab_numeq} provides the counts of the degrees of freedom and the number of equations contributed by formulae (52), (53), (56), (57), and (58) for schemes of second, third, and fourth orders of spatial convergence. Since there are more equations than unknowns, only the local and global divergence-free conditions (52) and (53) are strictly enforced; the remaining conditions can only be satisfied approximately, in the least squares sense (in principle, at third and fourth order of accuracy the constraints (56) can also be strictly imposed). This constitutes a constrained linear least square (CLSQ) problem. Figure \ref{fig_matstr} illustrates the structure of the LLS and constraints matrices at fourth order. From Table 3, the rank of the Karush-Kuhn-Tucker (KKT) matrix of the CLSQ problem is 29, 62, and 114 at second, third, and fourth order of accuracy. Note that despite the sparsity of the LLS and the constraint matrices, the KKT matrix is largely dense. 

Based on the results of the previous section, it may be expected that only a single KKT matrix needs to be constructed and inverted per t-face. Unfortunately, the difficulty here is with the global divergence-free condition (Constraint 1), which is, in general, incompatible with the reconstruction (31). Coordinate factorization is still possible, but only if the mesh is directly exponentially rationed, i.e., $r=r'$ and $\mathbf{W}=\ln(r_\text{max}/r_\text{min})\mathbf{U}$. This is the mesh that was used for all MHD applications discussed below.

At the end of the magnetic field reconstruction step, the previously obtained unconstrained modes are discarded and replaced with the constrained version. This ensures synchronization between the primary and supplementary magnetic field variables used by the code.


\section{Time advance and boundary exchange}
\label{sec_impl}
The complete finite volume method is implemented as follows. First, the zone-based variables (including $\mathbf{B}'$) are reconstructed to the quadrature points on the faces as described in the previous two sections. Pairs of states from each side of the interface are fed into a Riemann solver. We employ the HLL family of nonlinear solvers \citep{harten_upstream_1983, einfeldt_godunov-type_1988} that are very robust and usually positivity preserving as long as the speeds of the extremal waves are properly estimated. The popular HLLC solver \citep{batten_choice_1997, gurski_hllc-type_2004, li_hllc_2005} consists of four states separated by two fast shocks and a tangential discontinuity. The HLLD solver \citep{miyoshi_multi-state_2005} adds a pair of rotational discontinuities, and is therefore less dissipative than HLLC, but is somewhat less robust and can fail for certain combinations of input states. Our approach is to start with the least diffusive solver, downgrading to the more dissipative solver when the former fails to deliver a positive resolved state. The fluxes are evaluated at each quadrature point and combined together to obtain the total flux through a face. These fluxes update the conserved variables in the zone using a TVD Runge-Kutta scheme \citep{shu_efficient_1988}. A version of the code is also available that uses the so-called arbitrary derivative (ADER) update technique \citep{dumbser_unified_2008, balsara_efficient_2009}. The ADER implementation on the geodesic mesh has been reported elsewhere \citep{balsara_efficient_2019}.

Unlike the zone variables, the magnetic fields are reconstructed to the quadrature points lying on the \textit{edges}. A single point is sufficient at second order and two points at third and fourth orders. The constrained magnetic field is used here in place of the non-constrained reconstruction. Each t-edge receives four states and each r-face five or six states depending on whether it is a penta-corner or not. These states are fed into a multi-state two-dimensional Riemann solver \citep{balsara_multidimensional_2010, balsara_two-dimensional_2012, balsara_multidimensional_2014, balsara_multidimensional_2015} generating the electric field at the edges (the remaining flux components are discarded). The 2D Riemann solver used here is of the HLLI type \citep{dumbser_new_2016} that can include every MHD wave, including the Alfv\'en and slow magnetosonic waves. The face-based magnetic field is updated via the same Runge-Kutta procedure. This operation conserves the divergence of $\mathbf{B}$ to the machine precision.

A correct implementation of the above scheme must ensure that all variables are properly synchronized at the block boundaries. Each block can have up to 38 neighbors which at some point in the calculation must send some of their zone, face, or edge based boundary data and received equivalent data in return to fill in the ghost mesh element or synchronize the common boundaries. The implementation described here does not use neighbor lists, instead delegating all bookkeeping tasks to the message passing library.

Figure \ref{fig_cutout} demonstrates the typical mesh topology. Ten out of 20 blocks are shown in this cutout view, shaded using different colors. This corresponds to the smallest decomposition of the computational domain, confined between two concentric spheres with $r_\text{max}/r_\text{min}=2$, and using a single slab.

To formalize our communication strategy we define the concept of ``exchange site'' that could be a face, edge, or corner (vertex) of the block. Each exchange site maintains a list of blocks that share the site. The list consists of two elements for any face site, four for a t-edge, six or five for an r-edge, and twelve or ten for a vertex site. Each site further defines a number of exchanges that occur at the site as lists of participant blocks. A block maintains a list of exchanges it needs to perform during a time step and its own order in that exchange. All exchanges of the same kind are started at once on every participating block in the non-blocking regime; we therefore rely on the message passing library's own scheduling facilities to achieve optimal utilization of the interconnection network.

A block maintains a set of buffers and corresponding rules to pack a part of the block destined for exchange into contiguous memory of the buffer as well as the inverse (unpack) operation. Because neighboring blocks can have any of the three possible orientations, packing and unpacking must be done in a way that is independent of the choice of the base vertex. Care must also be taken to synchronize the variables at locations that are shared among blocks, such as face-based magnetic field values and fluxes, and edge-based electric fields. This synchronization is needed to eliminate possible divergence between neighboring blocks owing to roundoff errors.

Figure \ref{fig_nbrs} shows the block surrounded by twelve neighbors that belong to the same slab. The large yellow triangle is the interior area of the block, while the surrounding smaller trapezoidal or triangular areas represent the \textit{receive buffers} that correspond to the ghost zones of the block. To extract the data received from each neighbor from the corresponding buffer requires rotated TAS coordinates that are represented in Figure \ref{fig_nbrs} by pairs of black arrows showing the directions of the first and the second TAS coordinates, respectively. The convention for packing and unpacking a trapezoid is that the principal vertex is in the lower left corner with the trapezoid resting on its wider base. For small triangles, the principal vertex is the vertex shared with the block's interior. This convention automatically ensures that unpacking of a buffer is done in the same order as it was packed by the neighbor block.

Figure \ref{fig_loop} shows the structure of the complete simulation loop based on the Runge-Kutta time advance. The initial setup involving pre-computing the geometry matrices is time consuming, but the subsequent computation is sped up dramatically as a result.

The production version of the code was written in Fortran, and a development version writen in C++ is also available. The input and output is handled by the open source SILO library (https://wci.llnl.gov/simulation/computer-codes/silo). The library features a simple parallel I/O implementation that groups the blocks (which could number in the hundreds of thousands) into a smaller number of SILO files. An assembly file is then generated describing the relationship between the blocks for the visualizer. We use the VisIt visualizer (https://wci.llnl.gov/simulation/computer-codes/visit) for 3D rendering of the model output; several of the figures in this paper were produced with VisIt.


\section{Numerical tests}
\label{sec_test}
Here we present two simple tests validating the accuracy of the model. The first test in the ``manufactured'' solution of \citet{ivan_multi-dimensional_2013} that describes an interaction between a point source and a uniform flow; which is the most basic model of an astrosphere (an interface produced by a stellar wind expanding into a moving interstellar medium). This steady state, current-free field-aligned flow is given by
\begin{equation}
\rho=\rho_0\left(\frac{r_0}{r}\right)^{5/2},
\end{equation}
\begin{equation}
\mathbf{u}=\frac{u_0\mathbf{x}}{(r_0r)^{1/2}}+u_1\left(\frac{r}{r_0}\right)^{5/2}\hat{\mathbf{e}}_3,
\end{equation}
\begin{equation}
p=p_0\left(\frac{r_0}{r}\right)^{5/2},
\end{equation}
\begin{equation}
\mathbf{B}=\frac{B_0r_0^2\mathbf{x}}{r^3}+\frac{B_0u_1}{u_0}\hat{\mathbf{e}}_3.
\end{equation}
The source terms corresponding to Eqs. (59)--(62) that appear in the conservative MHD equations are
\begin{equation}
Q_\rho=0,
\end{equation}
\begin{equation}
\mathbf{Q}_u=\left[\rho_0 u_0\left(\frac{u_0}{r}-\frac{u_1z}{r^2}\right)-\frac{5p_0r_0}{r^2}\right]\frac{r_0^{3/2}\mathbf{x}}{2r^{5/2}}+\left(7u_0+\frac{5u_1zr}{r_0^2}\right)\frac{\rho_0u_1\hat{\mathbf{e}}_3}{2(r_0r)^{1/2}},
\end{equation}
\begin{equation}
Q_e=\frac{\rho_0u_0^2}{2r}\left(\frac{u_0r_0}{r}+\frac{7u_1z}{r_0}\right)+\frac{\rho_0u_0u_1^2(7r^2+4z^2)}{r_0^3}+\frac{5\rho_0u_1^3zr^3}{2r_0^5},
\end{equation}
\begin{equation}
\mathbf{Q}_B=0.
\end{equation}

The analytic solution is independent of the adiabatic index $\gamma$. Following \citet{ivan_multi-dimensional_2013}, the zero-subscripted constants are all set to unity, $u_1/u_0=0.017$, $\gamma=1.4$, and the simulation is performed in the region between $r_\text{min}=2$ and $r_\text{max}=3.5$. The radial shell width has an exponential dependence on $r$. Figure \ref{fig_conv} shows the rate of convergence for the $L_1$ and $L_\infty$ norms of the error in the density, total energy and one component of the magnetic field. Simulations were performed on division 5 through division 8 meshes with 32 through 256 radial shells and the same number of zones per sector side.

As is demonstrated in Figure \ref{fig_conv}, the $L_1$ error decreases at the nominal convergence rate of the scheme in each case. The $L_\infty$ norm displays the nominal convergence rate at second order, but decreases slower than the nominal rate and third and fourth order. This is further quantified in Tables \ref{tab_order2} through \ref{tab_order4} that show the numerical values of the order of convergence for the manufactured solution problem. The rates for density and energy (both zone based variables) are very similar, while magnetic field shows a different behavior. The imposition of constraints described in Section \ref{sec_constr} affects the accuracy of reconstruction. It seems to be detrimental at lower order of accuracy but is surprisingly beneficial at fourth order.

Figure \ref{fig_wind} shows the velocity magnitude (left panel) and the distribution of the error on the sphere at $r=2.75$ for a simulation on a division 6 mesh using the fourth order scheme. The flow geometry resembles that of a potential flow of gas around a point source, although the velocity field is not irrotational here. The right panel shows the error distribution in one spherical layer. In common with other geodesic meshes, the error distribution shows a distinctive imprint of the mesh. The errors are the largest near the penta-corners and at the boundaries of division 0 and division 1 sectors. Similar phenomena have been reported by \citet{tomita_shallow_2001}, \citet{weller_computational_2012} and \citet{peixoto_analysis_2013} in the context of the shallow water equations.

It is expected that the error becomes more concentrated near singular points with increasing refinement. For any division mesh, only 60 zones have a singular point as a vertex. Because the ratio of the number of large error zones to the total number of zones decreases with increased resolution, the $L_1$ norm is not affected even if the convergence order in high error zones is one lower than elsewhere in the mesh; this is supported by the numbers from Table \ref{tab_order4}.

The second test is a time-dependent blast problem from \citep{florinski_magnetohydrodynamic_2013}. The initial conditions are piecewise constant within each of the two concentric shells, $r_\text{min}\leq r\leq r_1$ and $r_1\leq r\leq r_\text{max}$. Both states have $\rho_0=1$ and $\mathbf{u}_0=0$, while the pressure is set to $p_0=10\;(r<r_1)$ and $p_0=0.1\;(r>r_1)$. The initial magnetic field is a superposition of a dipole and a uniform fields,
\begin{equation}
\mathbf{B}=\mathbf{B}_0\left(1+\frac{r_0^3}{2r^3}\right)-\frac{3r_0^3(\mathbf{B}_0\cdot\mathbf{x})\mathbf{x}}{2r^5}.
\end{equation}
We set $\mathbf{B}_0=10/\sqrt{3}(\hat{\mathbf{e}}_1+\hat{\mathbf{e}}_2+\hat{\mathbf{e}}_3)$, $\gamma=1.4$, and $r_1=0.1$. The simulation was performed in the region between $r_\text{min}=r_0=0.01$ and $r_\text{max}=0.5$ until the time $t=0.07$ with third order reconstruction on a division 6 mesh with 256 exponentially spaced radial shells. A reflective condition was used at the internal boundary and the fixed initial state maintained at the external boundary.

The magnetic field obtained for this problem is shown in Figure \ref{fig_blast}. The flow consists of a fast shock wave and two dense shells of material elongated along the magnetic field. The result is in excellent agreement with that reported in \citet{florinski_magnetohydrodynamic_2013}.


\section{Summary}
\label{sec_sum}
This paper documents many of the original techniques and innovations that were incorporated into our newly developed computational model for MHD equations based on an icosahedral triangular geodesic mesh. The new geodesic framework features numerous improvements compared with our earlier icosahedral hexagonal model reported in \citet{florinski_magnetohydrodynamic_2013} that was used successfully to simulate the interaction between the solar wind and the surrounding interstellar medium \citep{guo_galactic_2016, guo_effects_2018}. These improvements are summarized below.

\textit{Triangle based mesh.} Using triangles instead of hexagons paved the way to efficient decomposition of the computational domain into sectors in addition to radial shells (shell decomposition was the only one available in \citet{florinski_magnetohydrodynamic_2013}). As a result, the new code scales up to tens of thousands of CPU cores with almost linear weak scaling \citep{balsara_efficient_2019}. The second advantage of the TGM is that it is amenable to adaptive mesh refinement, which is not possible with a hexagonal mesh.

\textit{Increased order of accuracy.} The new framework provides second, third, and fourth orders of accuracy for the MHD equations. High accuracy was achieved by using larger stencils and multiple families of stencils including symmetric central and asymmetric directional stencils. This is a major improvement over our earlier geodesic model that was only second order capable. The only other fourth order geodesic mesh MHD model we are aware of was reported in \citet{ivan_high-order_2015}; however it was based on a cube-sphere rather than a TGM.

\textit{Accurate representation of spherical geometry.} The new framework uses linear, quadratic, and cubic Lagrangian basis function to perform coordinate transformations from a reference element (a right prism) to the physical computational zone. These maps are based on the serendipity family of triangular finite elements with three, six, and ten nodes, respectively. This approach allows a very accurate representation of the spherical surface without the drawback of dealing directly with spherical coordinates. While the accuracy of the geometry representation does not improve the convergence order of the scheme, it could be potentially very important for the models of thin shells, such as planetary atmospheres.

\textit{Divergence free MHD.} The new model features the first implementation of the constraint transport method on a geodesic mesh. In this approach the magnetic field is maintained divergence free because of exact cancellation of all contributions to divergence in a zone during the time update. In addition, the model features pointwise and functional divergence-free reconstruction of the magnetic field.

\textit{Implementation flexibility.} All geodesic meshes are based on the same set of underlying principles used in mesh generation and spatial reconstruction. We have found that the present framework can be adapted, with a very limited number of changes, to build a model around any of the five regular polyhedra. We have developed an initial implementation of the geometry framework component for the hexahedral QGM. This will eventually permit a direct comparison between geodesic meshes of different types.


\section*{Appendix A: Basis functions for triangular faces}
\label{sec_apa}

\noindent
Linear elements: nodal points at vertices,
\begin{eqnarray}
\psi_1&=&1-\xi-\frac{1}{\sqrt{3}}\eta, \nonumber \\
\psi_2&=&\xi-\frac{1}{\sqrt{3}}\eta, \nonumber \\
\psi_3&=&\frac{2}{\sqrt{3}}\eta, \\
\phi_1&=&1-\delta, \nonumber \\
\phi_2&=&\delta. \nonumber
\end{eqnarray}

\noindent
Quadratic elements: nodal points at vertices and edge midpoints,
\begin{eqnarray}
\psi_1&=&1-3\xi-\sqrt{3}\eta+2\xi^2+\frac{4}{\sqrt{3}}\xi\eta+\frac{2}{3}\eta^2, \nonumber \\
\psi_2&=&-\xi+\frac{1}{\sqrt{3}}\eta+2\xi^2-\frac{4}{\sqrt{3}}\xi\eta+\frac{2}{3}\eta^2, \nonumber \\
\psi_3&=&-\frac{2}{\sqrt{3}}\eta+\frac{8}{3}\eta^2, \nonumber \\
\psi_4&=&\frac{8}{\sqrt{3}}\xi\eta-\frac{8}{3}\eta^2, \nonumber \\
\psi_5&=&\frac{8}{\sqrt{3}}\eta-\frac{8}{\sqrt{3}}\xi\eta-\frac{8}{3}\eta^2, \\
\psi_6&=&4\xi-\frac{4}{\sqrt{3}}\eta-4\xi^2+\frac{4}{3}\eta^2, \nonumber \\
\phi_1&=&1-3\delta+2\delta^2, \nonumber \\
\phi_2&=&-\delta+2\delta^2, \nonumber \\
\phi_3&=&4\delta-4\delta^2. \nonumber
\end{eqnarray}

\noindent
Cubic elements: nodal points at vertices, edge thirds, and geometric center,
\begin{eqnarray}
\psi_1&=&1-\frac{11}{2}\xi-\frac{11}{2\sqrt{3}}\eta+9\xi^2+6\sqrt{3}\xi\eta+3\eta^2-\frac{9}{2}\xi^3-\frac{9\sqrt{3}}{2}\xi^2\eta-\frac{9}{2}\xi\eta^2-\frac{\sqrt{3}}{2}\eta^3, \nonumber \\
\psi_2&=&\xi-\frac{1}{\sqrt{3}}\eta-\frac{9}{2}\xi^2+3\sqrt{3}\xi\eta-\frac{3}{2}\eta^2+\frac{9}{2}\xi^3-\frac{9\sqrt{3}}{2}\xi^2\eta+\frac{9}{2}\xi\eta^2-\frac{\sqrt{3}}{2}\eta^3, \nonumber \\
\psi_3&=&\frac{2}{\sqrt{3}}\eta-6\eta^2+4\sqrt{3}\eta^3, \nonumber \\
\psi_4&=&-3\sqrt{3}\xi\eta+3\eta^2+9\sqrt{3}\xi^2\eta-18\xi\eta^2+3\sqrt{3}\eta^3, \nonumber \\
\psi_5&=&-3\sqrt{3}\xi\eta+3\eta^2+18\xi\eta^2-6\sqrt{3}\eta^3, \nonumber \\
\psi_6&=&-3\sqrt{3}\eta+3\sqrt{3}\xi\eta+21\eta^2-18\xi\eta^2-6\sqrt{3}\eta^3, \nonumber \\
\psi_7&=&6\sqrt{3}\eta-15\sqrt{3}\xi\eta-15\eta^2+9\sqrt{3}\xi^2\eta+18\xi\eta^2+3\sqrt{3}\eta^3, \\
\psi_8&=&9\xi-3\sqrt{3}\eta-\frac{45}{2}\xi^2+\frac{15}{2}\eta^2+\frac{27}{2}\xi^3+\frac{9\sqrt{3}}{2}\xi^2\eta-\frac{9}{2}\xi\eta^2-\frac{3\sqrt{3}}{2}\eta^3, \nonumber \\
\psi_9&=&-\frac{9}{2}\xi+3\frac{3\sqrt{3}}{2}\eta+18\xi^2-9\sqrt{3}\xi\eta+3\eta^2-\frac{27}{2}\xi^3+\frac{9\sqrt{3}}{2}\xi^2\eta+\frac{9}{2}\xi\eta^2-\frac{3\sqrt{3}}{2}\eta^3, \nonumber \\
\psi_{10}&=&18\sqrt{3}\xi\eta-18\eta^2-18\sqrt{3}\xi^2\eta-6\sqrt{3}\eta^3, \nonumber \\
\phi_1&=&1-\frac{11}{2}\delta+9\delta^2-\frac{9}{2}\delta^3, \nonumber \\
\phi_2&=&\delta-\frac{9}{2}\delta^2+\frac{9}{2}\delta^3, \nonumber \\
\phi_3&=&9\delta-\frac{45}{2}\delta^2+\frac{27}{2}\delta^3, \nonumber \\
\phi_4&=&-\frac{9}{2}\delta+18\delta^2-\frac{27}{2}\delta^3. \nonumber
\end{eqnarray}


\section*{Appendix B: Reconstruction accuracy vs. stencil size}
\label{sec_apb}

To determine whether the stencil configuration affects the accuracy of the reconstructed solution, we conducted a statistical test by initializing a single grid block with an ensemble of $N_w=10$ waves with isotropically distributed wavevectors $\mathbf{k}_j$ and random phases $\varphi_j$,
\begin{equation}
U(\mathbf{x})=\frac{1}{N_w}\sum_{j=1}^{N_w}\cos(\mathbf{k}_j\cdot\mathbf{x}+\varphi_j),
\end{equation}
where $U$ is the scalar variable to be reconstructed. The wavelengths $\lambda_j=2\pi/k_j$ were logarithmically distributed between $\lambda_\mathrm{min}=1$ and $\lambda_\mathrm{max}=10$. For this test we used fourth order of accuracy because it offers the largest choice of stencils. A single division 1 block with 16 shells, $r_\text{min}=1.71$, $r_\text{max}=2.92$, and division 5 t-faces was used. Shell spacing was exponential as given by Eq. (27).

Figure \ref{fig_stencilcomp} shows the average $L_1$ error, its standard deviation, and the largest error over all trials, as a function of the number of zones in the stencil that varied between $D(3)=20$ and $3D(3)=60$. Interestingly, both the average and the maximum errors are slowly increasing with the stencil size, although a zero trend would also be consistent with the error bars. This trend was previously observed by \citet{ivan_high-order_2015}, who suggested that smaller stencils provide higher accuracy because the reconstruction data is more local to the zone. The only exception is the first point corresponding to the stencil of the smallest possible size. It would seem reasonable, then, to use smaller stencils as long as they have a few extra zones to benefit from the least square procedure. One has to remember, however, that this result may not hold for every problems or mesh configuration.


\section*{List of abbreviations}
\label{sec_abbr}

2D -- Two dimensional \\
3D -- Three dimensional \\
ADER -- Arbitrary high order scheme using Riemann problem to advect high-order Derivatives \\
CFD -- Computational Fluid Dynamics \\
CLSQ -- Constrained Least Square (system) \\
CPU -- Central Processing Unit \\
E -- East \\
EF -- Edge-face (connectivity) \\
ENC -- Exponential Normalized Coordinates \\
EV -- Edge-vertex (connectivity) \\
FE -- Face-edge (connectivity) \\
FF -- Face-face (connectivity) \\
FV -- Face-vertex (connectivity) \\
HLL -- Harten-Lax-van Leer (Riemann solver) \\
HLLC -- Harten-Lax-van Leer with Contact discontinuity (Riemann solver) \\
HLLD -- Harten-Lax-van Leer with multiple Discontinuities (Riemann solver) \\
HLLI -- Harten-Lax-van Leer with Intermediate characteristic fields (Riemann solver) \\
KKT -- Karush-Kuhn-Tucker (matrix) \\
LHS -- Left Hand Side \\
LLS -- Linear Least Square (system) \\
LU -- Lower-Upper \\
MHD -- Magneto-hydrodynamic \\
MPI -- Message Passing Interface \\
N -- North \\
NE -- North-East \\
NW -- North-West \\
PDE -- Partial Differential Equation \\
QGM -- Quadrilateral Geodesic Mesh \\
RHS -- Right Hand Side \\
S -- South \\
SE -- South-East \\
SI -- Syst\`eme International \\
SW -- South-West \\
TAS -- Triangular Addressing Scheme \\
TGM -- Triangular Geodesic Mesh \\
TVD -- Total Variation Diminishing \\
VE -- Vertex-edge (connectivity) \\
VF -- Vertex-face (connectivity) \\
VV -- Vertex-vertex (connectivity) \\
W -- West \\
WENO -- Weighted Essentially Non-Oscillatory \\
WENO-AO -- Weighted Essentially Non-Oscillatory -- Adaptive Order


\begin{backmatter}

\section*{Declarations}
\label{sec_decl}

\subsection*{Availability of data and materials}
Data used to produce the figures in this paper are available from the authors on request.

\subsection*{Competing interests}
None.

\subsection*{Funding}
This work was supported, in part, by NSF grant DMS-1361219 and by NASA grant NNX17AB85G. DBS acknowledges support via NSF grants ACI-1533850, DMS-1622457, ACI-1713765 and DMS-1821242. Support from a grant by Notre Dame International is also acknowledged.

\subsection*{Authors' contributions}
V. Florinski co-developed the mathematical models, implemented them in a C++ code, and wrote most of the paper. D. S. Balsara co-developed the mathematical models and designed the Fortran version of the code. S. Garain wrote parts of the Fortran code and performed test simulations. K. F. Gurski developed numerical quadrature algorithms used by the code.

\subsection*{Acknowledgements}
Not applicable.

\bibliographystyle{bmc-mathphys}
\bibliography{vf_cac_geodesic}
\nocite{label}


\section*{Figures}
\label{sec_fig}

\begin{figure}
\includegraphics[width=2.0in, clip=]{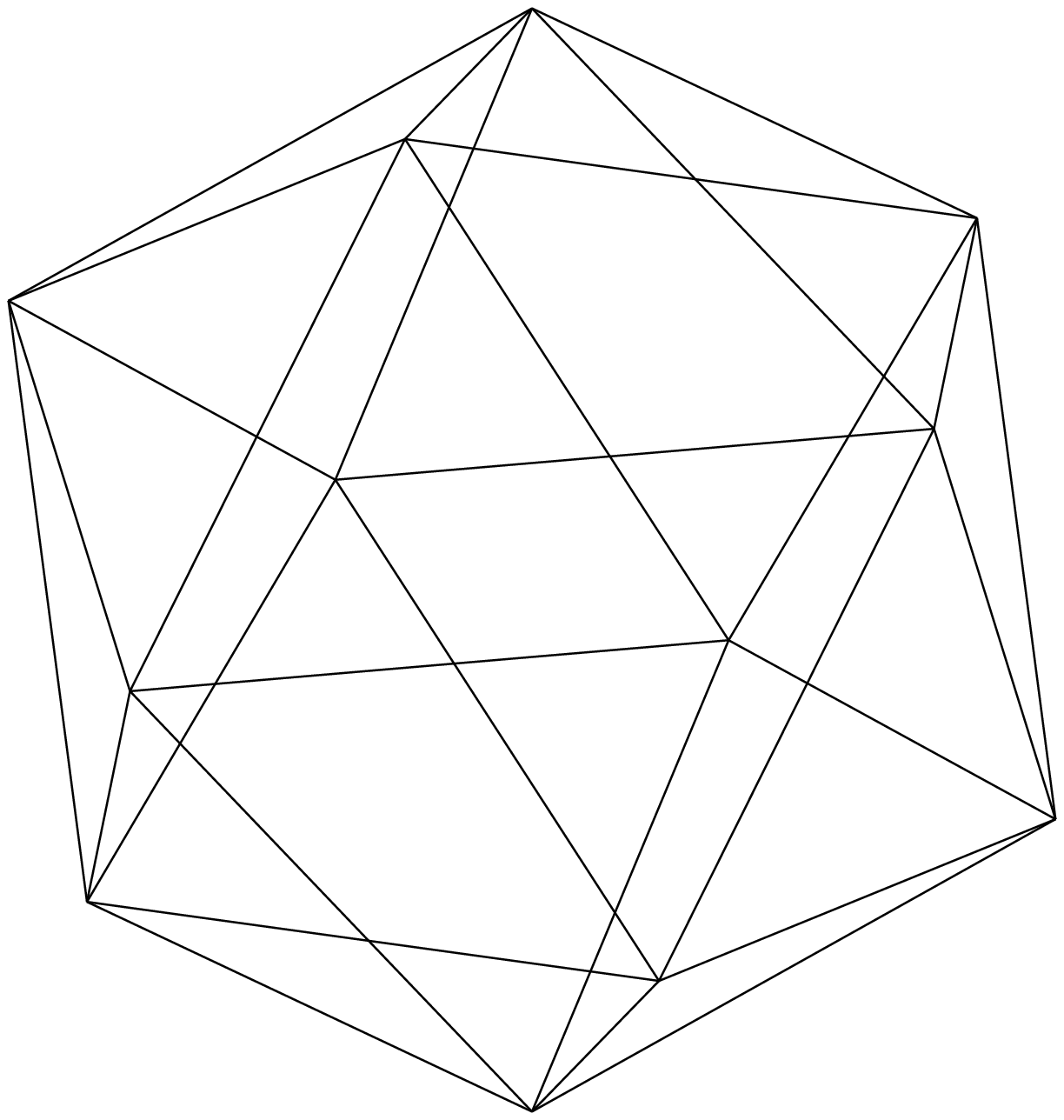}
\includegraphics[width=2.0in, clip=]{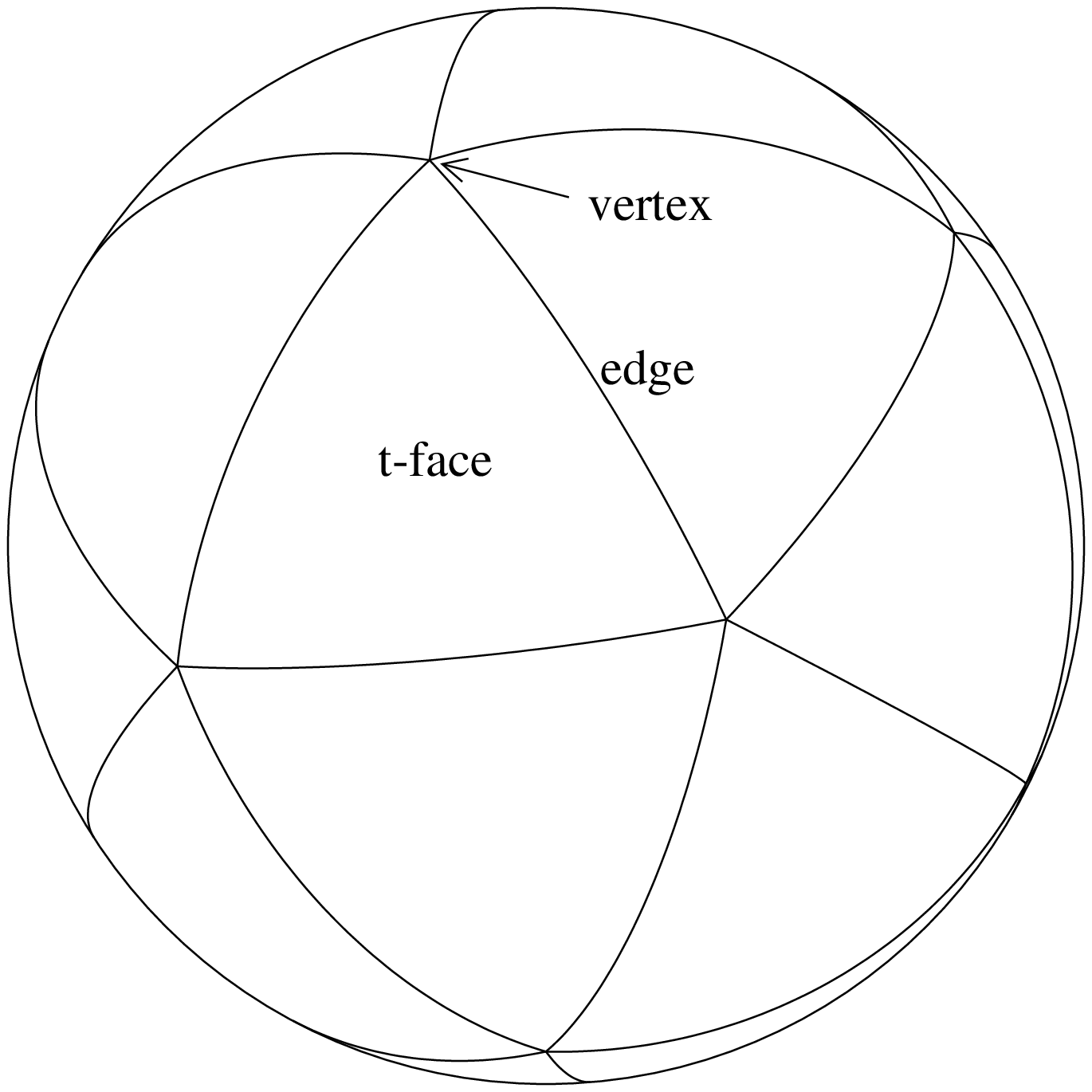}
\includegraphics[width=2.0in, clip=]{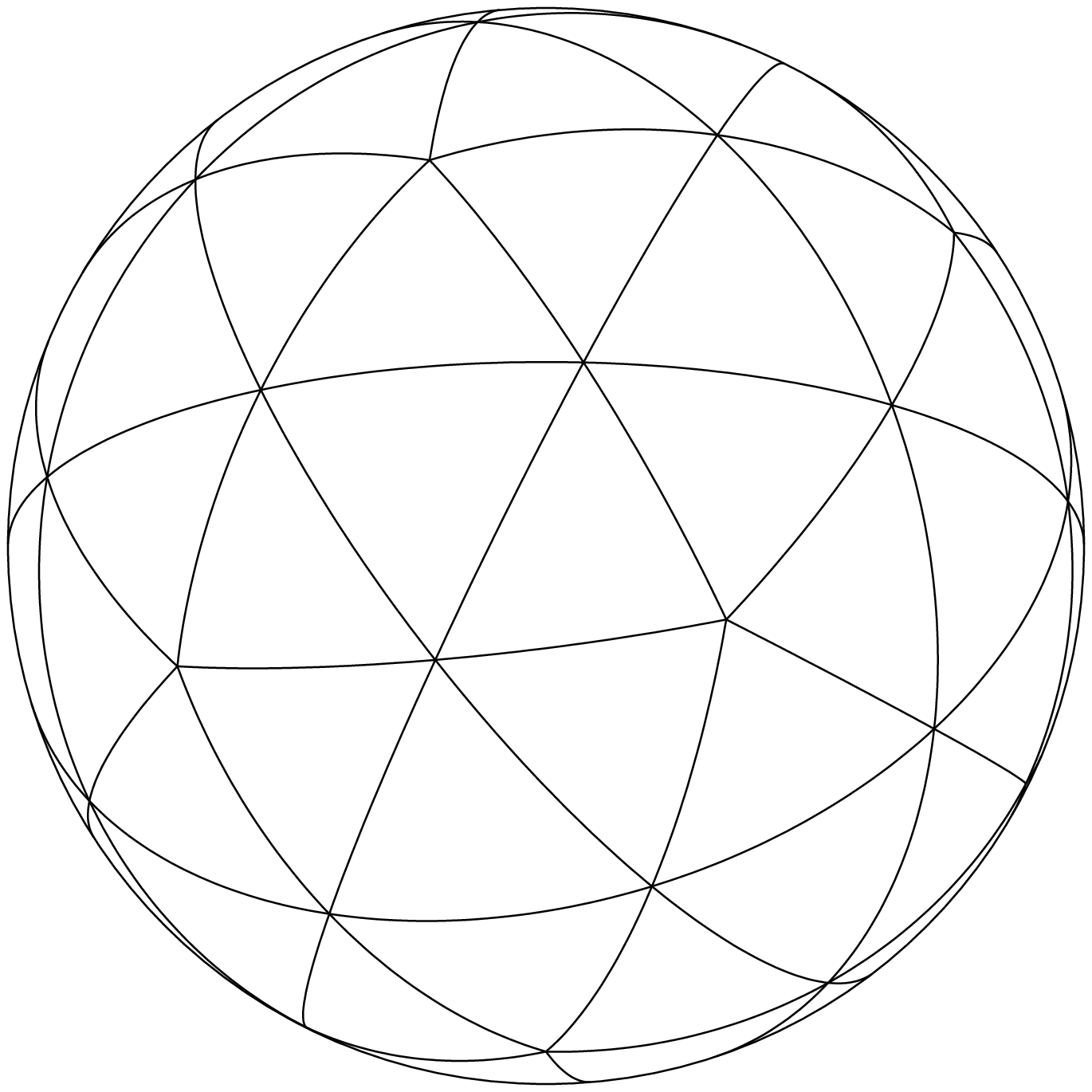}
\includegraphics[width=2.0in, clip=]{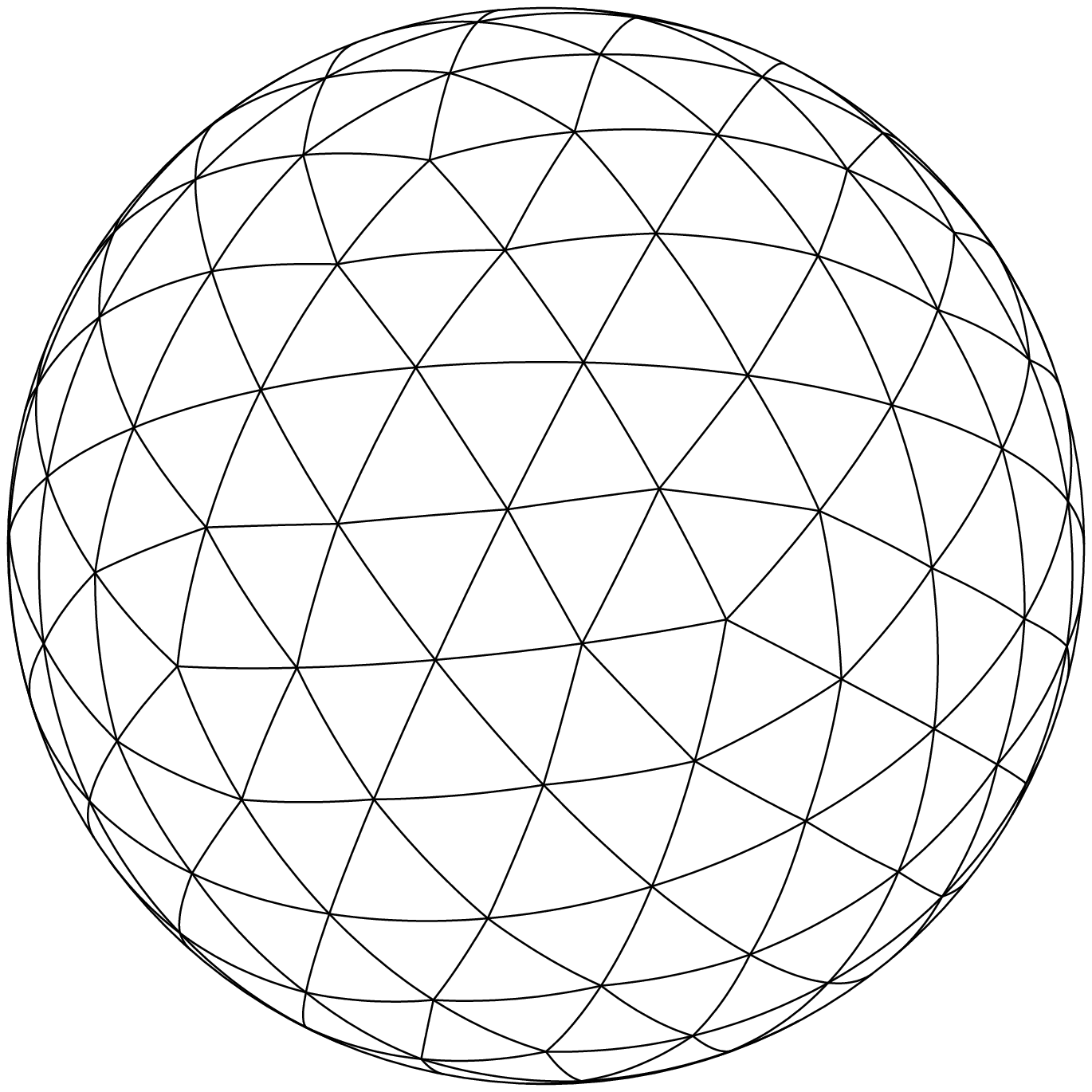}
\caption{Recursive icosahedral mesh generation. Shown are the inscribed icosahedron (top left) and the division 0 (top right), 1(bottom left), and 2(bottom right) triangular tesselations. The edges of the tesselation are repeatedly bisected until a desired level of refinement is reached.}
\label{fig_ico}
\end{figure}

\begin{figure}
\begin{center}
\includegraphics[width=3.0in, clip=]{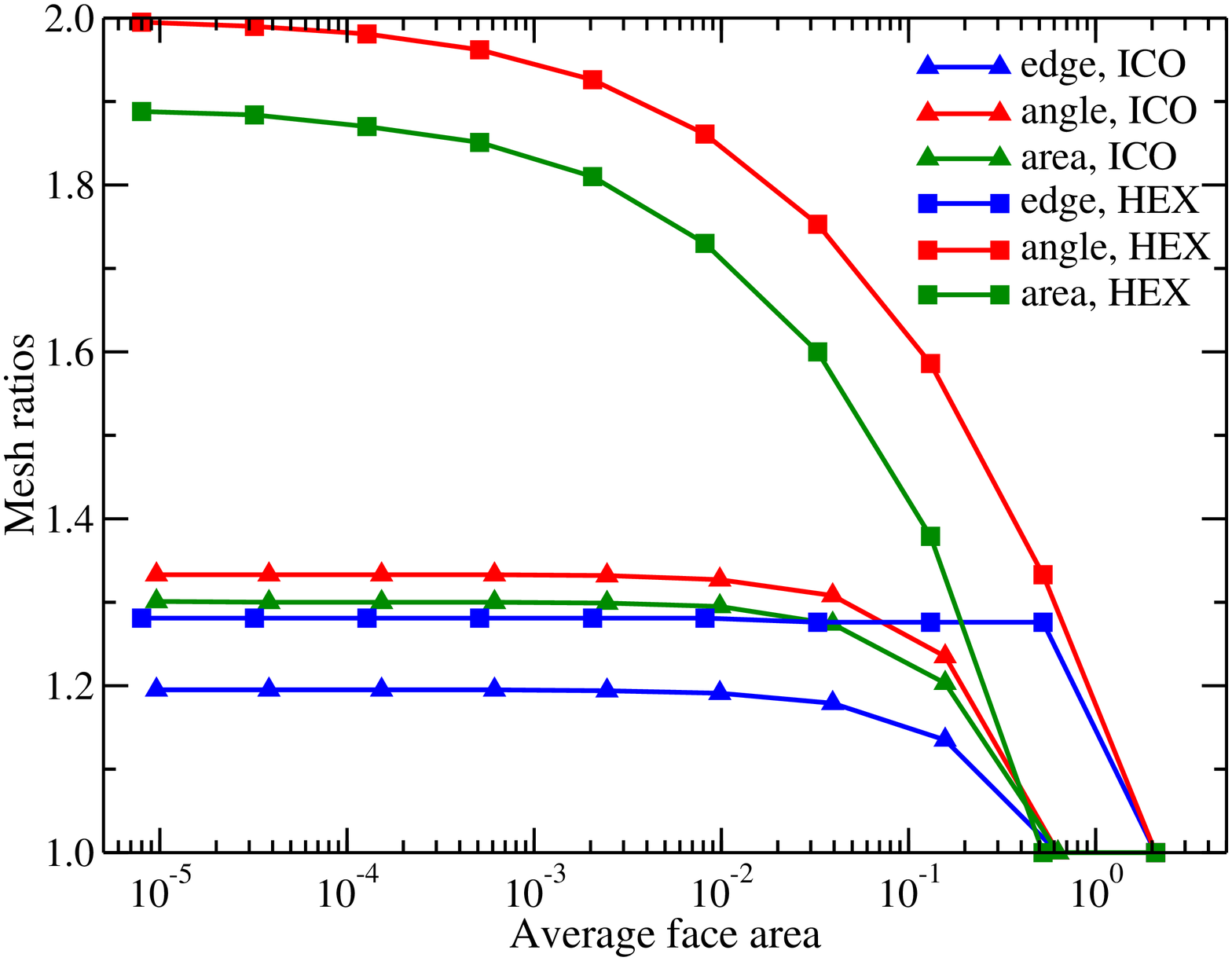}
\end{center}
\caption{Uniformity measures of the icosahedral mesh (triangles) and the hexahedral mesh (squares) for division 0--8 (former) and 0--9 (latter). Blue lines show edge ratios, red angle ratios, and green area ratios. The icosahedral TGM is measurably more uniform than the gnomonic cube sphere.}
\label{fig_icohex}
\end{figure}

\begin{figure}
\begin{center}
\includegraphics[width=4.0in, clip=]{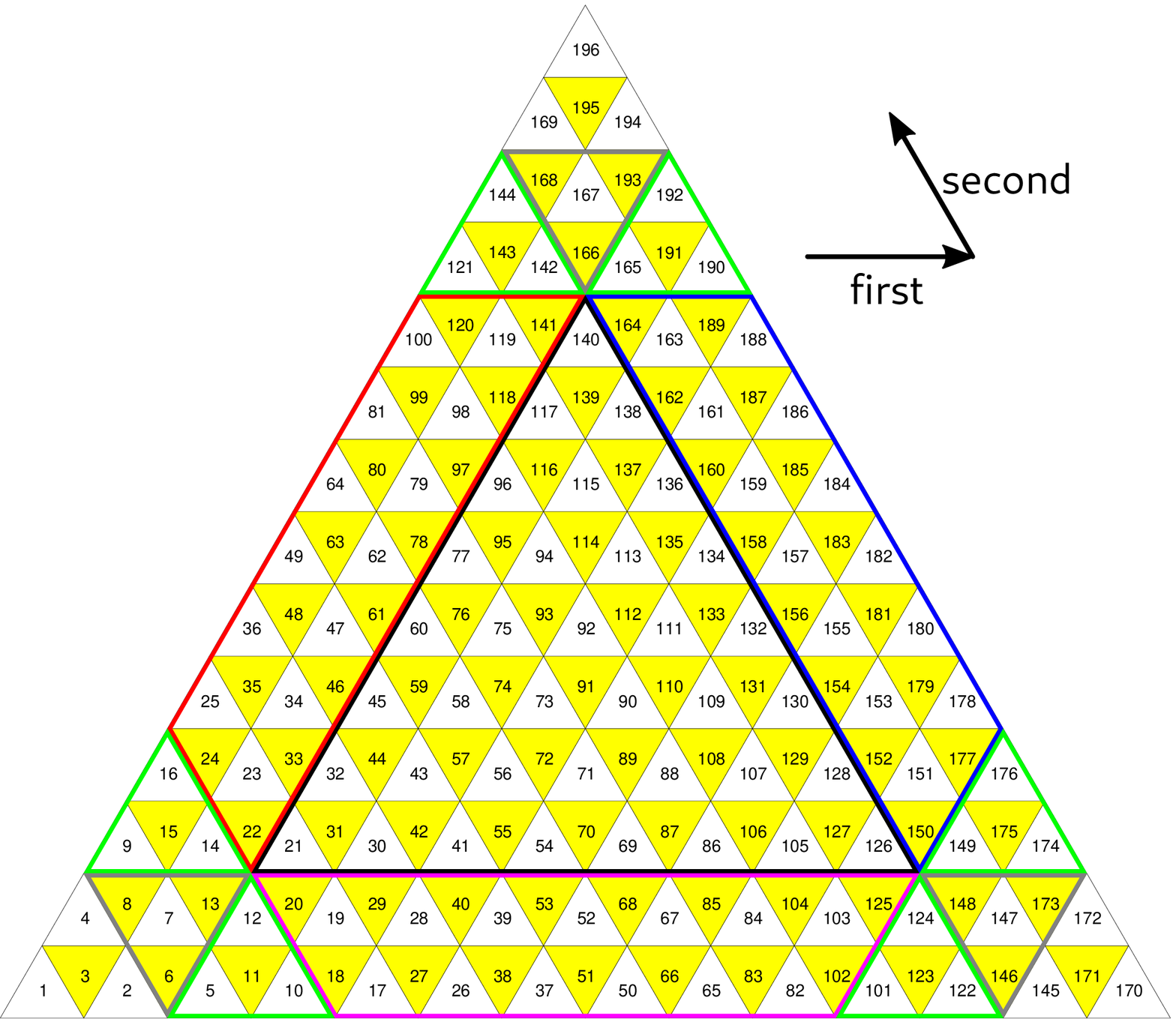}
\end{center}
\caption{A single sector of the mesh. In this example the face division is equal to the sector division plus three. The black arrows show the directions of the first and second TAS coordinates. Two layers of ghost faces are visible. The three trapezoidal and nine small triangular pieces marked with different border colors are the areas subject to boundary exchange with neighboring blocks.}
\label{fig_faces}
\end{figure}

\begin{figure}
\begin{center}
\includegraphics[width=3.5in, clip=]{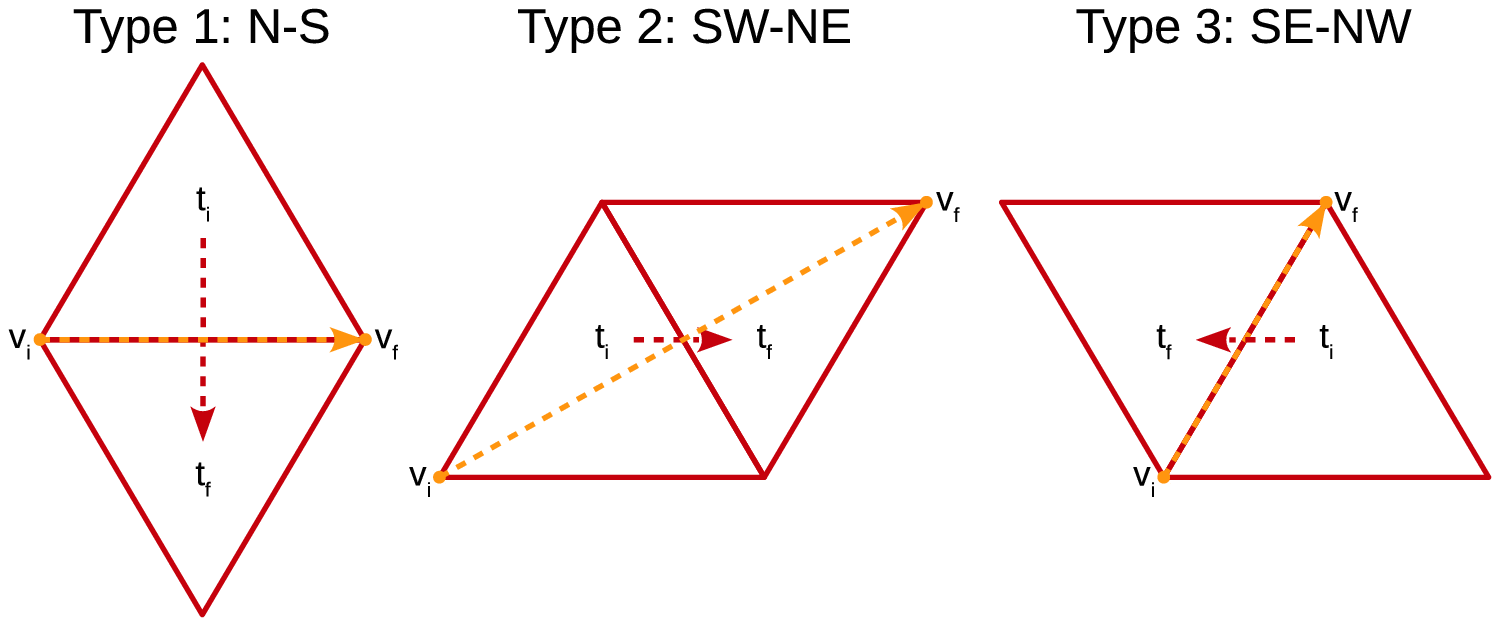}
\end{center}
\caption{Step operators on the mesh. The three unshaded to shaded step operators shown are used to walk the sector with its ghost t-faces. The step is from the vertex-face pair $v_i,t_i$ to $v_f,t_f$. The shaded to unshaded steps are obtained by switching the origin and destination vertex-face pairs.}
\label{fig_steps}
\end{figure}

\begin{figure}
\begin{center}
\includegraphics[width=2.0in, clip=]{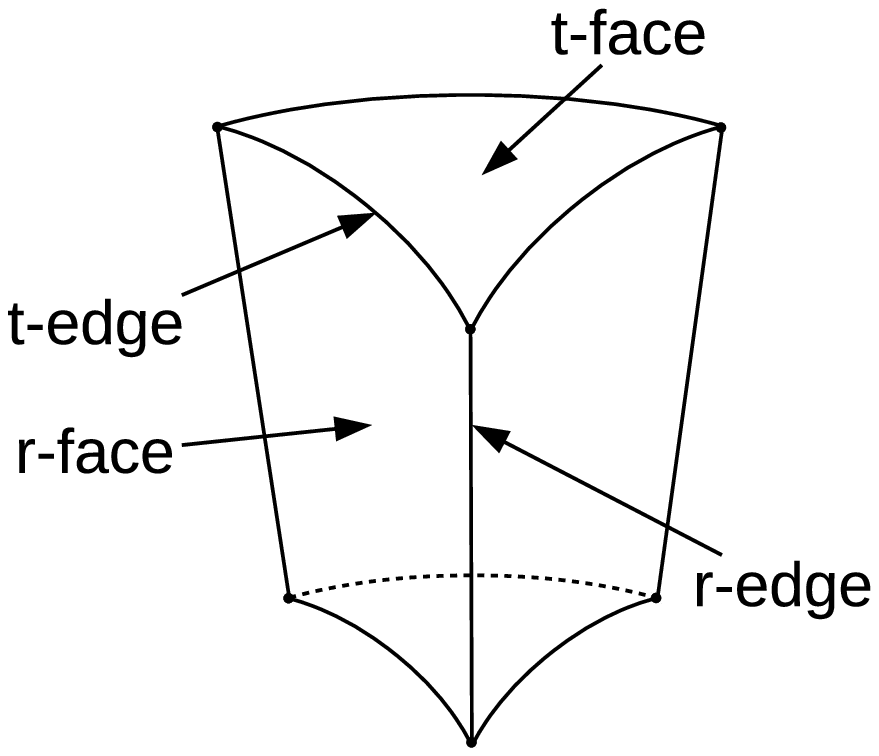}
\includegraphics[width=2.0in, clip=]{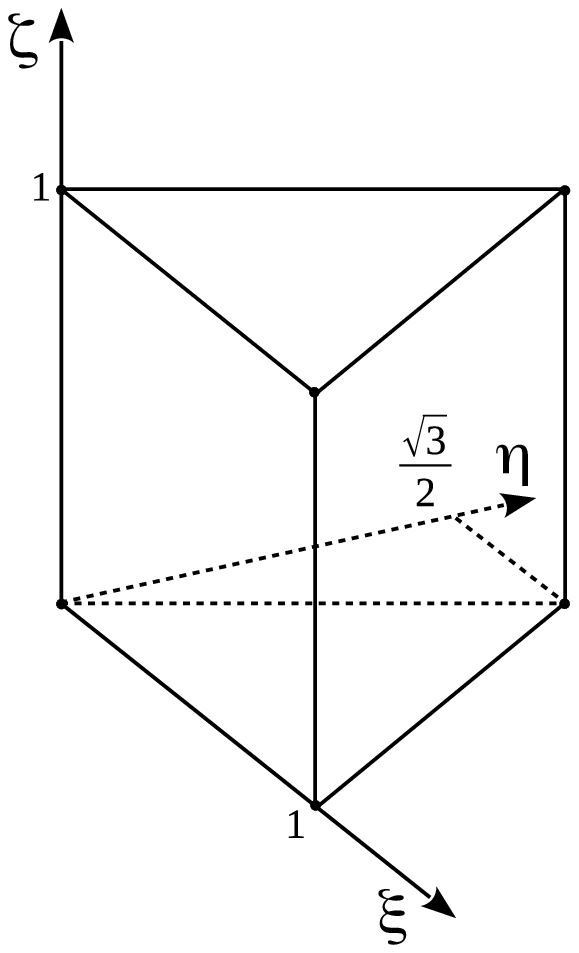}
\end{center}
\caption{A computational zone (left) and the reference element (right). The reference shape is a right equilateral triangular prism which is mapped into the frustum in physical space.}
\label{fig_prism}
\end{figure}

\begin{figure}
\includegraphics[width=1.5in, clip=]{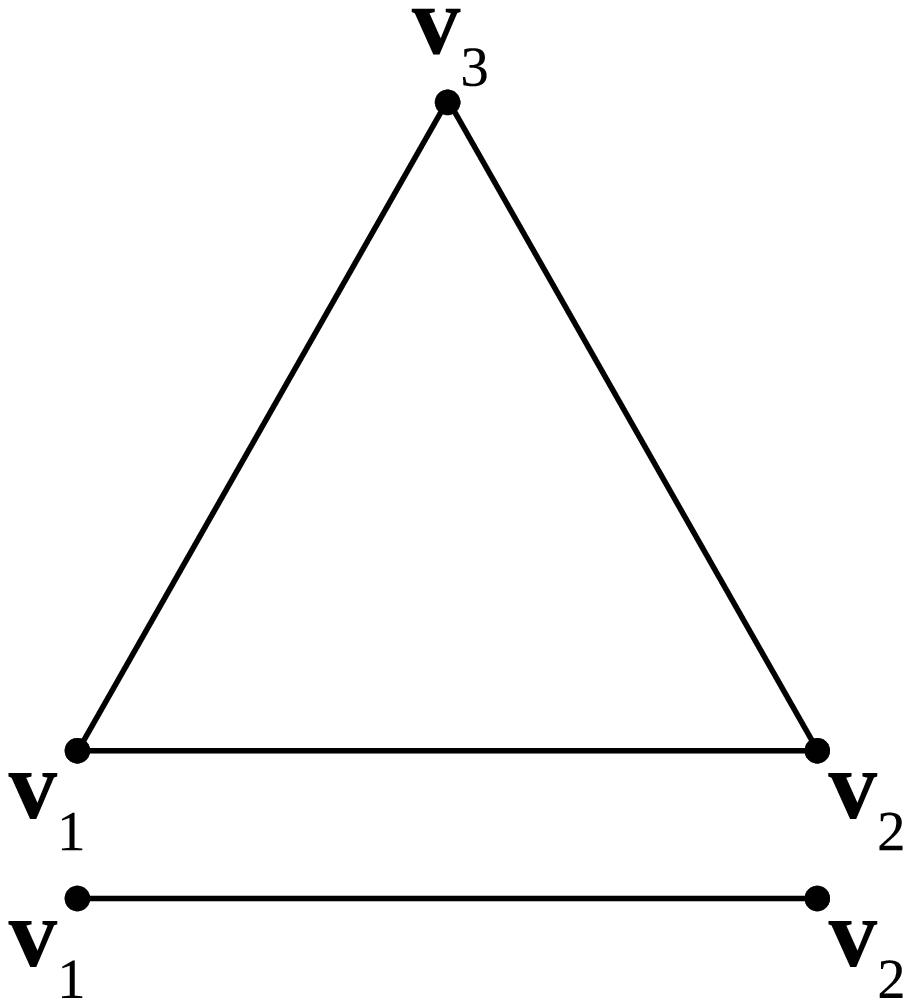}
\includegraphics[width=1.5in, clip=]{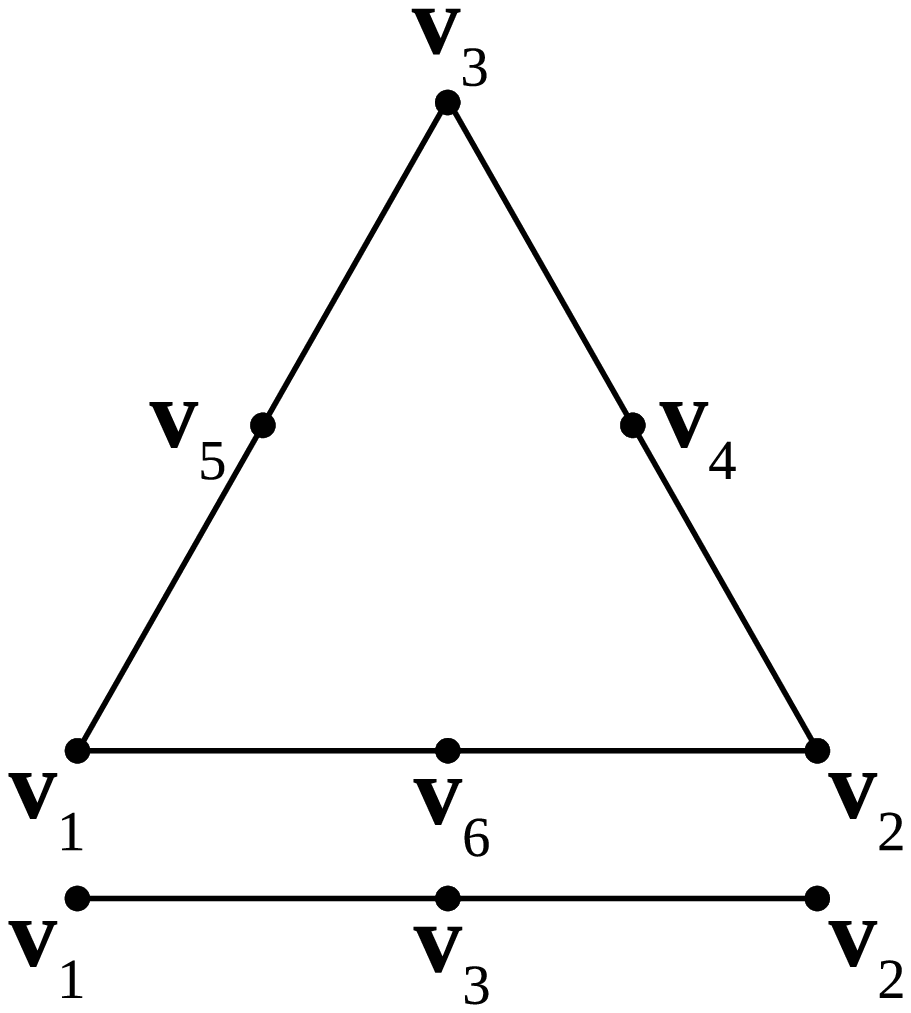}
\includegraphics[width=1.5in, clip=]{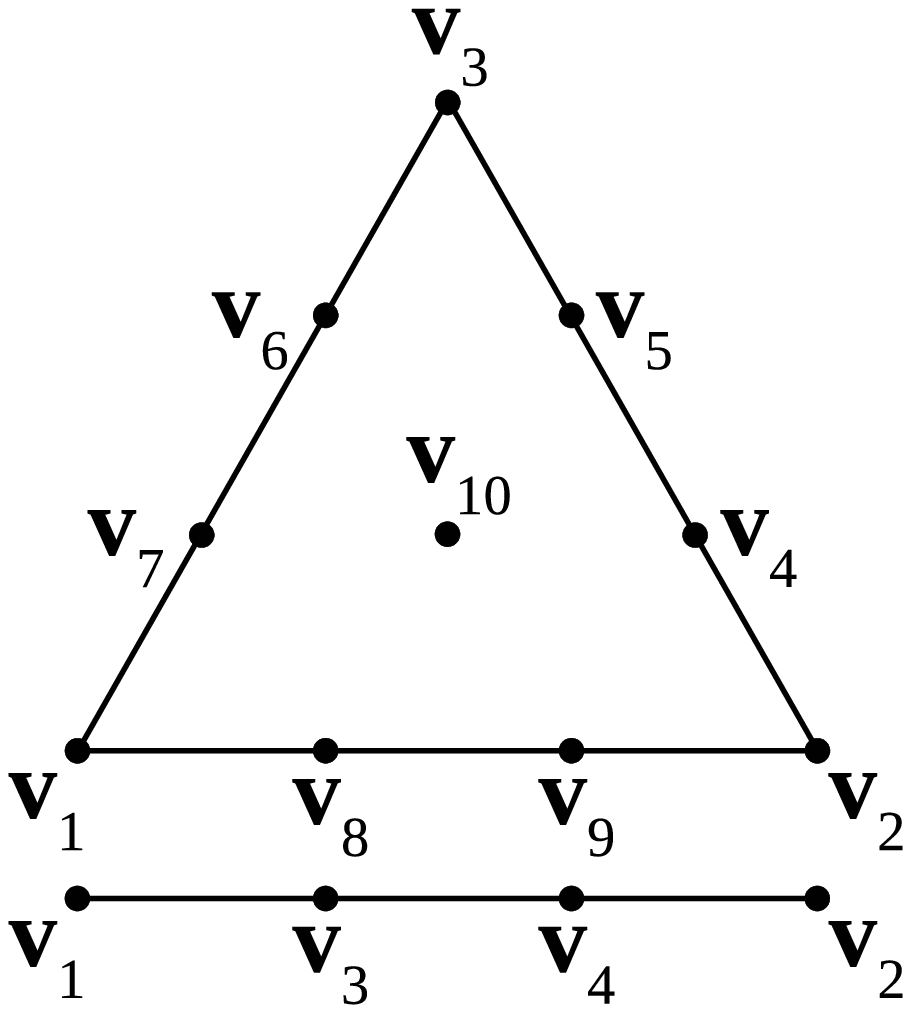}
\caption{Node locations on t-faces and t-edges for linear (left), quadratic (middle), and cubic (right) mapping. The first order surface element contains three nodes coincident with the vertices. The second order surface element includes all nodes from the first order elements plus the edge midpoints for the total of 6 nodes. The third order surface element includes all nodes from the first order elements plus the points at the thirds of each edge and the centroid, 10 nodes in total. The nodes of line elements representing t-edges are shown below each surface element.}
\label{fig_elements}
\end{figure}

\begin{figure}
\includegraphics[width=1.5in, clip=]{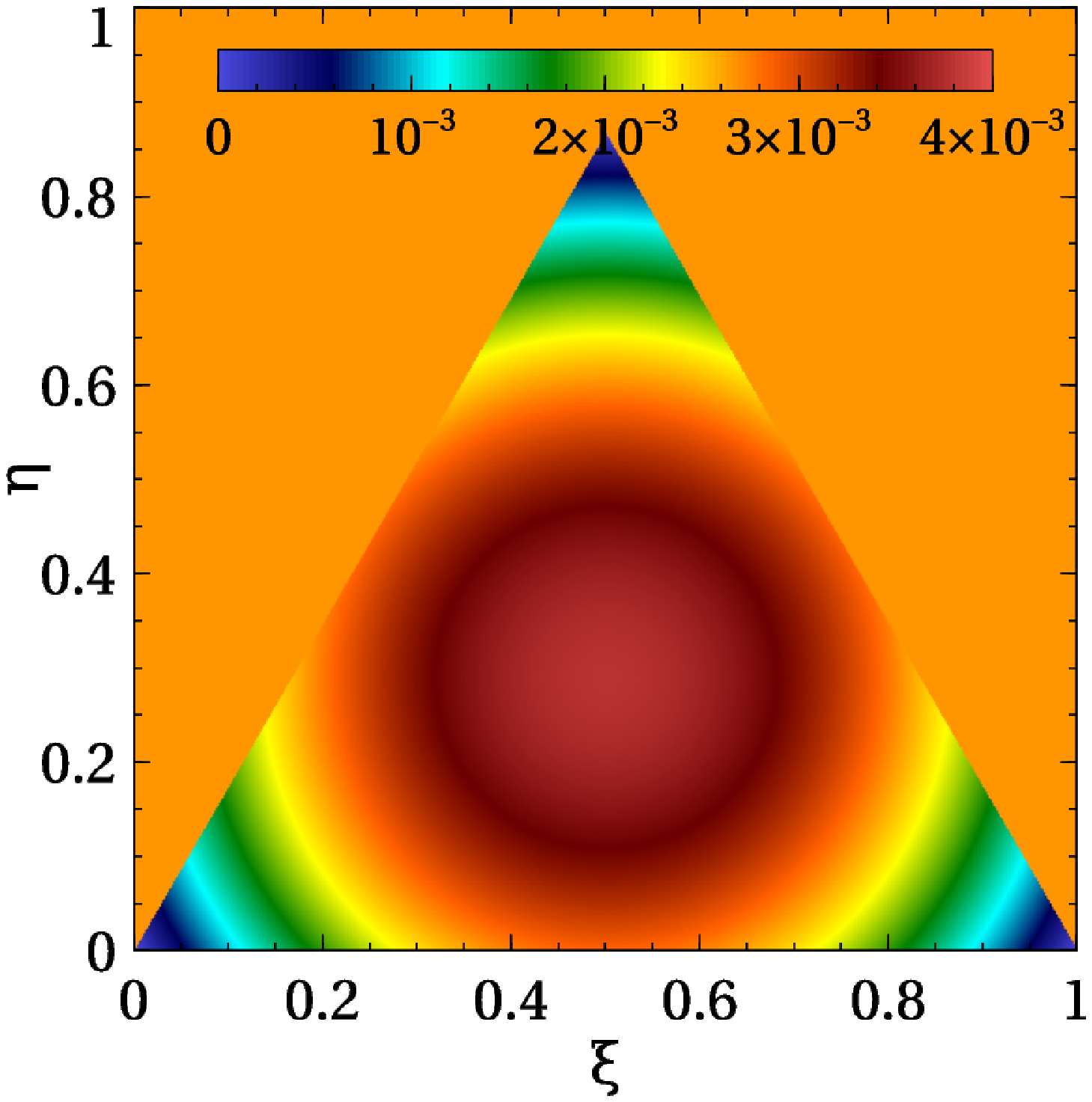}
\includegraphics[width=1.5in, clip=]{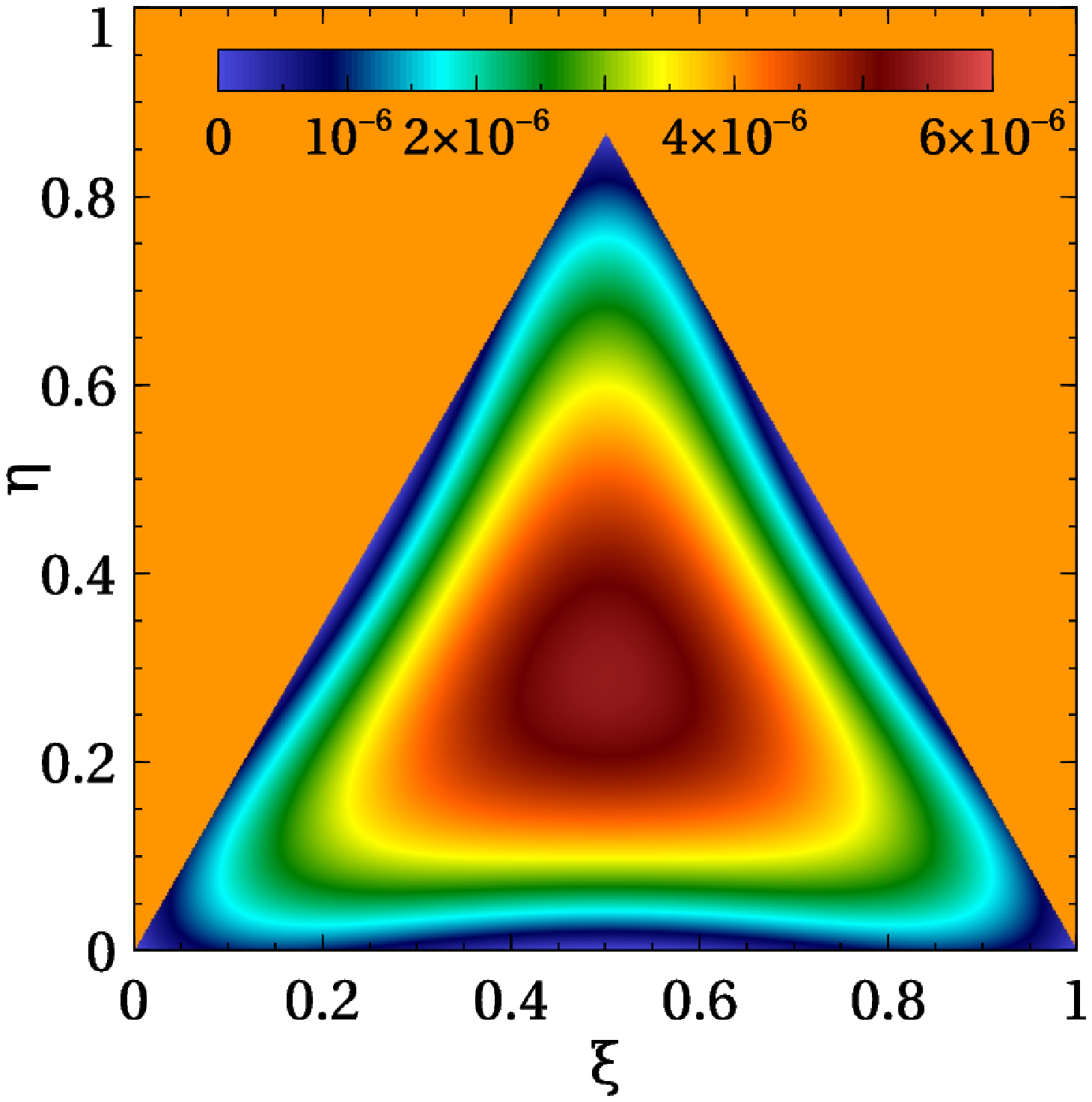}
\includegraphics[width=1.5in, clip=]{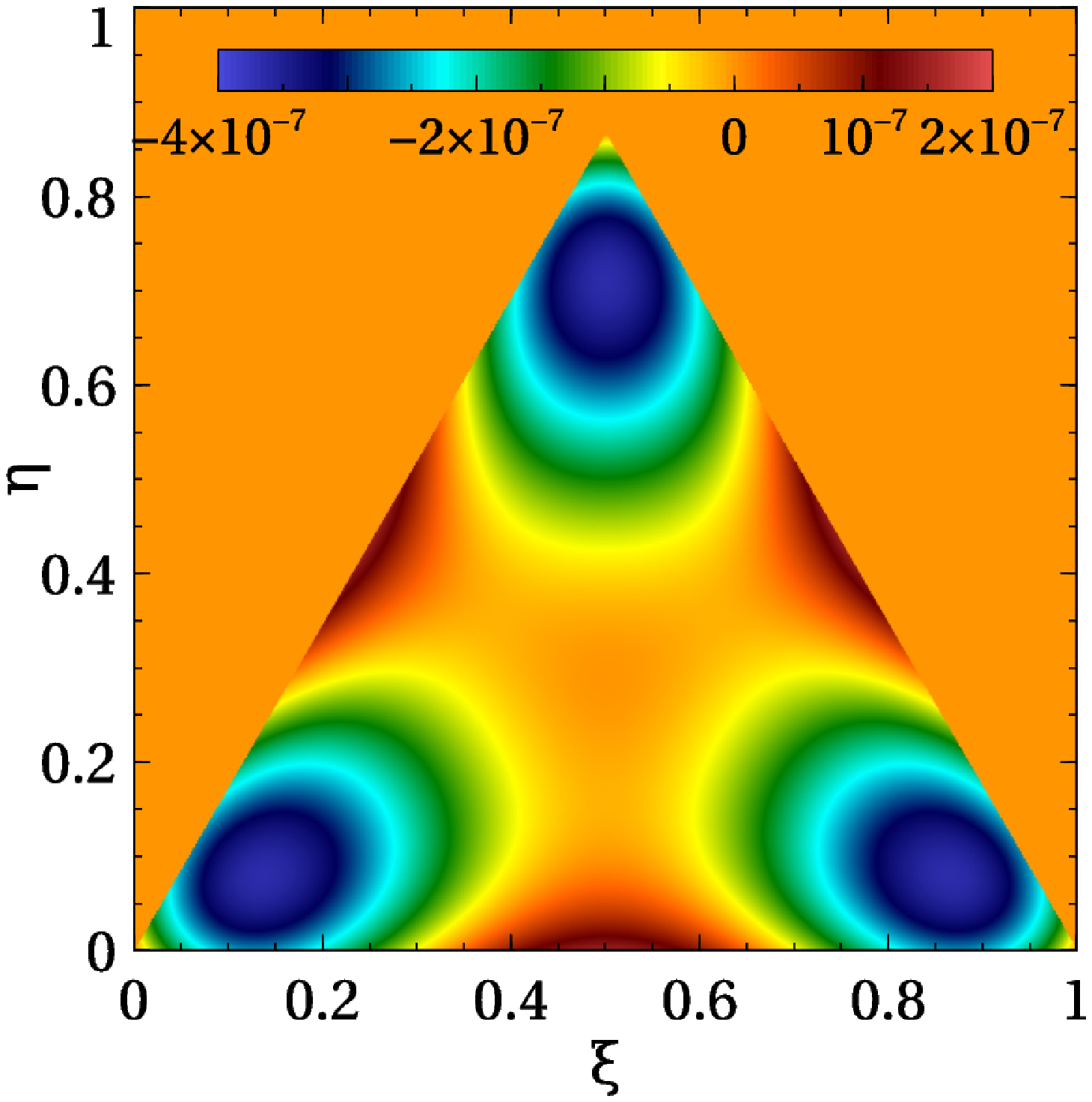}
\caption{Linear (left), quadratic (center), and cubic (right) mapping accuracy. Shown is the distance between the mapped surface and the perfect sphere computed for an equilateral triangle with a circumcircle radius of $5^\circ$.}
\label{fig_maperr}
\end{figure}

\begin{figure}
\begin{center}
\includegraphics[width=2.0in, clip=]{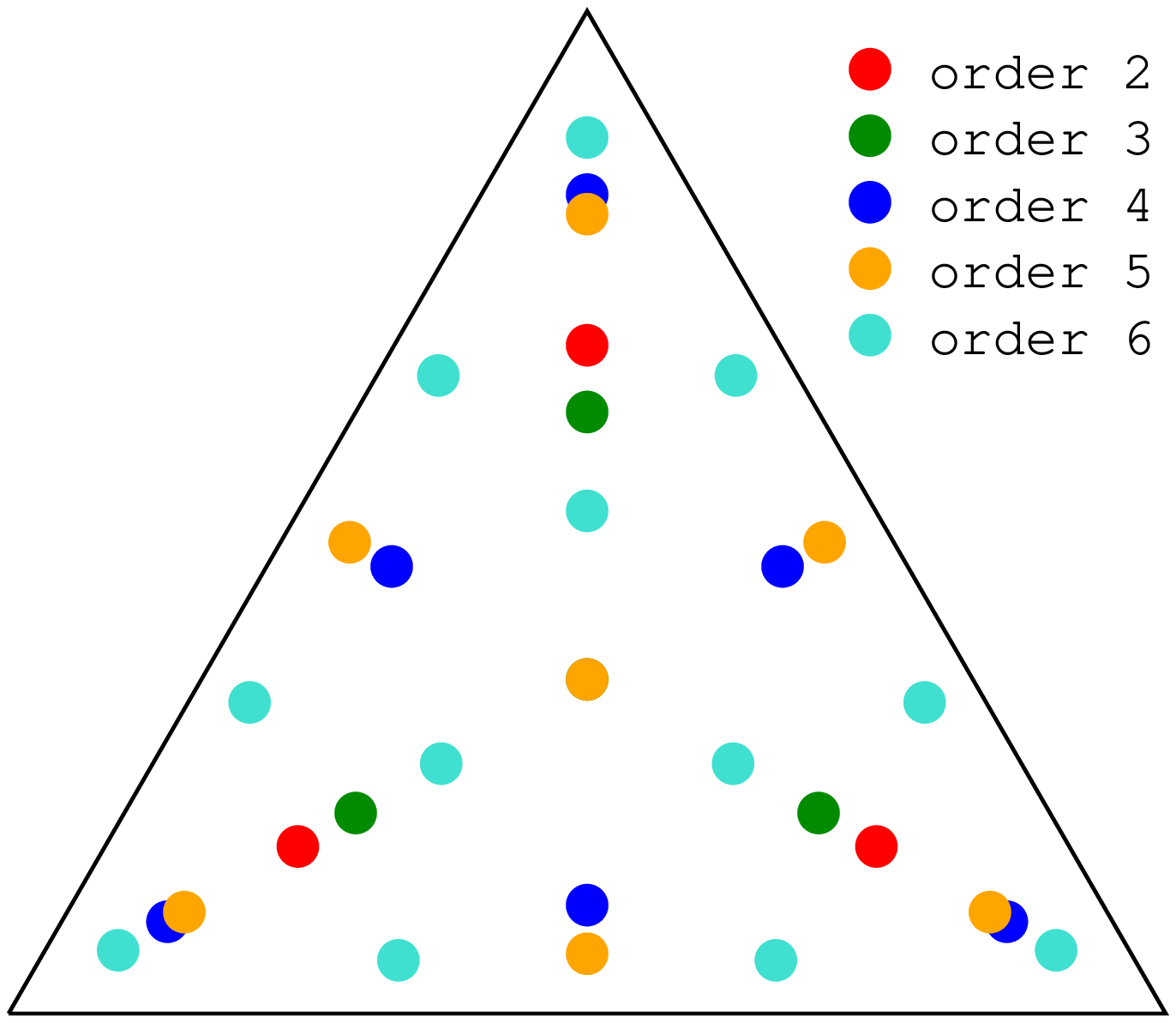}
\includegraphics[width=2.0in, clip=]{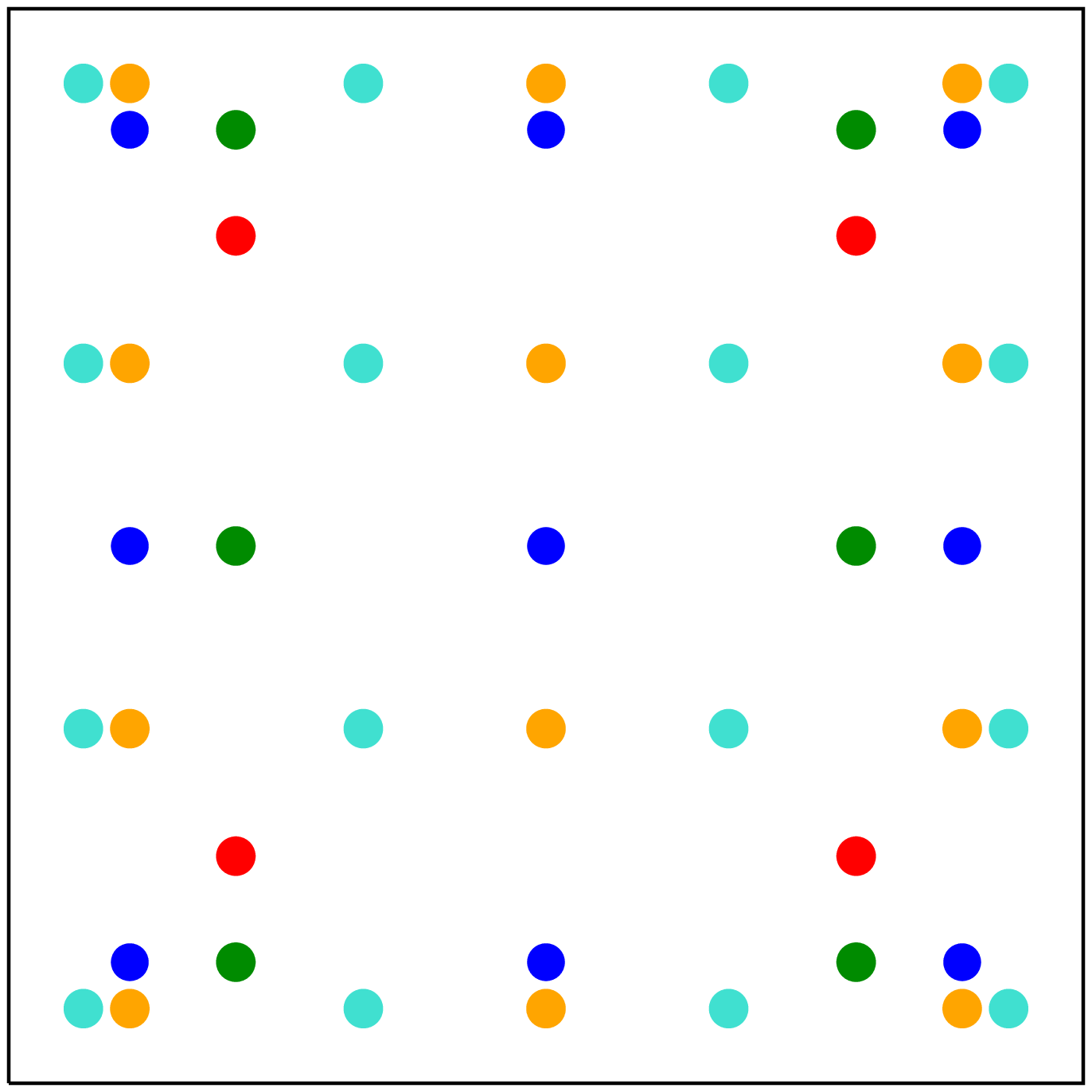}
\end{center}
\caption{Quadrature points on t-faces (left) and r-faces (right) of the prism. Three and four points are used to integrate a quadratic function exactly on t-faces and r-faces, respectively, four and six for cubic, six and nine for quartic, seven and twelve for quintic, and twelve and sixteen for sextic functions. The four point rule is not used on t-faces; instead the six point rule is used for third degree polynomials.}
\label{fig_quadpts}
\end{figure}

\begin{figure}
\includegraphics[width=0.5in, clip=]{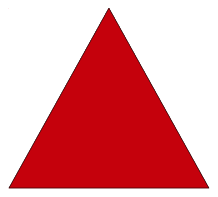}
\includegraphics[width=0.5in, clip=]{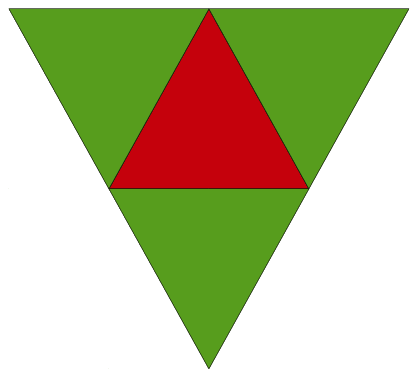}
\includegraphics[width=0.5in, clip=]{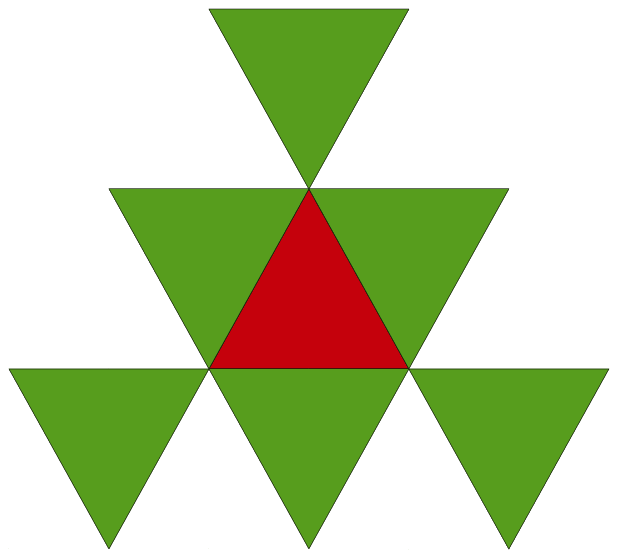}
\includegraphics[width=0.5in, clip=]{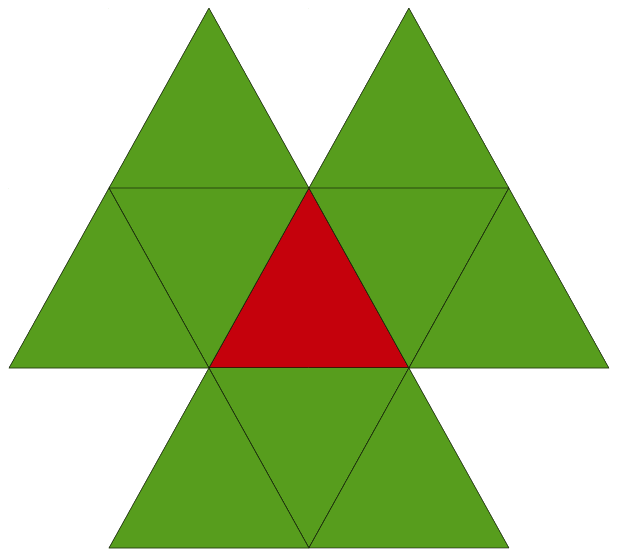}
\includegraphics[width=0.5in, clip=]{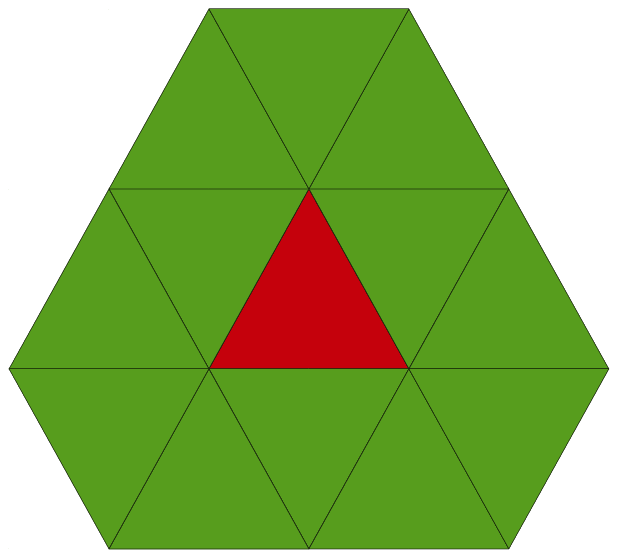}
\includegraphics[width=0.5in, clip=]{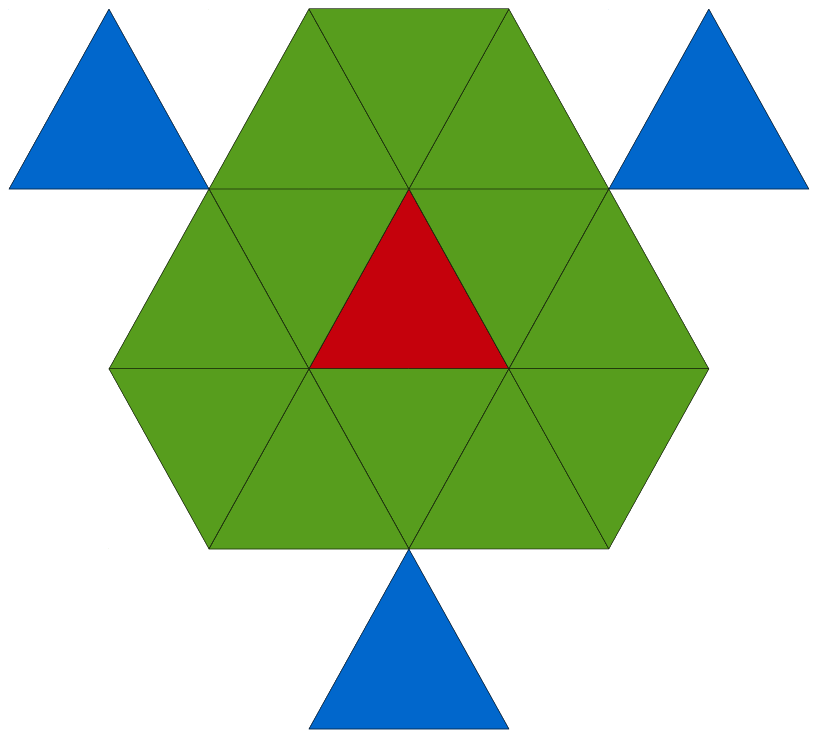}
\includegraphics[width=0.5in, clip=]{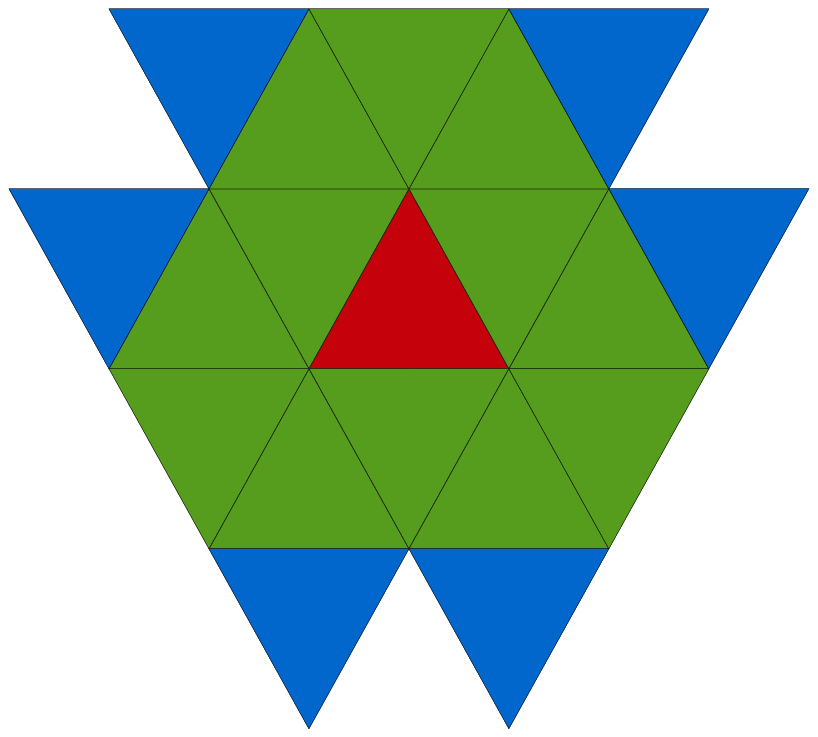}
\includegraphics[width=0.5in, clip=]{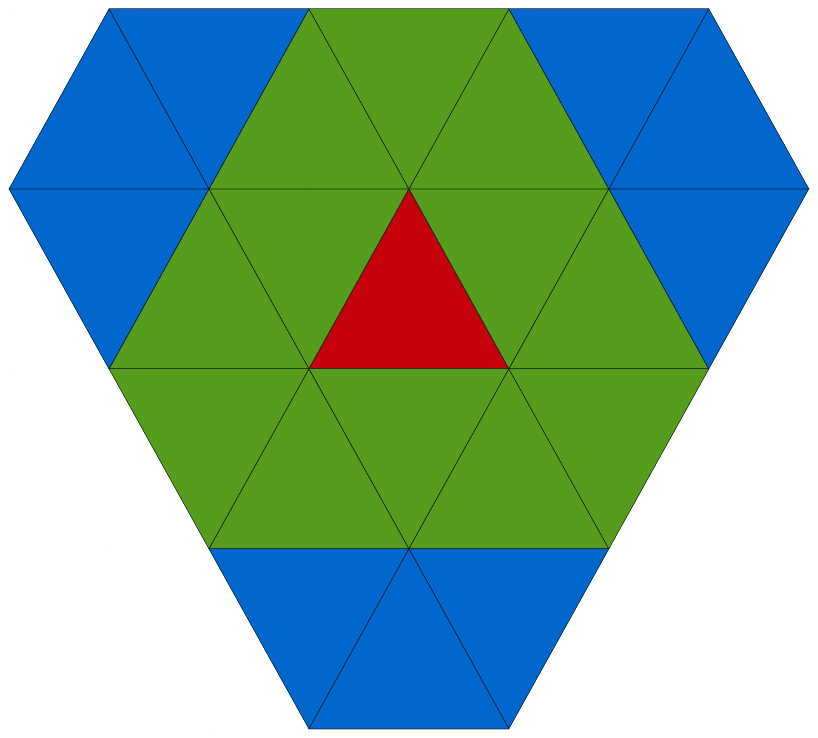}

\includegraphics[width=0.5in, clip=]{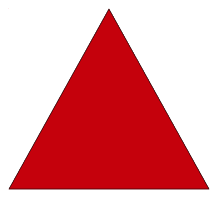}
\includegraphics[width=0.5in, clip=]{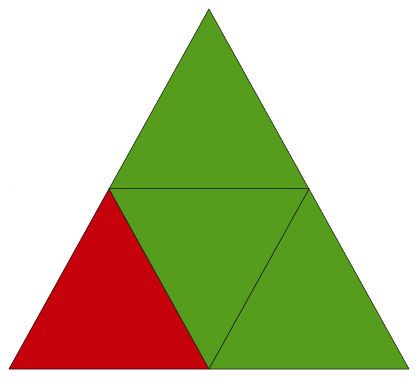}
\includegraphics[width=0.5in, clip=]{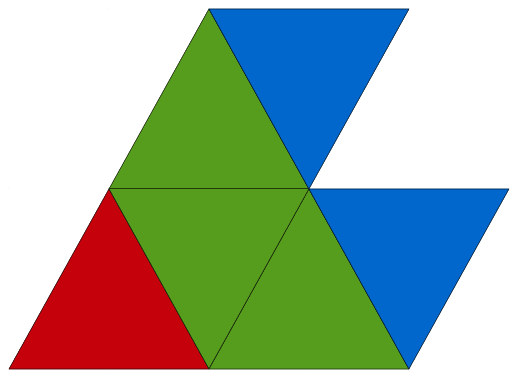}
\includegraphics[width=0.5in, clip=]{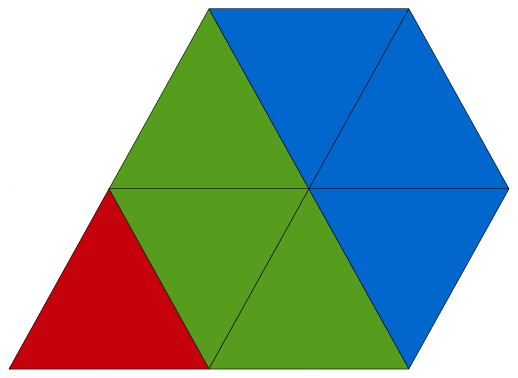}
\includegraphics[width=0.5in, clip=]{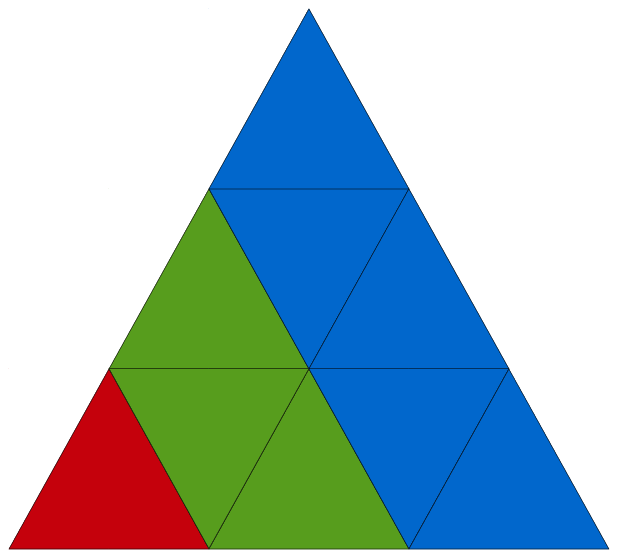}
\includegraphics[width=0.5in, clip=]{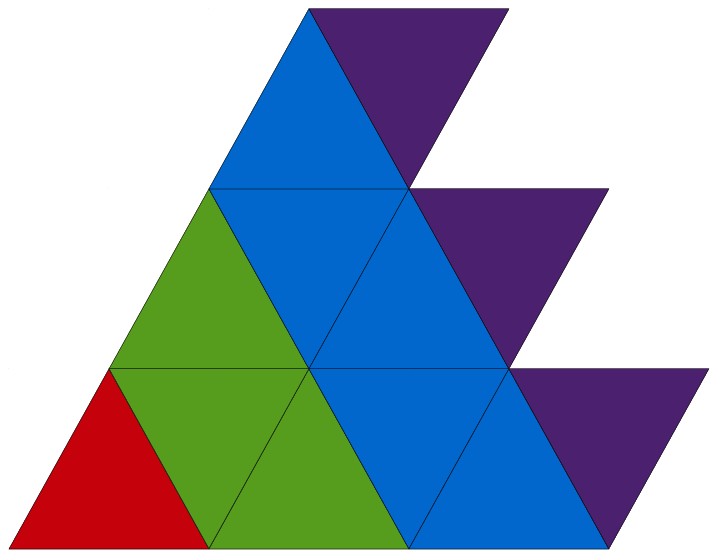}
\includegraphics[width=0.5in, clip=]{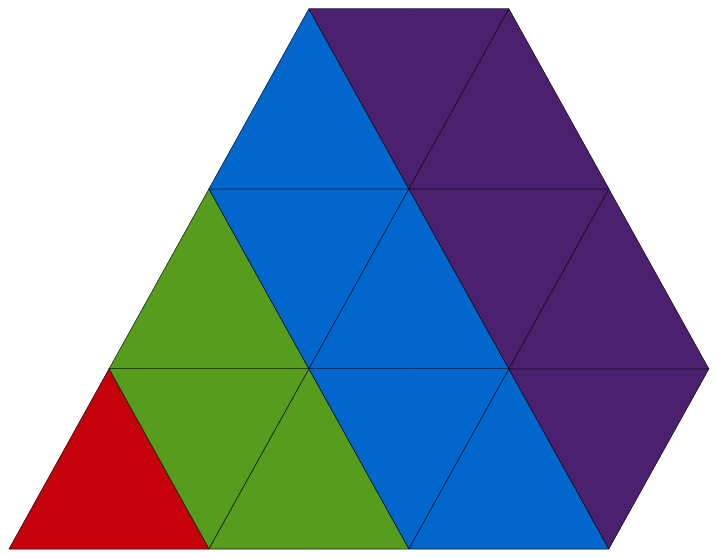}
\includegraphics[width=0.5in, clip=]{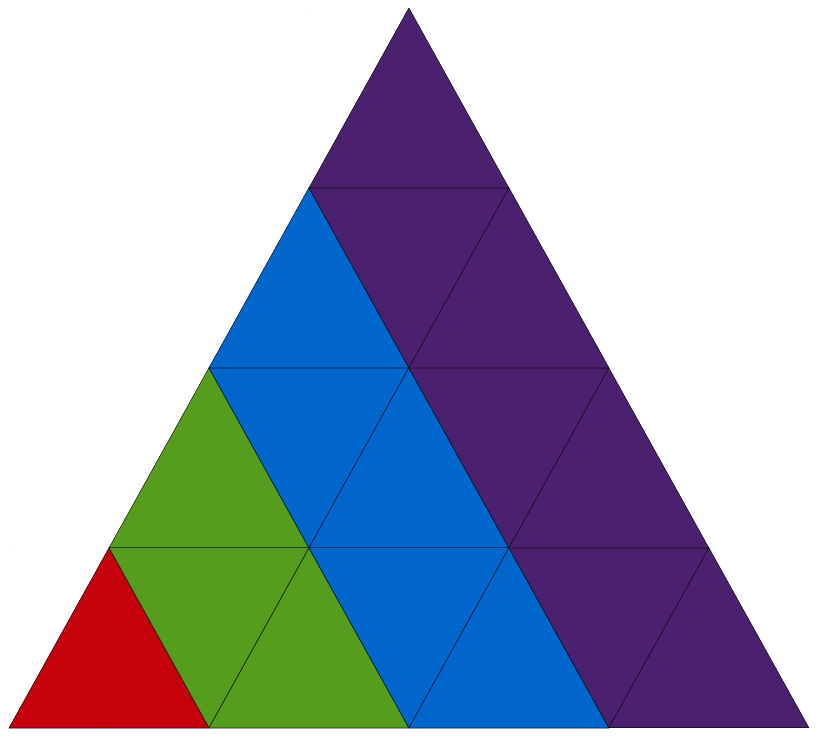}

\includegraphics[width=0.5in, clip=]{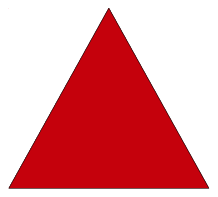}
\includegraphics[width=0.5in, clip=]{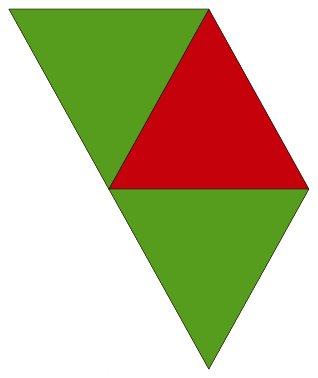}
\includegraphics[width=0.5in, clip=]{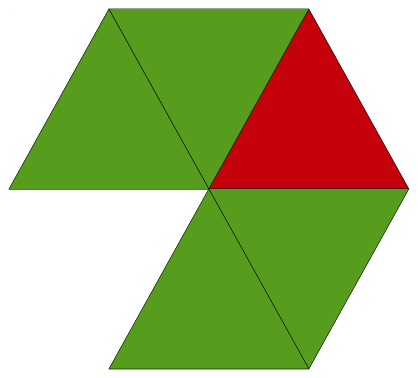}
\includegraphics[width=0.5in, clip=]{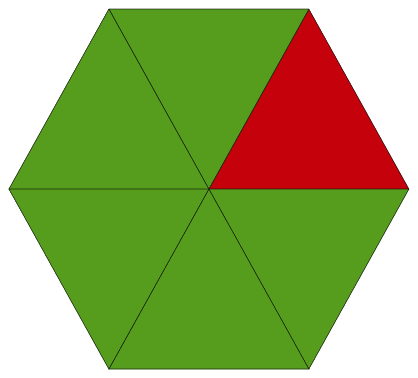}
\includegraphics[width=0.5in, clip=]{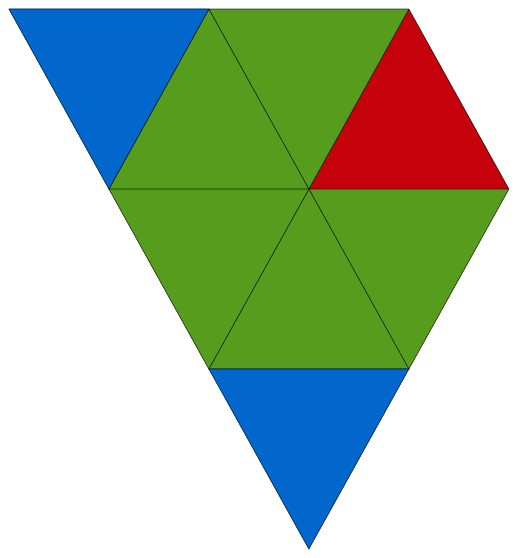}
\includegraphics[width=0.5in, clip=]{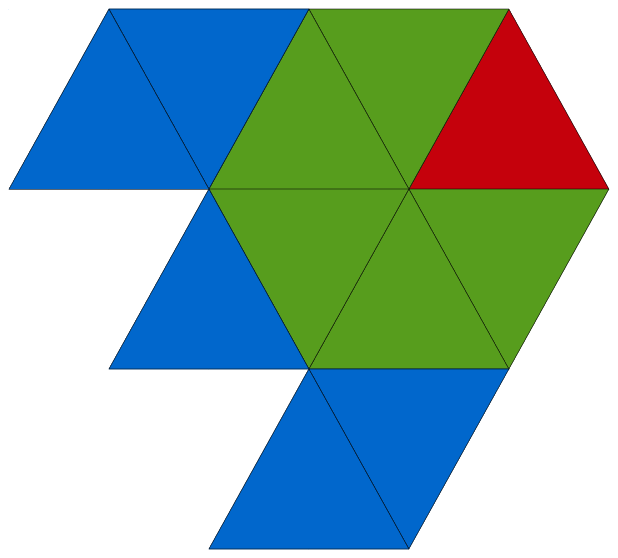}
\includegraphics[width=0.5in, clip=]{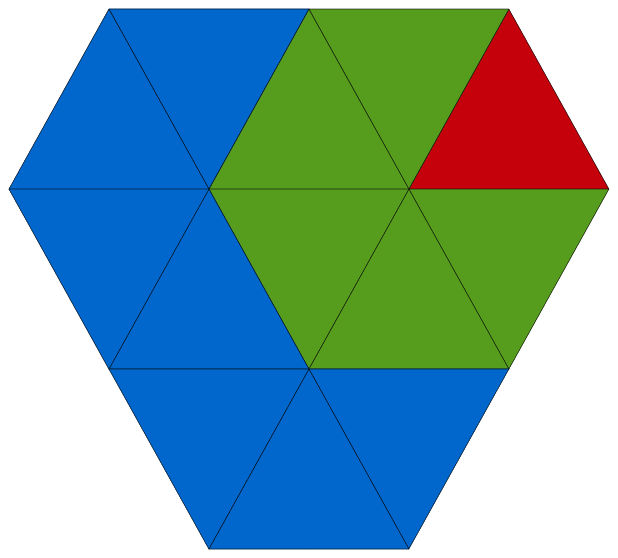}
\includegraphics[width=0.5in, clip=]{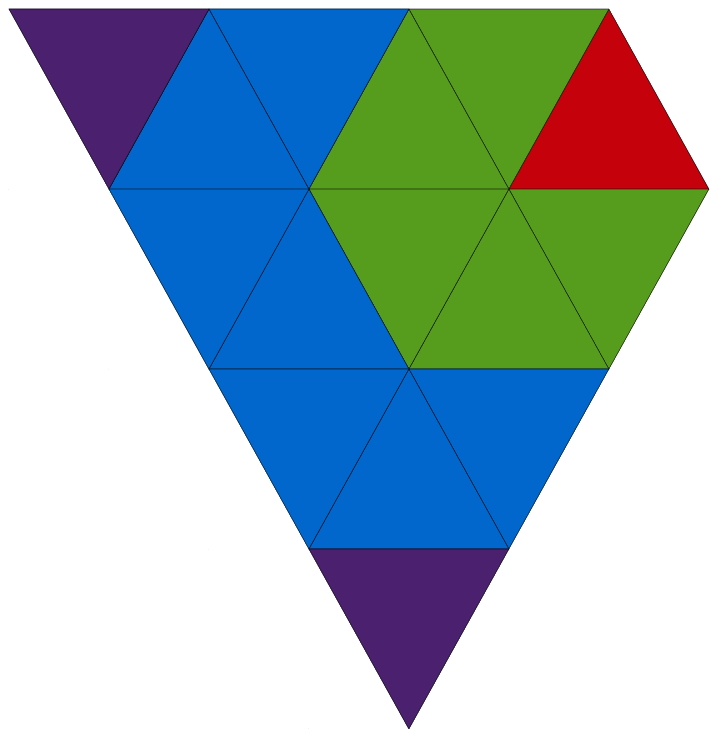}
\caption{Stencil shapes in the plane. The top, middle, and bottom rows show the central, forward-biased, and backward-biased stencils, respectively. The principal zone is shown in red. Green, blue, and purple colors represent first, second, and third von Neumann neighbors of the principal zone, respectively.}
\label{fig_stshapes}
\end{figure}

\begin{figure}
\begin{center}
\includegraphics[width=4in, clip=]{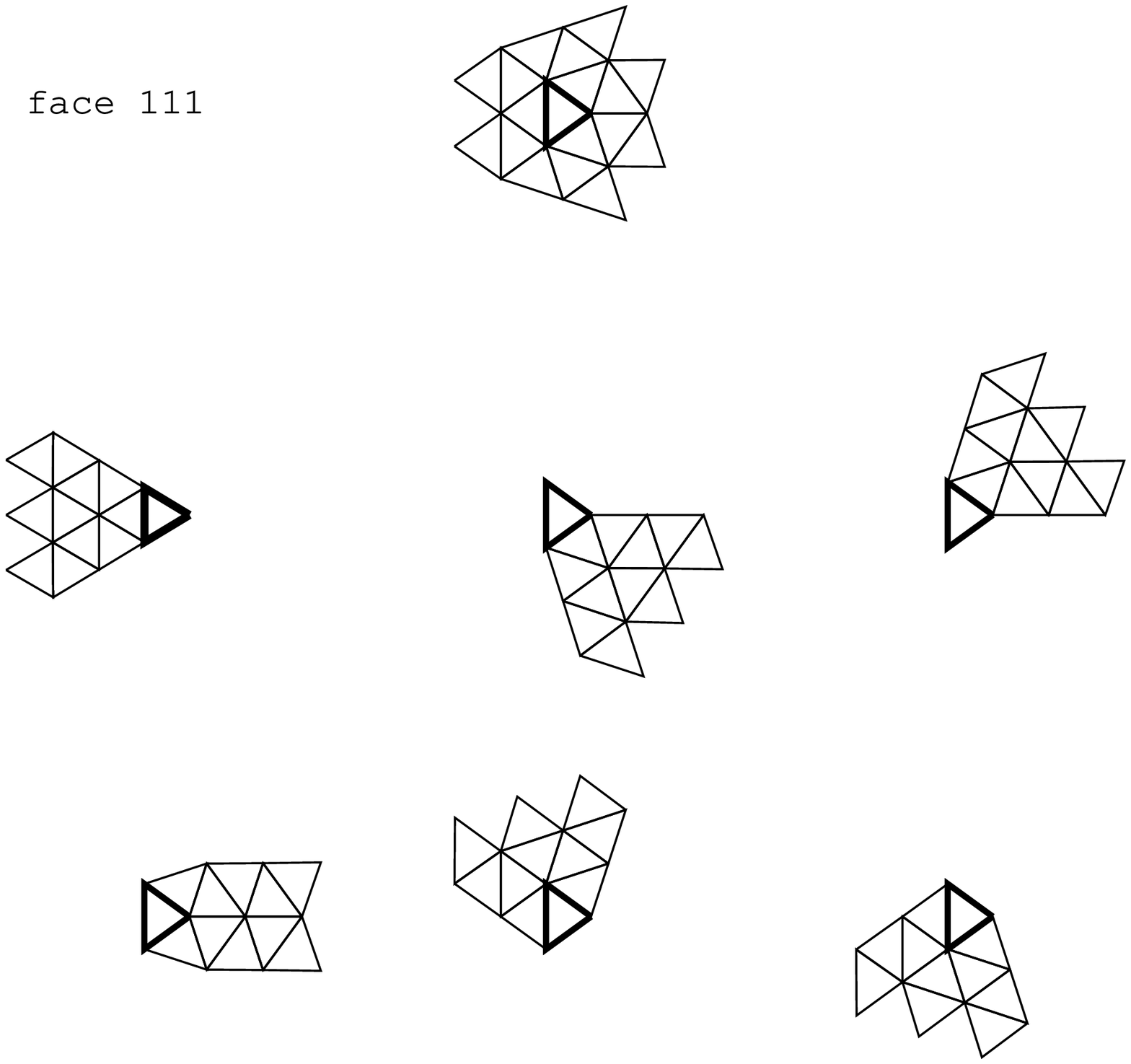}
\end{center}
\caption{Stencils of one selected face at division four near a penta-corner. Shown are the central stencil (top), the forward stencils (middle row) and the backward stencils (bottom row). The principal face is drawn with thick lines. Because of the penta-corner to the right of the principal face, some of the stencils have a different shape.}
\label{fig_7stencils}
\end{figure}

\begin{figure}
\begin{center}
\includegraphics[width=2.0in, clip=]{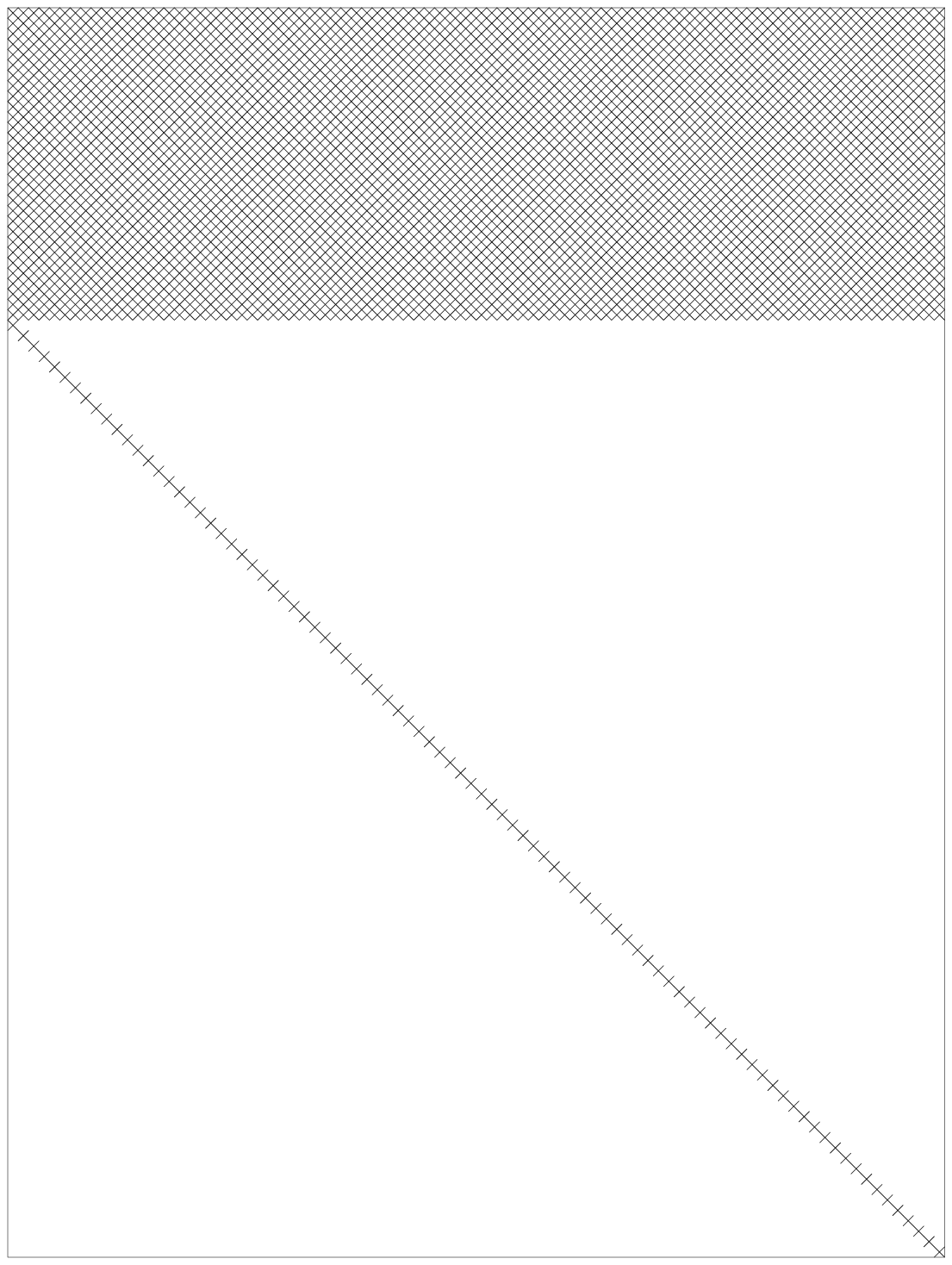}
\includegraphics[width=2.0in, clip=]{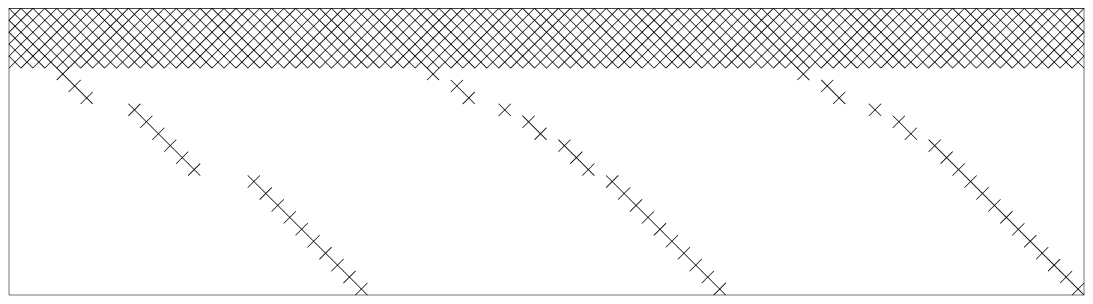}
\end{center}
\caption{Matrix structure for divergence-free reconstruction. The left panel shows the nonzero elements (marked with x's) of the least squares matrix and the right panel shows the same for the constraint matrix for fourth order reconstruction. The corresponding KKT matrix is $114\times 114$.}
\label{fig_matstr}
\end{figure}

\begin{figure}
\begin{center}
\includegraphics[width=4.0in, clip=]{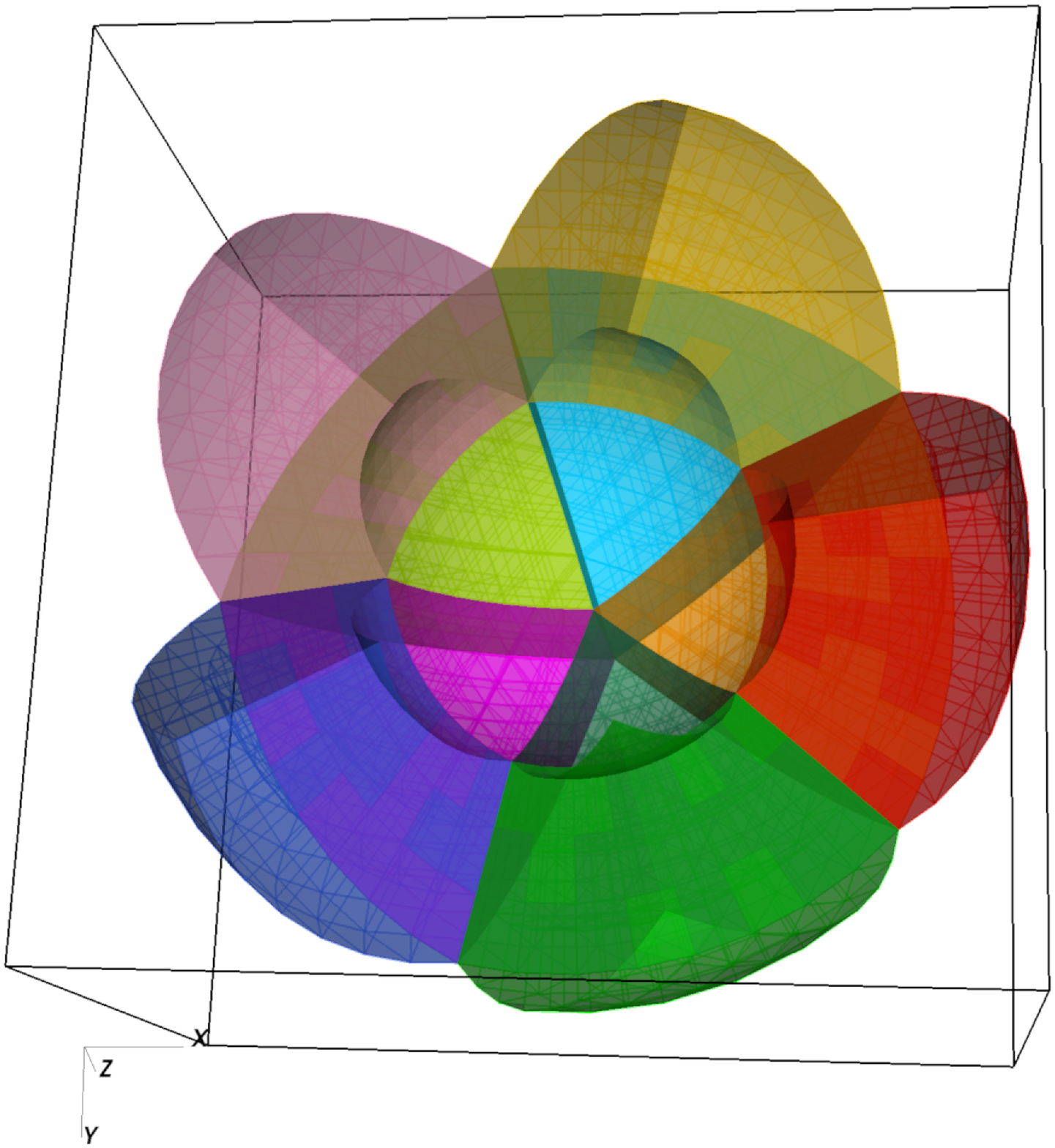}
\end{center}
\caption{Cutout view of the mesh showing ten blocks out of 20 (the smallest number of blocks on this mesh, consisting of a single slab and 20 division 0 sectors). Each block is shaded using a unique color.}
\label{fig_cutout}
\end{figure}

\begin{figure}
\begin{center}
\includegraphics[width=3.0in, clip=]{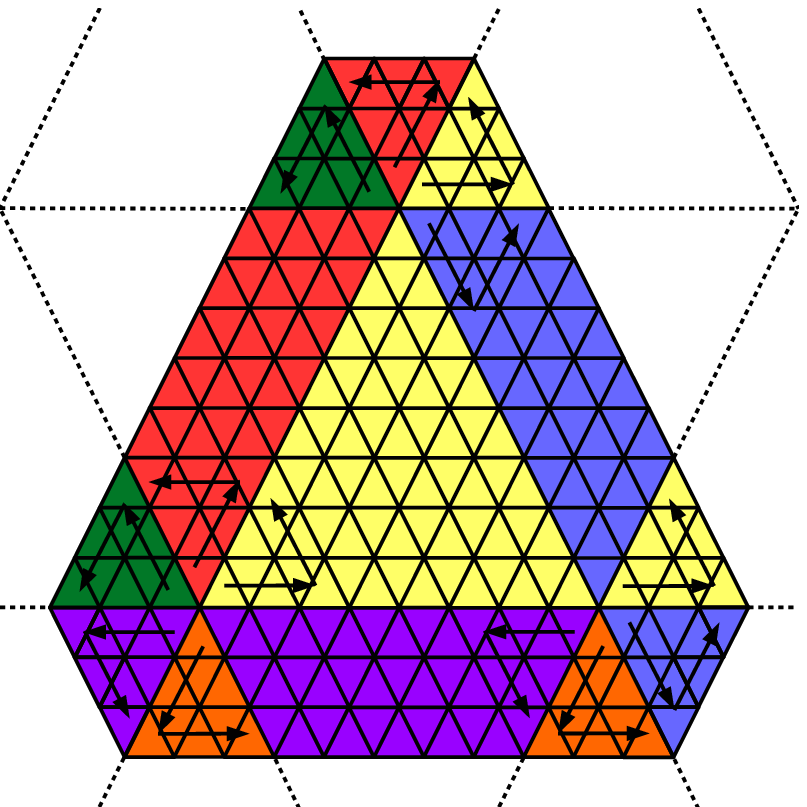}
\end{center}
\caption{Neighbor arrangement around a block. The pairs of arrows indicate the directions of the first and the second TAS coordinates, respectively. The different rotations of the TAS are shown with different color shading.}
\label{fig_nbrs}
\end{figure}

\begin{figure}
\begin{center}
\includegraphics[width=4.5in, clip=]{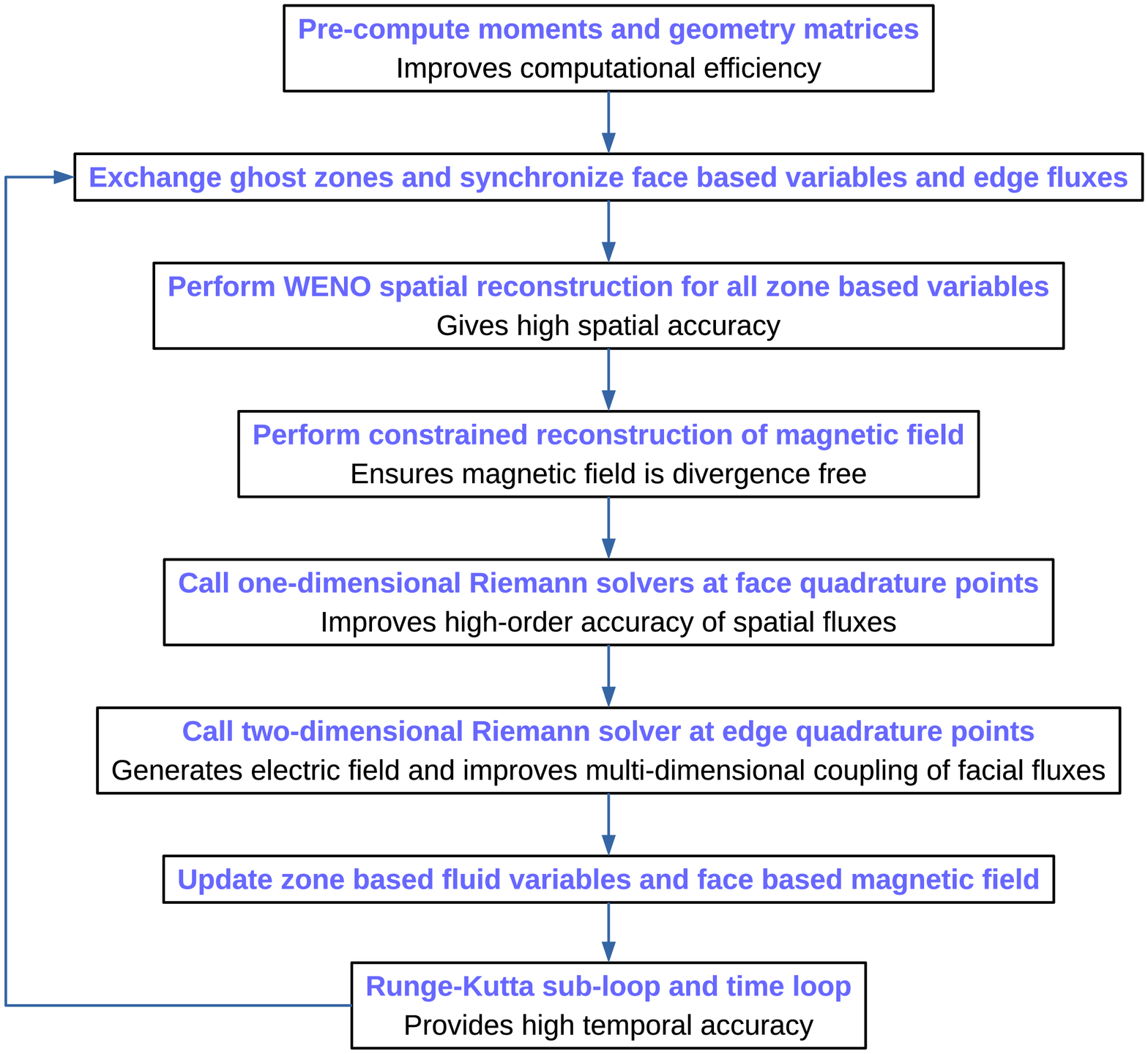}
\end{center}
\caption{Sequence of steps in one iteration of the time loop.}
\label{fig_loop}
\end{figure}

\begin{figure}
\includegraphics[width=2.25in, clip=]{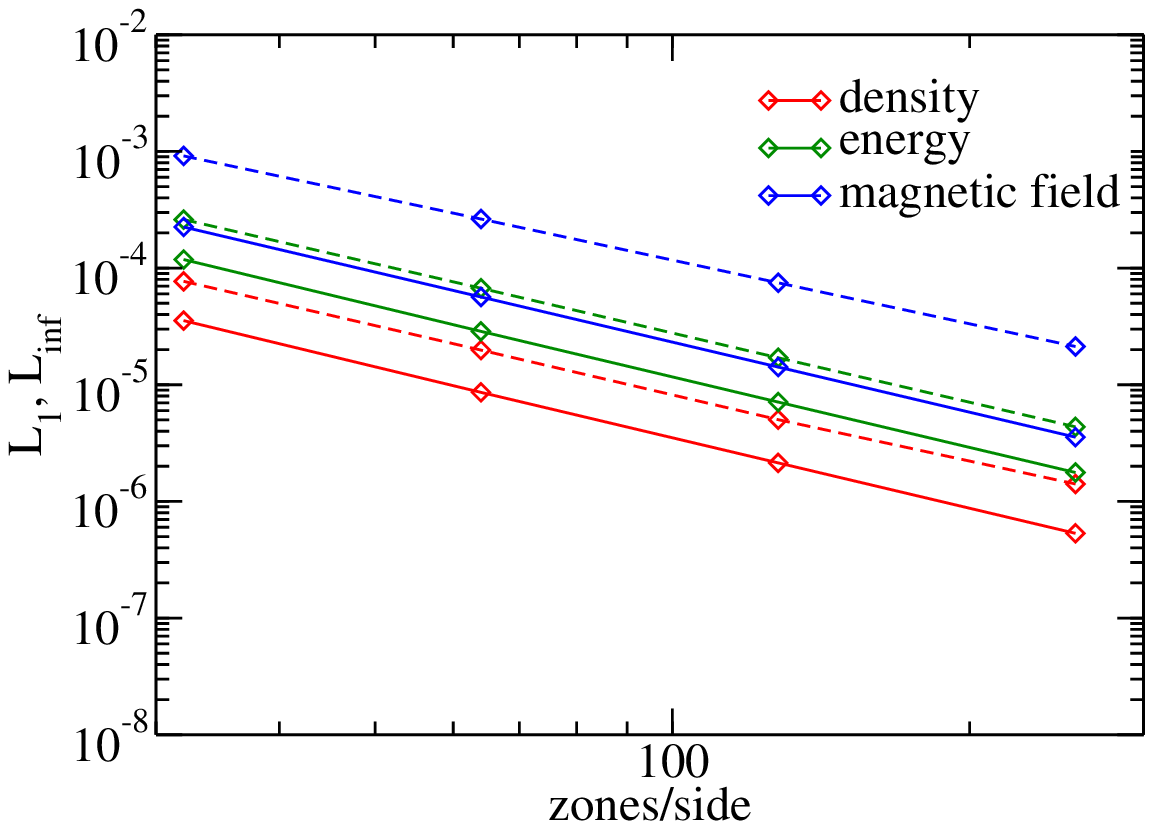}
\includegraphics[width=2.25in, clip=]{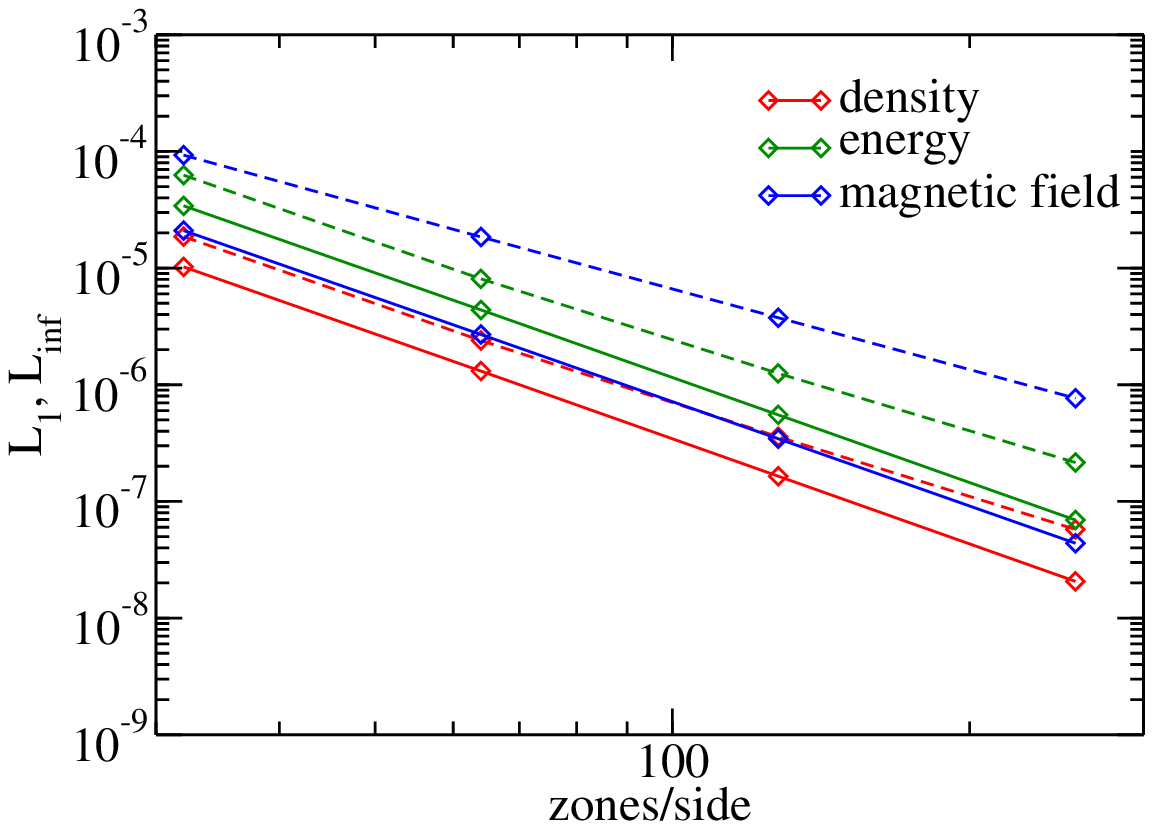}
\includegraphics[width=2.25in, clip=]{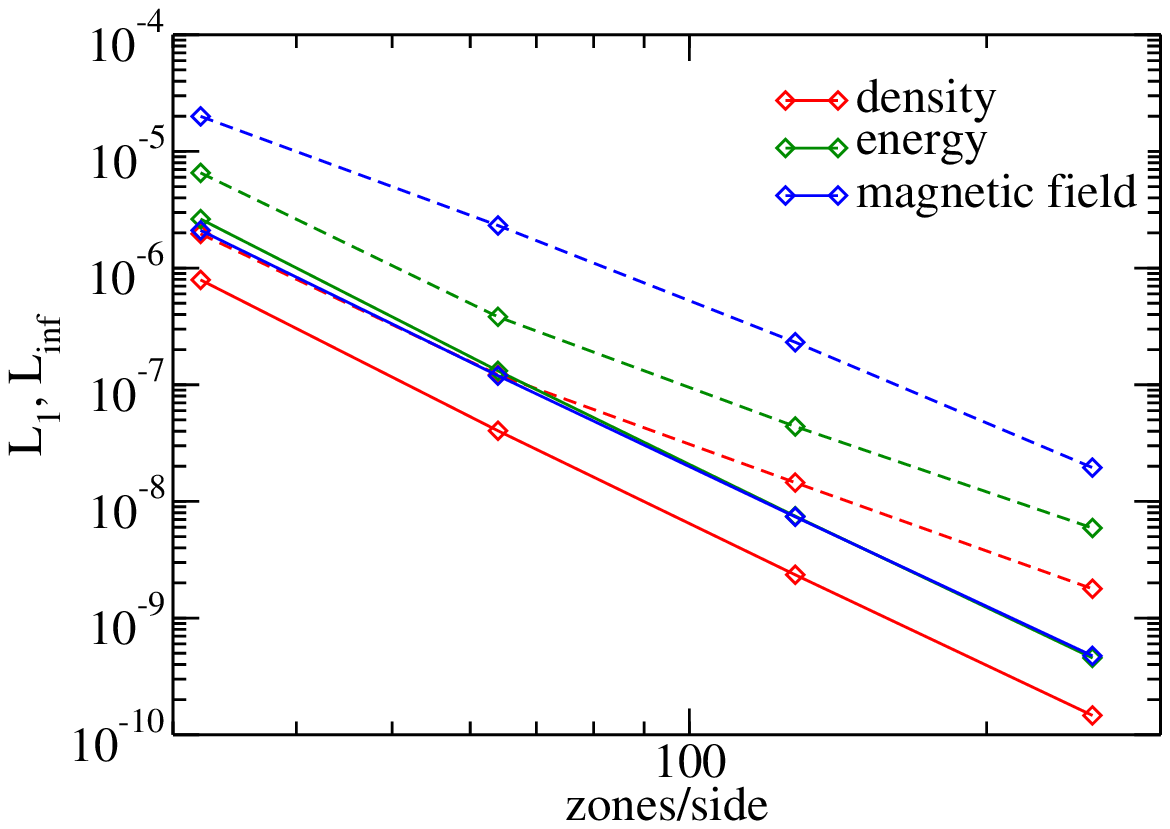}
\caption{$L_1$ (solid lines) and $L_\infty$ (dashed lines) norms of the error of density, total energy, and the $x$ component of the magnetic field for the manufactured solution on division 4--7 TGM. The three panels correspond to second through fourth order schemes.}
\label{fig_conv}
\end{figure}

\begin{figure}
\includegraphics[width=2.25in, clip=]{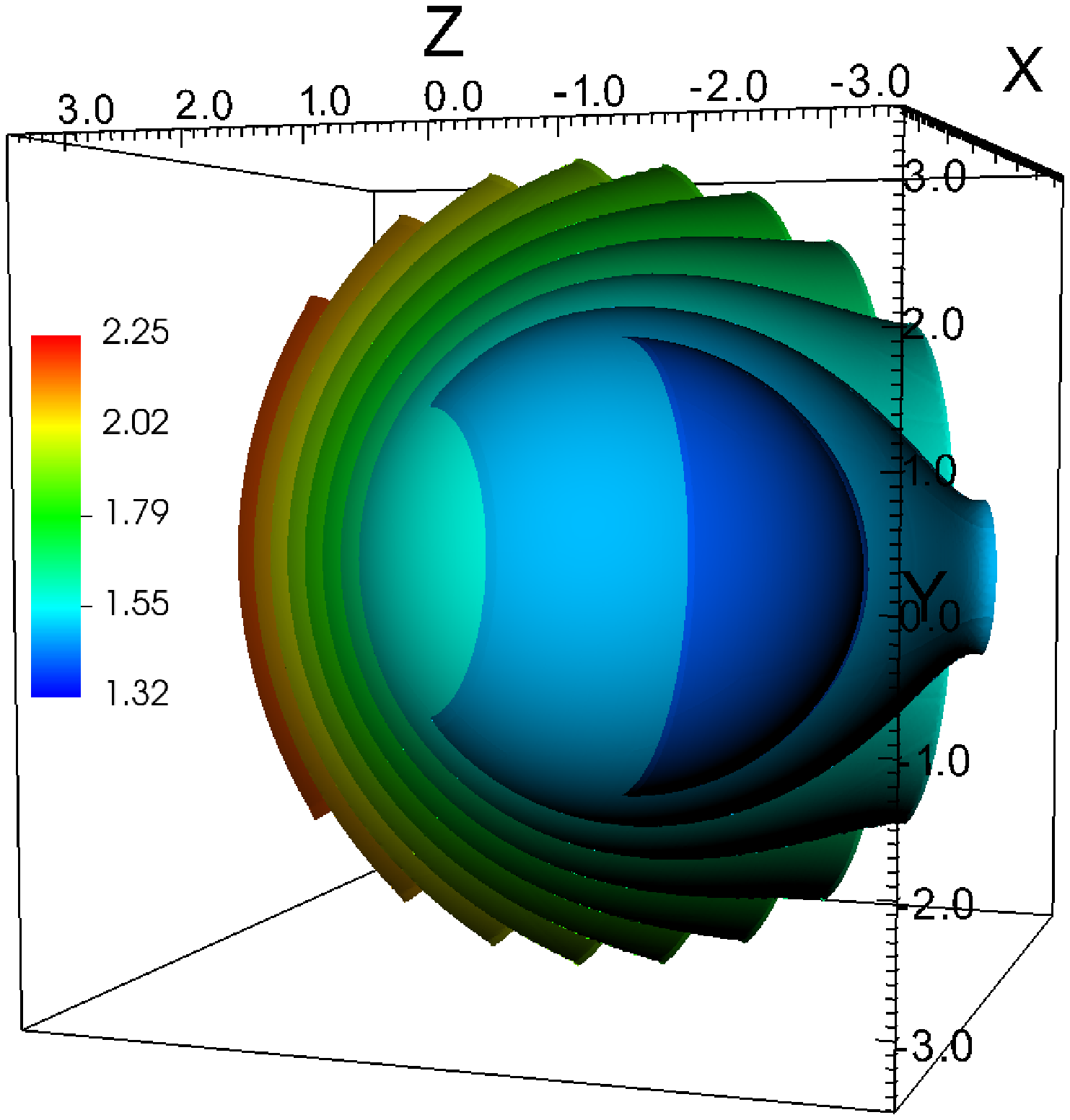}
\includegraphics[width=2.25in, clip=]{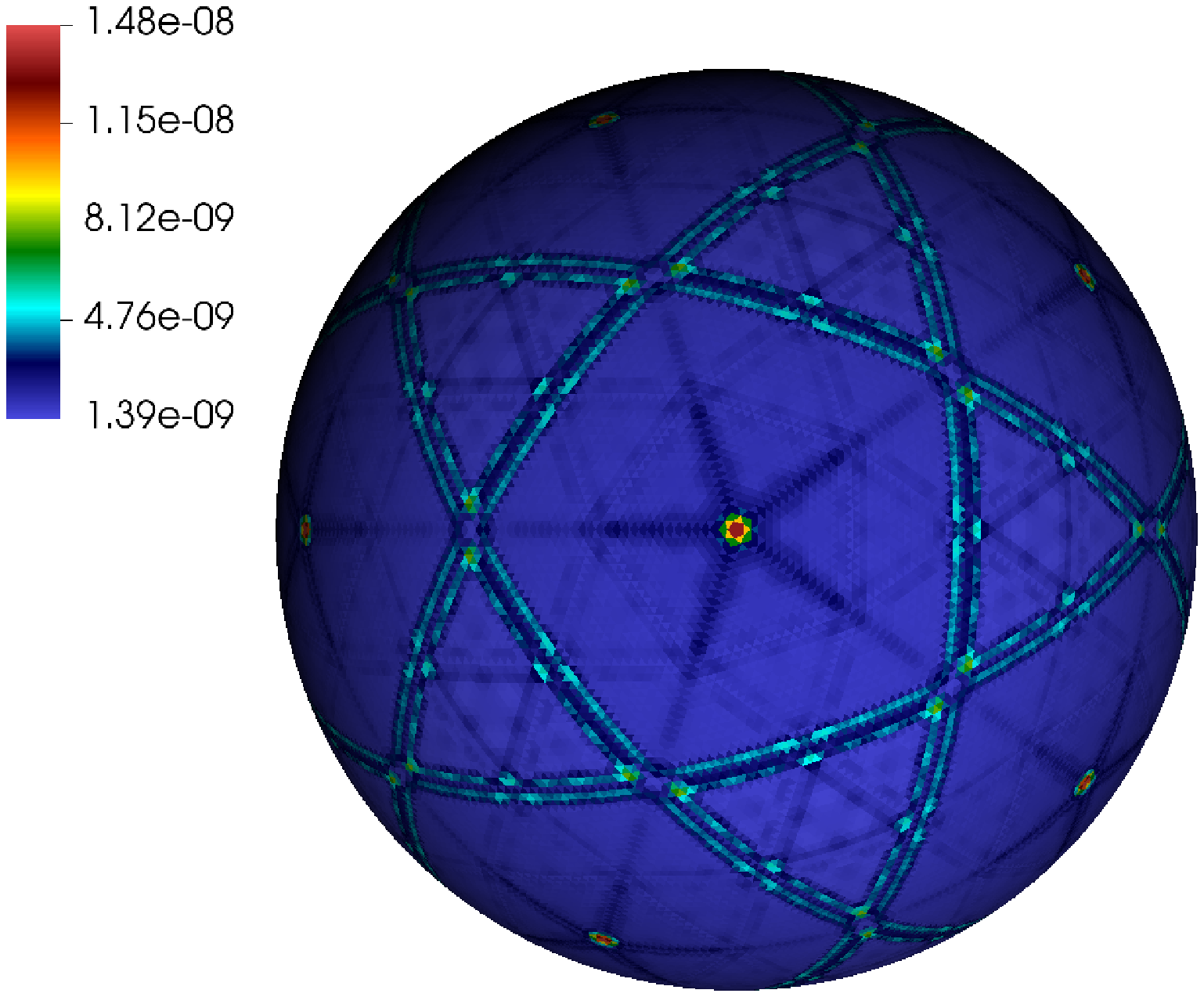}
\caption{Left: velocity magnitude isosurfaces from the manufactured solution problem. The solution has a resemblance to an interaction between a stellar wind an a uniform flow. A fourth order scheme was used on division 6 mesh with 64 radial shells. Right: density error distribution in a spherical layer at $r=2.75$.}
\label{fig_wind}
\end{figure}

\begin{figure}
\begin{center}
\includegraphics[width=3.5in, clip=]{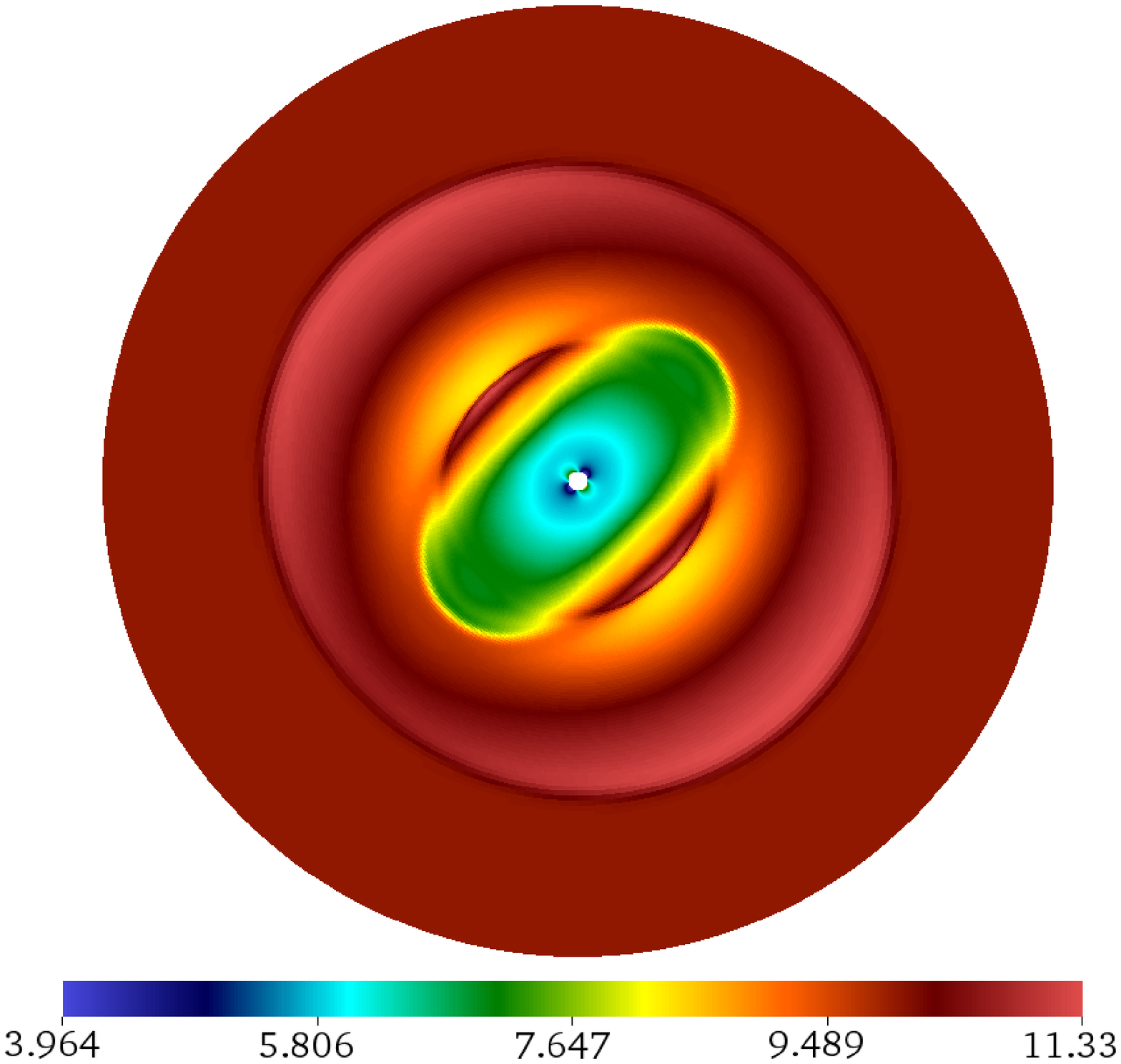}
\end{center}
\caption{Magnitude of the magnetic field from the solution to the blast wave problem at $t=0.07$. A linear color scale is used.}
\label{fig_blast}
\end{figure}

\begin{figure}
\begin{center}
\includegraphics[width=3.0in, clip=]{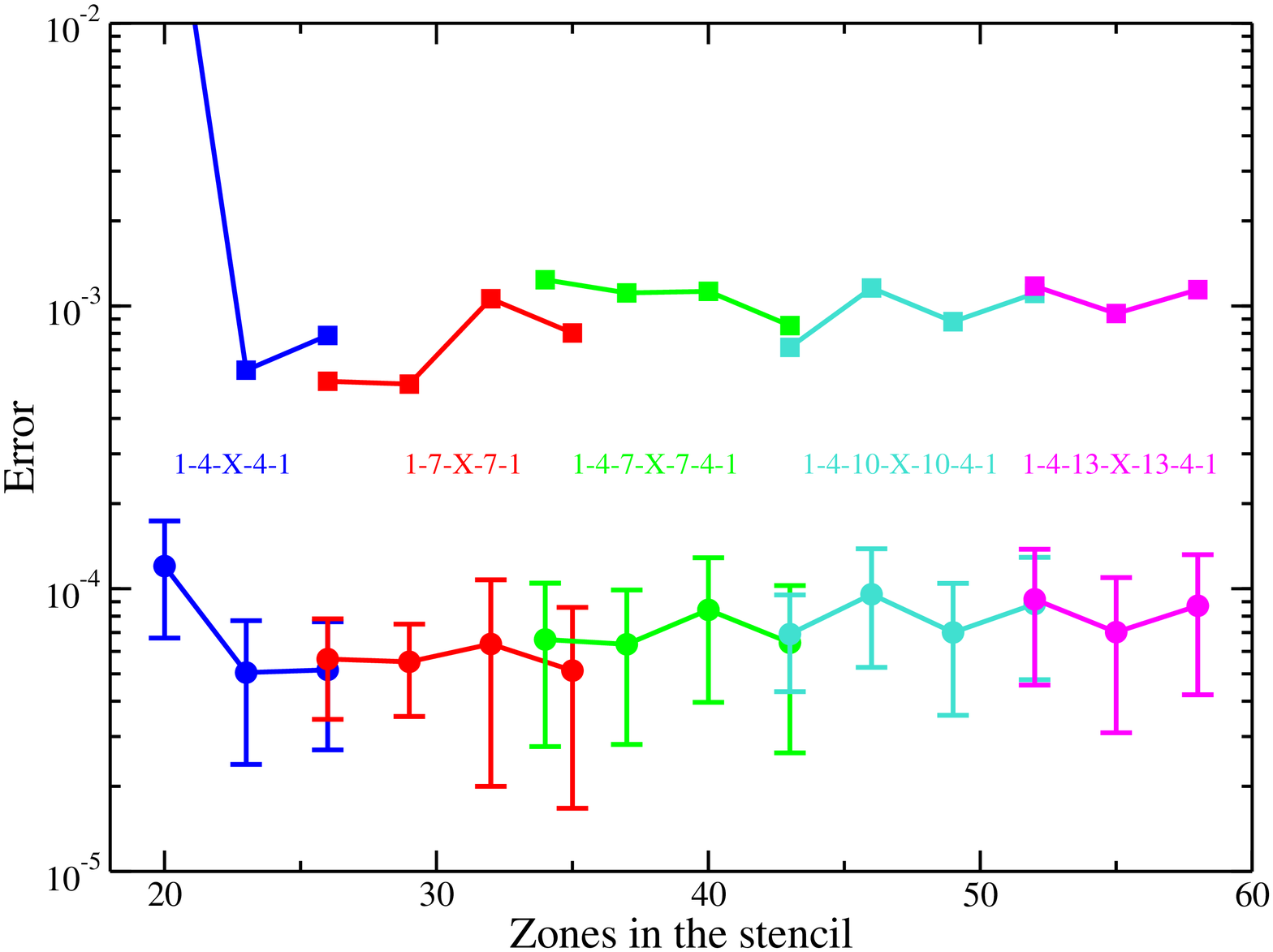}
\end{center}
\caption{Fourth order reconstruction error for different size stencils. Average of $L_1$ reconstruction error (circles) with standard deviation, shown as error bars, and the largest error, shown with square symbols, on a division 1 block with division 5 faces and 16 radial shells. Averaging was done over 1000 trials, each initialized with a random ensemble of ten waves with isotropically distributed wavevectors. Stencil configuration is color-coded; the value of X can be deduced by subtracting the sum of zones in all planes, save the principal plane, from the total number of zones on the horizontal axis.}
\label{fig_stencilcomp}
\end{figure}


\section*{Tables}
\label{sec_tab}

\begin{table}
\caption{Connectivity table construction methods}
\begin{tabular}{ccll}
\hline
Step & Table & Prerequisite    & Method of construction \\
\hline
  1  &   VV  & parent division & Based on numbering scheme \\
  2  &   EV  & VV              & Insert edge per VV entry with no duplicates  \\
  3  &   FV  & parent division & Based on numbering scheme \\
  4  &   VE  & EV              & Inverse of EV \\
  5  &   VF  & FV              & Inverse of FV \\
  6  &   EF  & EV, VF          & Match two faces sharing this edge's vertices \\
  7  &   FE  & EF              & Inverse of EF \\
  8  &   FF  & EF, FE          & Find the other face sharing each edge \\
\hline
\end{tabular}
\label{tab_connectivity}
\end{table}

\begin{table}
\caption{Triangular geodesic mesh properties at divisions 0--8.}
\begin{tabular}{cccccccccc}
\hline
Div &Vertices &  Edges  &  Faces  &     Avg.     &    Avg.      &       Average        & Edge   & Angle  & Area   \\
     &        &         &         &     edge     &    angle     &         area         & ratio  & ratio  & ratio  \\
\hline
  0  &     12 &      30 &      20 & $63.4^\circ$ & $72.0^\circ$ & $6.28\times 10^{-1}$ &  1.00  &  1.00  &  1.00  \\
  1  &     42 &     120 &      80 & $33.9^\circ$ & $63.0^\circ$ & $1.57\times 10^{-1}$ &  1.14  &  1.24  &  1.20  \\
  2  &    162 &     480 &     320 & $17.2^\circ$ & $60.8^\circ$ & $3.93\times 10^{-2}$ &  1.18  &  1.31  &  1.28  \\
  3  &    642 &    1920 &    1280 & $8.64^\circ$ & $60.2^\circ$ & $9.82\times 10^{-3}$ &  1.19  &  1.33  &  1.29  \\
  4  &   2562 &    7680 &    5120 & $4.33^\circ$ & $60.0^\circ$ & $2.45\times 10^{-3}$ &  1.19  &  1.33  &  1.30  \\
  5  &  10242 &   30720 &   20480 & $2.16^\circ$ & $60.0^\circ$ & $6.14\times 10^{-4}$ &  1.19  &  1.33  &  1.30  \\
  6  &  40962 &  122880 &   81920 & $1.08^\circ$ & $60.0^\circ$ & $1.53\times 10^{-4}$ &  1.19  &  1.33  &  1.30  \\
  7  & 163842 &  491520 &  327680 & $0.54^\circ$ & $60.0^\circ$ & $3.84\times 10^{-5}$ &  1.19  &  1.33  &  1.30  \\
  8  & 655362 & 1966080 & 1310720 & $0.27^\circ$ & $60.0^\circ$ & $1.16\times 10^{-5}$ &  1.19  &  1.33  &  1.30  \\
\hline
\end{tabular}
\label{tab_properties}
\end{table}

\begin{table}
\caption{The choice of the auxilliary variable $\delta$}
\begin{tabular}{c|ccc}
\hline
t-edge/r-face & $\Lambda_1$ & $\Lambda_2$ & $\Lambda_3$ \\
\hline
1 & 0 & $1-\delta$ & $\delta$ \\
2 & $\delta$ & 0 & $1-\delta$ \\
3 & $1-\delta$ & $\delta$ & 0 \\
\hline
\end{tabular}
\label{tab_aux}
\end{table}

\begin{table}
\caption{The number of unknowns (the ``Unknowns'' column) vs. the number of conditions of each type (C1 through C5) in magnetic field reconstruction.}
\begin{tabular}{c|cccccccc}
\hline
Order & $D(M)$ & $\bar{D}(M)$ & Unknowns & C1 & C2 & C3 & C4 & C5 \\
\hline
2 &  4 &  7 & 21 &  3 & 5 & 15 & 12 &  9 \\
3 & 10 & 16 & 48 &  9 & 5 & 24 & 30 & 18 \\
4 & 20 & 30 & 90 & 19 & 5 & 30 & 60 & 30 \\
\hline
\end{tabular}
\label{tab_numeq}
\end{table}

\begin{table}
\caption{Actual order of convergence for the manufactured solution problem using the nominally second order scheme. Density ($\rho$), total energy ($e$), and one component of magnetic field ($B_x$) are shown.}
\begin{tabular}{c|ccc|ccc|ccc}
\hline
       & \multicolumn{3}{c|}{$\rho$} & \multicolumn{3}{c|}{$e$} & \multicolumn{3}{c}{$B_x$} \\
division   & 5 & 6 & 7 & 5 & 6 & 7 & 5 & 6 & 7 \\
\hline
$L_1$      & 2.0 & 2.0 & 2.0 & 2.0 & 2.0 & 2.0 & 2.0 & 2.0 & 2.0 \\
$L_\infty$ & 2.0 & 2.0 & 1.8 & 2.0 & 2.0 & 2.0 & 1.8 & 1.8 & 1.8 \\
\hline
\end{tabular}
\label{tab_order2}
\end{table}
\begin{table}
\caption{Actual order of convergence for the manufactured solution problem using the nominally third order scheme. Density ($\rho$), total energy ($e$), and one component of magnetic field ($B_x$) are shown.}
\begin{tabular}{c|ccc|ccc|ccc}
\hline
       & \multicolumn{3}{c|}{$\rho$} & \multicolumn{3}{c|}{$e$} & \multicolumn{3}{c}{$B_x$} \\
division   & 5 & 6 & 7 & 5 & 6 & 7 & 5 & 6 & 7 \\
\hline
$L_1$      & 3.0 & 3.0 & 3.0 & 3.0 & 3.0 & 3.0 & 3.0 & 3.0 & 3.0 \\
$L_\infty$ & 3.0 & 2.7 & 2.6 & 3.0 & 2.7 & 2.5 & 2.3 & 2.3 & 2.3 \\
\hline
\end{tabular}
\label{tab_order3}
\end{table}

\begin{table}
\caption{Actual order of convergence for the manufactured solution problem using the nominally fourth order scheme. Density ($\rho$), total energy ($e$), and one component of magnetic field ($B_x$) are shown.}
\begin{tabular}{c|ccc|ccc|ccc}
\hline
       & \multicolumn{3}{c|}{$\rho$} & \multicolumn{3}{c|}{$e$} & \multicolumn{3}{c}{$B_x$} \\
division   & 5 & 6 & 7 & 5 & 6 & 7 & 5 & 6 & 7 \\
\hline
$L_1$      & 4.3 & 4.1 & 4.0 & 4.3 & 4.1 & 4.0 & 4.1 & 4.0 & 4.0 \\
$L_\infty$ & 4.0 & 3.1 & 3.0 & 4.1 & 3.1 & 2.9 & 3.1 & 3.3 & 3.6 \\
\hline
\end{tabular}
\label{tab_order4}
\end{table}

\end{backmatter}
\end{document}